\documentclass[twoside,english]{article}
\usepackage[T1]{fontenc}
\usepackage[latin9]{inputenc}
\usepackage{geometry}
\usepackage{color}
\usepackage{babel}
\usepackage{array}
\usepackage{verbatim}
\usepackage{prettyref}
\usepackage{rotfloat}
\usepackage{booktabs}
\usepackage{units}
\usepackage{textcomp}
\usepackage{multirow}
\usepackage{amsmath}
\usepackage{amssymb}
\usepackage{stmaryrd}
\usepackage{graphicx}
\usepackage[unicode=true,pdfusetitle,bookmarks=true,bookmarksnumbered=false,bookmarksopen=false breaklinks=false,pdfborder={0 0 1},backref=false,colorlinks=true]{hyperref}

\makeatletter

\newcommand{\noun}[1]{\textsc{#1}}
\providecommand{\tabularnewline}{\\}

\newenvironment{lyxlist}[1]
	{\begin{list}{}
		{\settowidth{\labelwidth}{#1}
		 \setlength{\leftmargin}{\labelwidth}
		 \addtolength{\leftmargin}{\labelsep}
		 }}
	{\end{list}}

\usepackage{libertine}
\usepackage{libertinust1math}
\usepackage[T1]{fontenc}

\newrefformat{fig}{Fig \hyperref[#1]{\ref{#1}}}
\newrefformat{tab}{Tab \hyperref[#1]{\ref{#1}}}
\newrefformat{eq}{Eq (\hyperref[#1]{\ref{#1}})}

\renewcommand\[{\begin{equation}}
\renewcommand\]{\end{equation}}

\@ifundefined{showcaptionsetup}{}{%
 \PassOptionsToPackage{caption=false}{subfig}}
\usepackage{subfig}
\makeatother

\begin{document}
\title{Analysis and comparisons of various models in cold spray simulations
: towards high fidelity simulations}
\author{Louis-Vincent Bouthier\thanks{Ecole Polytechnique, Route de Palaiseau, 91120 Palaiseau, France,
\protect \\
email: \texttt{louis.bouthier@polytechnique.edu}}, Elie Hachem\thanks{MINES ParisTech, PSL - Research University, CEMEF - Centre for material
forming, CNRS UMR 7635, CS 10207, rue Claude Daunesse, 06904 Sophia-Antipolis
Cedex, France}}
\maketitle
\begin{abstract}
Cold spray technology is a quickly growing manufacturing technology
which impacts lots of industries. Despite many years of studies about
the comprehension of the phenomena and the improvements of the performance
of the system, ensuring high fidelity simulations remains a challenge.
We propose in this work a detailed high fidelity modeling and simulations
giving more insight of the phenomena appearing in cold spray such
as turbulence, oblique shocks, bow shocks, fluctuations, particles
motion and particles impacts. It is mainly based on a richer model
known as the Detached Eddy Simulation (DES) model. Moreover, we present
several analysis of various existing models for both validations and
comparisons purposes. Finally, this high fidelity framework will allow
us to deal with a new configuration showing an improved performance
assessed with the previous models.
\end{abstract}

\section*{Notations}
\begin{lyxlist}{00.00.0000}
\item [{$\left\Vert \cdot\right\Vert $}] Norm of a vector
\item [{$\overline{\cdot}$}] Mean of a quantity in RANS
\item [{$C$}] Correction factor taking into account compressible effects
(-)
\item [{$C_{D}$}] Drag coefficient (-)
\item [{$c$}] Speed of sound ($\unit{m.s^{-1}}$)
\item [{$c_{p}$}] Mass thermal capacity of solid particles ($\unit{J.kg^{-1}.K^{-1}}$)
\item [{$D$}] Nozzle diameter ($\unit{m}$)
\item [{$d_{p}$}] Solid particle diameter ($\unit{m}$)
\item [{$\frac{\mathrm{d}}{\mathrm{d}t}$}] Particle derivative for time-
and space-dependent quantities ($\unit{s^{-1}}$)
\item [{$d_{w}$}] Distance to the nearest wall ($\unit{m}$)
\item [{$e$}] Internal energy per unit of mass of the gas ($\unit{J.kg^{-1}}$)
\item [{$\varepsilon$}] Turbulent dissipation rate ($\unit{m^{2}.s^{-3}}$)
\item [{$\eta$}] Ratio of the length of the stagnation chamber to the
length of the convergent (-)
\item [{$\gamma$}] Ratio of the specific heats (-)
\item [{$h_{\max}$}] Maximum edge length of a cell ($\unit{m}$)
\item [{$k$}] Turbulent kinetic energy ($\unit{m^{2}.s^{-2}}$)
\item [{$\kappa$}] Volume viscosity relative to expansion ($\unit{kg.m^{-1}.s^{-1}}$)
\item [{$L$}] Total nozzle length ($\unit{m}$)
\item [{$\lambda$}] Thermal conductivity of gas ($\unit{W.m^{-1}.K^{-1}}$)
\item [{$\lambda_{p}$}] Thermal conductivity of solid particles ($\unit{W.m^{-1}.K^{-1}}$)
\item [{$M$}] Mach number (-)
\item [{$\mu$}] Dynamic gas viscosity ($\unit{kg.m^{-1}.s^{-1}}$)
\item [{$\mu_{t}$}] Turbulent gas viscosity ($\unit{kg.m^{-1}.s^{-1}}$)
\item [{$\mathrm{Nu}$}] Nusselt number (-)
\item [{$\omega$}] Specific turbulent dissipation rate ($\unit{s^{-1}}$)
\item [{$\Omega$}] Magnitude of the vorticity tensor ($\unit{s}^{-1}$)
\item [{$p$}] Gas pressure ($\unit{Pa}$)
\item [{$p_{c}$}] Critical gas pressure ($\unit{Pa}$)
\item [{$\mathrm{Pr}$}] Prandtl number (-)
\item [{$r$}] Ideal gas specific constant ($\unit{J.kg^{-1}.K^{-1}})$
\item [{$R_{ij}$}] Reynolds tensor ($\unit{kg.m^{-1}.s^{-2}}$)
\item [{$\mathrm{Re}$}] Reynolds number (-)
\item [{$\rho$}] Density of gas ($\unit{kg.m^{-3}}$)
\item [{$\rho_{c}$}] Critical density of gas ($\unit{kg.m^{-3}}$)
\item [{$\rho_{p}$}] Density of solid particles ($\unit{kg.m^{-3}}$)
\item [{$S$}] Magnitude of the strain rate tensor ($\unit{s^{-1}}$)
\item [{$t$}] Time variable ($\unit{s}$)
\item [{$T$}] Gas temperature ($\unit{K}$)
\item [{$T_{c}$}] Critical gas temperature ($\unit{K}$)
\item [{$T_{p}$}] Particle temperature ($\unit{K}$)
\item [{$\boldsymbol{u}$}] Gas velocity ($\unit{m.s^{-1}}$)
\item [{$\boldsymbol{u_{p}}$}] Solid particle velocity ($\unit{m.s^{-1}}$)
\item [{$\upsilon$}] Acentric factor (-)
\item [{$\xi_{i}$}] Ratio of the inlet diameter to the throat diameter
(-)
\item [{$\xi_{o}$}] Ratio of the outlet diameter to the throat diameter
(-)
\item [{$\zeta$}] Ratio of the length of the divergent to the total length
of the nozzle (-)
\end{lyxlist}

\section{Introduction}

\subsection{About cold spray and its operating principles}

Cold spray is a manufacturing process that began at the end of the
20\textsuperscript{th} century. It consists of manufacturing objects
with a wide range of materials, from plastics to metals and more complex
materials. Its fields of application are vast and include start-ups
and fab-labs for prototyping, aeronautics, construction or the automotive
industry. The stakes of this technology are multiple because it allows
the design of complex geometry parts relatively quickly and often
at a lower cost. Nevertheless, limits appear with regard to the cost
of machines and materials, the surface state of the parts after the
operation or the internal state of the built structure in terms of
residual stresses, porosity, solidity and control of reproducibility.

In this work, we provide a rich overview on the cold spray technology
in terms of experiments, numerical simulations and challenges to tackle,
presented by the following sub sections. While it is focusing mainly
on high-pressure cold spray but the review remains rather general.

\begin{figure}
\begin{centering}
\includegraphics[width=8cm]{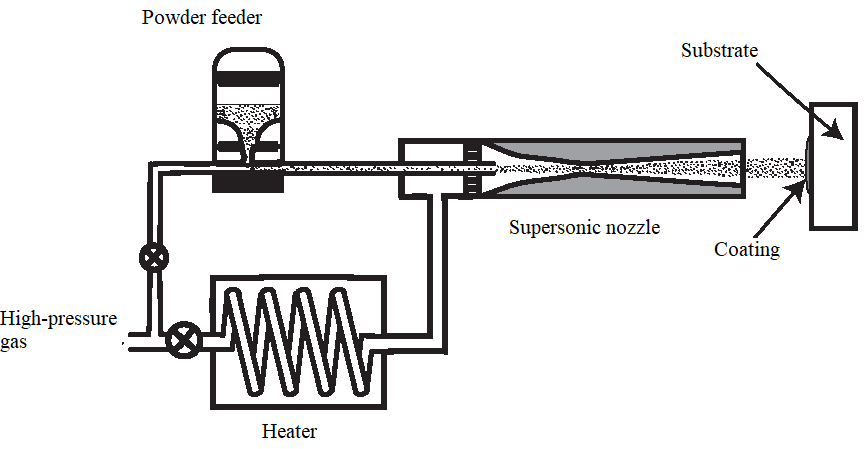}
\par\end{centering}
\caption{Diagram of a convergent-divergent channel De Laval (cf (Ref \cite{WikiColdspray}))\label{fig:Sch=0000E9ma-d'un-canal}}
\end{figure}

We recall first that the operating principle of dynamic cold spray
is relatively simple and has been studied since the 1980s in (Ref
\cite{Papyrin2007}). The system requires the use of a convergent-divergent
channel (also called De Laval) as shown in \prettyref{fig:Sch=0000E9ma-d'un-canal}
in order to reach supersonic velocities. The main gas, which can be
air, nitrogen or even helium, is introduced at the inlet of the channel
with a high pressure $p\approx\unit[50]{bar}$ and a high temperature
$T\approx\unit[800]{{^\circ}C}$ to prime the nozzle and obtain supersonic
flow. The inlet pressure and temperature conditions are highly dependent
on the nozzle geometry, the substrate to be coated, the particles
to be sprayed and the desired coating condition.

The previously introduced gas will be used to accelerate solid particles
with an average size $d_{p}\approx\unit[30]{\text{\textmu}m}$ made
of a metal alloy, ceramic or polymer (see (Ref \cite{Alhulaifi2012})).
Despite the variability in application conditions, most authors agree
that there is a critical particle impact velocity beyond which adhesion
to the substrate can occur (see (Ref \cite{Faizan_ur_rab,MacDonald2016,SinghalMurtazaParvej,yin2014,Liebersbach2020,Yin2016})).
The value of this critical velocity under experimental conditions
is still unknown and the previous authors each mention an empirical
model to estimate it as a function of the temperature at impact and
the characteristics of the material. Moreover, it is essential to
recall that the adhesion of particles to the substrate is achieved
in the solid state, in particular thanks to the strong plastic deformation
of the particles due to their impact velocity. In (Ref \cite{Leitz2017}),
for example, the authors simulated the impact and adhesion of a particle
by taking into account the data from their previous study using a
finite element calculation.

In summary, the goal of the whole system is to accelerate solid particles
using a supersonic gas flow to make them adhere to a substrate due
to their high impact velocity. The challenge is then to understand
what are the relevant parameters to consider in order to optimize
this process according to the specifications. We will establish a
work similar to the one performed in (Ref \cite{Yin2016}) to assess
the current knowledge.

\subsection{Supersonic flow and phenomena}

\subsubsection{Early theoretical relationships}

As mentioned above, a necessary issue is the establishment of supersonic
flow within the nozzle. In an approximate way, in (Ref \cite{Yin2016,Leitz2017,CFDEMmodelling}),
the authors presented the relationships in the De Laval nozzle in
the case of a perfect fluid flow allowing to have a first order of
magnitude of the parameters involved. Knowing that a certain velocity
is desired for the solid particles, it thus fixes a gas velocity inside
the nozzle. By the preceding relations, this velocity is simply given,
as a first approximation, by the ratio of the diameter of the nozzle
outlet to the diameter of the throat. Then, all the thermodynamic
quantities are deduced from this speed and thus make it possible to
know in an approximate way the behavior of the gas inside the nozzle.
This method, although effective, neglects phenomena such as turbulence
illustrated in (Ref \cite{raoelison2020}) or shocks. This is why
the majority of the authors used in parallel numerical simulation
tools to account more finely for the phenomena involved. However,
it is worth mentioning that a large part of the authors modeled gases
in supersonic flow by the law of ideal gas. The relevance of this
law in many fields is no longer to be proven, however, in the context
of cold spray, it can vacillate. In the context of high pressures
and high temperatures, so-called real state laws would undoubtedly
be better able to predict the phenomena precisely, knowing also that
many calculations are made numerically. Examples of such laws are
Van der Waals' state law (see (Ref \cite{Berthelot})), the state
law presented in (Ref \cite{RedlichKwong,Handbook}) and its version
modified in (Ref \cite{Aungier})\footnote{This law is notably available in \emph{Ansys' Fluent} software (see
(Ref \cite{FluentManual}))} or the state law presented in (Ref \cite{Handbook,BenedictWebbRubin}).

\subsubsection{Numerical processing of phenomena}

In fact, the numerical modeling of turbulence plays a major role in
all previous works dealing with the cold spray simulation. While some
approached the subject by laminar flow as in (Ref \cite{Leitz2017})
or chose to directly simulate the Navier-Stokes equations (i.e. DNS)
as in (Ref \cite{raoelison2020})\footnote{Even if the authors considered their simulation as DNS using \emph{Fluent},
the \emph{Ansys }developers do not encourage the customers to use
their softwares for DNS (see (Ref \cite{FluentManual})).}, many models were used such as the standard $k-\varepsilon$ model
in the Reynolds averaged equations (or RANS) with (Ref \cite{Alhulaifi2012,Rab2015}),
the realizable $k-\varepsilon$ model with (Ref \cite{SinghalMurtazaParvej}),
the Reynolds Stress model (see (Ref \cite{Yin2016})) or others
notably in (Ref \cite{Gerschenfeld}). These models have the advantage
over a DNS to reduce considerably the computation time: during a DNS,
the number of nodes of the mesh is proportional to $\mathrm{Re}^{9/4}$
and the simulation is necessarily unsteady and three-dimensional whereas
a turbulence model with RANS can be driven in two dimensions and in
steady state, which limits considerably the computation time. The
pieces of advice in (Ref \cite{Yin2016}) on comparing these turbulence
models for each application are therefore relevant.

However, the models are not limited to turbulence modeling alone.
The viscosity of the gas, for example, can be considered constant
or temperature dependent according to the law proposed in (Ref \cite{Sutherland})
and used in (Ref \cite{VaradaraajanMohanty}) or Maxwell's law
(see (Ref \cite[p.25]{Bird})).

An additional remark must be made concerning Stokes's hypothesis.
Indeed, this hypothesis imposing the nullity of the volume viscosity,
notably used in (Ref \cite{raoelison2020}), is relevant in the
context of incompressible flows, i.e. with a very small Mach number
in front of the unit and low density monoatomic gas where it is even
a property (see (Ref \cite{Handbook,Bird,WikiBulkviscosity})).
However, in (Ref \cite{Volumeviscosity,ShangWuWangYangYeHuTaoHe,GuUmbachs}),
the authors presented room temperature measurements of $\kappa/\mu$
for polyatomic gases such as nitrogen, used in cold spray. All these
measurements show that this ratio is often of the order of the unit
or even much higher in the case of dihydrogen and invite to question
the Stokes hypothesis in the case of gases at high temperatures and
high pressures used in cold spray.

Also, thermomechanical coupling is unavoidable in compressible flow
which adds more models to take into account all relevant phenomena.
In (Ref \cite{Gerschenfeld}), the author cited for example the
$k_{\theta}-\varepsilon_{\theta}$ model, the $\varepsilon_{t}$ model
of turbulent dissipation or the $R_{ij}-\varepsilon$ models with
the Reynolds tensor.

\subsection{Interaction with Solids}

The complexity of the flow along the nozzle is then a challenge but
one of the objectives of the cold spray technology is to accelerate
solid particles to adhere onto the substrate. Therefore, the gas flow
in the nozzle is not single-phase and the supply of solid particles
to be accelerated is not insignificant. The coupling between the particles
and the gas can be taken into account or not. In (Ref \cite{Li2005}),
the authors considered that the particles have no influence on the
flow of the fluid due to the very low mass proportion of particles
in the flow. On the contrary, in (Ref \cite{samareh2009}), the
authors set up a fully coupled model between compressible flow and
particle displacement.

\subsubsection{Drag Force}

This same displacement, calculated by numerical simulations, has variations
according to the numerical scheme used. Particle tracking can be carried
out according to the Lagrangian description (see (Ref \cite{Leitz2017}))
or according to the Eulerian description (see (Ref \cite{CFDEMmodelling})).
This displacement is also subject to an equation involving the compressible
drag force of the flow. From this force comes a large number of models
seeking to evaluate the drag coefficient. These models take into account,
for example, the fact that the particles under consideration may have
irregular shapes, that the Mach number has a dominant influence especially
at the location of the shock waves, or that the Reynolds number is
also relevant in the study. For an overview, in (Ref \cite{Liebersbach2020,Yin2016,Leitz2017,CFDEMmodelling,Li2005,Faizan_Ur_Rab2016,Ozdemir_Widener2017,Sova2018}),
each author presented a number of models for the drag coefficient
of a particle. In (Ref \cite{samareh2009}), the authors also proposed
to consider the effect of shock waves by adding an additional force
to a particle related to the local expansion of the fluid.

\subsubsection{Particle thermodynamics}

Second, since we are considering a cold spray process, thermal effects
within the particle are paramount. Almost all authors agreed to consider
the temperature within a particle as uniform. This fact is related
to the very low value of the Biot number which compares the exchange
coefficient with the flow and the internal conductivity of the particle.
However, despite this great simplification, the exchange coefficient
between the flow and a particle must be evaluated to account for the
heat exchanges between the two bodies. In contrast to drag coefficient
models, few exchange coefficient models are used by the authors and
the main one is the Ranz-Marshall model (see (Ref \cite{RanzMarshall}))
involving the Nusselt number, the Prandtl number and the Reynolds
number.

\subsubsection{Particle-nozzle wall interaction}

Furthermore, the interaction between the nozzle walls and the particles
is still unknown. In (Ref \cite{Liebersbach2020}), the authors
illustrated the effects of nozzle clogging which are strongly dependent
on boundary conditions, especially pressure. Together with the authors
of (Ref \cite{raoelison2020}), they suggested cooling the nozzle
walls to limit clogging. In (Ref \cite{MacDonald2016}), the authors
demonstrated experimentally that the material covering the nozzle
walls has a significant impact on the performance of cold gas spraying,
all else being equal. They interpreted this effect through variations
in the thermal conductivity of the wall material. In addition, many
authors assumed elastic rebound in three-dimensional simulations,
but, in (Ref \cite{Ozdemir_Widener2017}), the authors proposed
to use elastic-plastic impact restitution coefficients.

\subsection{Summary of the numerical treatment of the problem}

{\tiny{}}
\begin{sidewaystable}
\begin{centering}
{\tiny{}}%
\begin{tabular}{cccccccccccc}
\toprule 
{\tiny{}Source} & {\tiny{}Turbulence model} & {\tiny{}Type of simulation} & {\tiny{}Geometry} & {\tiny{}Particle-flow interaction} & {\tiny{}Thermal flow boundary condition} & {\tiny{}Gas model} & {\tiny{}Viscosity model} & {\tiny{}Order of the elements} & {\tiny{}Solver} & {\tiny{}Particle tracking scheme} & {\tiny{}Drag coefficient}\tabularnewline
\midrule
\midrule 
{\tiny{}(Ref \cite{Alhulaifi2012})} & {\tiny{}$k-\varepsilon$} & {\tiny{}-} & {\tiny{}2D} & {\tiny{}-} & {\tiny{}-} & {\tiny{}I} & {\tiny{}-} & {\tiny{}-} & {\tiny{}D} & {\tiny{}L} & {\tiny{}-}\tabularnewline
\midrule 
{\tiny{}(Ref \cite{Faizan_ur_rab})} & {\tiny{}$k-\varepsilon$} & {\tiny{}St} & {\tiny{}3D} & {\tiny{}2} & {\tiny{}A} & {\tiny{}-} & {\tiny{}-} & {\tiny{}-} & {\tiny{}D} & {\tiny{}L} & {\tiny{}SN}\tabularnewline
\midrule 
{\tiny{}(Ref \cite{MacDonald2016})} & {\tiny{}$k-\varepsilon$} & {\tiny{}St} & {\tiny{}2D} & {\tiny{}-} & {\tiny{}A} & {\tiny{}I} & {\tiny{}-} & {\tiny{}-} & {\tiny{}P} & {\tiny{}-} & {\tiny{}-}\tabularnewline
\midrule 
{\tiny{}(Ref \cite{SinghalMurtazaParvej})} & {\tiny{}$k-\varepsilon$/RSM} & {\tiny{}St} & {\tiny{}2D} & {\tiny{}-} & {\tiny{}A} & {\tiny{}I} & {\tiny{}C} & {\tiny{}-} & {\tiny{}P} & {\tiny{}L} & {\tiny{}-}\tabularnewline
\midrule 
{\tiny{}(Ref \cite{yin2014})} & {\tiny{}$k-\varepsilon$} & {\tiny{}St} & {\tiny{}2D} & {\tiny{}1} & {\tiny{}-} & {\tiny{}I} & {\tiny{}C} & {\tiny{}Q} & {\tiny{}D} & {\tiny{}L} & {\tiny{}-}\tabularnewline
\midrule 
{\tiny{}(Ref \cite{Liebersbach2020})} & {\tiny{}$k-\varepsilon$} & {\tiny{}T and St} & {\tiny{}2D} & {\tiny{}1} & {\tiny{}A} & {\tiny{}I} & {\tiny{}-} & {\tiny{}F} & {\tiny{}D} & {\tiny{}L} & {\tiny{}Cl}\tabularnewline
\midrule 
{\tiny{}(Ref \cite{Leitz2017})} & {\tiny{}Laminar} & {\tiny{}T} & {\tiny{}3D} & {\tiny{}-} & {\tiny{}-} & {\tiny{}-} & {\tiny{}C} & {\tiny{}-} & {\tiny{}P} & {\tiny{}L} & {\tiny{}SN/WY/NS/PM}\tabularnewline
\midrule 
{\tiny{}(Ref \cite{CFDEMmodelling})} & {\tiny{}Laminar} & {\tiny{}T} & {\tiny{}3D} & {\tiny{}2} & {\tiny{}-} & {\tiny{}I} & {\tiny{}C} & {\tiny{}-} & {\tiny{}P} & {\tiny{}E} & {\tiny{}SN/G/DF/B/KH}\tabularnewline
\midrule 
{\tiny{}(Ref \cite{raoelison2020})} & {\tiny{}Laminar} & {\tiny{}T} & {\tiny{}2D} & {\tiny{}-} & {\tiny{}-} & {\tiny{}I} & {\tiny{}Su} & {\tiny{}-} & {\tiny{}D} & {\tiny{}-} & {\tiny{}-}\tabularnewline
\midrule 
{\tiny{}(Ref \cite{VaradaraajanMohanty})} & {\tiny{}$k-\varepsilon$/RSM} & {\tiny{}St} & {\tiny{}3D} & {\tiny{}2} & {\tiny{}A} & {\tiny{}I} & {\tiny{}Su} & {\tiny{}Se/Q} & {\tiny{}P} & {\tiny{}L} & {\tiny{}-}\tabularnewline
\midrule 
{\tiny{}(Ref \cite{Li2005})} & {\tiny{}$k-\varepsilon$} & {\tiny{}St} & {\tiny{}2D} & {\tiny{}1} & {\tiny{}-} & {\tiny{}I} & {\tiny{}-} & {\tiny{}Se} & {\tiny{}-} & {\tiny{}L} & {\tiny{}M}\tabularnewline
\midrule 
{\tiny{}(Ref \cite{samareh2009})} & {\tiny{}RSM} & {\tiny{}-} & {\tiny{}2D} & {\tiny{}2} & {\tiny{}-} & {\tiny{}I} & {\tiny{}Su} & {\tiny{}Se:Q} & {\tiny{}P} & {\tiny{}L} & {\tiny{}Cr+G}\tabularnewline
\midrule 
{\tiny{}(Ref \cite{Ozdemir_Widener2017})} & {\tiny{}$k-\omega$} & {\tiny{}St} & {\tiny{}3D} & {\tiny{}1} & {\tiny{}-} & {\tiny{}-} & {\tiny{}-} & {\tiny{}-} & {\tiny{}-} & {\tiny{}L} & {\tiny{}Bi}\tabularnewline
\midrule 
{\tiny{}(Ref \cite{Sova2018})} & {\tiny{}$k-\varepsilon$} & {\tiny{}St} & {\tiny{}2D} & {\tiny{}1} & {\tiny{}-} & {\tiny{}I} & {\tiny{}-} & {\tiny{}-} & {\tiny{}P} & {\tiny{}L} & {\tiny{}H}\tabularnewline
\midrule 
{\tiny{}(Ref \cite{Yin2010})} & {\tiny{}$k-\varepsilon$} & {\tiny{}St} & {\tiny{}3D} & {\tiny{}1} & {\tiny{}A} & {\tiny{}I} & {\tiny{}-} & {\tiny{}Se} & {\tiny{}P} & {\tiny{}L} & {\tiny{}-}\tabularnewline
\midrule 
{\tiny{}(Ref \cite{Lupoi2011})} & {\tiny{}$k-\varepsilon$} & {\tiny{}St} & {\tiny{}2D} & {\tiny{}1} & {\tiny{}A} & {\tiny{}I} & {\tiny{}-} & {\tiny{}se} & {\tiny{}D} & {\tiny{}L} & {\tiny{}MC}\tabularnewline
\midrule 
{\tiny{}(Ref \cite{Samareh2008})} & {\tiny{}$k-\varepsilon$} & {\tiny{}-} & {\tiny{}2D} & {\tiny{}2} & {\tiny{}-} & {\tiny{}I} & {\tiny{}-} & {\tiny{}-} & {\tiny{}-} & {\tiny{}E} & {\tiny{}Cr}\tabularnewline
\midrule 
{\tiny{}(Ref \cite{Cui2019})} & {\tiny{}$k-\varepsilon$} & {\tiny{}-} & {\tiny{}3D} & {\tiny{}-} & {\tiny{}-} & {\tiny{}-} & {\tiny{}-} & {\tiny{}-} & {\tiny{}-} & {\tiny{}-} & {\tiny{}-}\tabularnewline
\midrule 
{\tiny{}(Ref \cite{HuangFukanuma})} & {\tiny{}$k-\varepsilon$} & {\tiny{}St} & {\tiny{}2D} & {\tiny{}-} & {\tiny{}A} & {\tiny{}I} & {\tiny{}-} & {\tiny{}-} & {\tiny{}-} & {\tiny{}L} & {\tiny{}-}\tabularnewline
\midrule
\midrule 
{\tiny{}Present work} & {\tiny{}DES} & {\tiny{}T} & {\tiny{}2D} & {\tiny{}2} & {\tiny{}Temperature fixed outside} & {\tiny{}Law in (Ref \cite{Aungier})} & {\tiny{}Su} & {\tiny{}Third order} & {\tiny{}D} & {\tiny{}L} & {\tiny{}MC+G+AM}\tabularnewline
\bottomrule
\end{tabular}{\tiny\par}
\par\end{centering}
{\tiny{}\caption{{\tiny{}List of the different nuemrical parameters used by the authors.
The \textquotedbl -\textquotedbl{} sign means the information was
not available in the article. ``RSM'' means Reynolds Stress Model,
``St'' means steady, ``T'' means transient, ``2'' means two-way
coupled, ``1'' means one-way coupled, ``A'' means adiabatic, ``I''
means ideal gas law, ``C'' means constant, ``Su'' means Sutherland's
law (Ref \cite{Sutherland}), ``F'' means first order elements,
``Se'' means second order elements, ``Q'' means QUICK (a discretization
scheme described in (Ref \cite{FluentManual})), ``P'' means
pressure-based solver, ``D'' means density-based solver, ``L''
means Lagrangian, ``E'' means Eulerian, ``DES'' means Detached
Eddy Simulation, ``SN'' means Schiller-Naumann model (see Ref \cite{CFDEMmodelling}),
``G'' means Gidaspow model (see (Ref \cite{CFDEMmodelling})),
``DF'' means Di Felice model (see (Ref \cite{CFDEMmodelling})),
``B'' means Beestra model (see (Ref \cite{CFDEMmodelling})),
``KH'' means Koch-Hill model (see (Ref \cite{CFDEMmodelling})),
``WY'' means Wen-Yu model (see (Ref \cite{Leitz2017})), ``NS''
means Non-Sphere model (see (Ref \cite{Leitz2017})), ``PM''
means Du Plessis and Masliyah model (see (Ref \cite{PlessisMasliyah})),
``Cr'' means Crowe model (see (Ref \cite{Crowe})), ``Cl''
means Clift et al. model (see (Ref \cite{CliftGraceWeber})), ``M''
means Morsi and Alexander model (see (Ref \cite{MorsiAlexander})),
``MC'' means Morsi and Alexander model with compressibility correction
(see (Ref \cite{CliftGraceWeber})), ``Bi'' means Bird et al.
model (see (Ref \cite{Bird})), ``G'' means gradient force, ``H''
means Henderson model (see (Ref \cite{Yin2016})), ``AD'' means
added mass.\label{tab:Liste-des-diff=0000E9rents-1}}}
}{\tiny\par}
\end{sidewaystable}
{\tiny\par}

We present in \prettyref{tab:Liste-des-diff=0000E9rents-1} a new
and an interesting summary on all the numerical details used until
now in the literature. We can notice some similarities in the previous
works. We have added our approach in the last row to clearly highlights
the novelty proposed in this work. Another comment must be made about
the drag coefficients. Indeed, except for the Morsi and Alexander
model with compressibility correction, the Crowe model and the Henderson
model (see (Ref \cite{Yin2016})), none of the previous drag coefficients
takes into account compressible effects which could be not relevant
in the shocks for example.

\subsection{Parametric study of the nozzle geometry}

\begin{table}
\begin{centering}
\begin{tabular}{ccccccc}
\toprule 
Source & $L$ (mm) & $D$ (mm) & $\xi_{o}$ & $\xi_{i}$ & $\zeta$ & Shape\tabularnewline
\midrule
\midrule 
(Ref \cite{Alhulaifi2012}) & $138.12$ & $2.66$ & $2.37$ & $3.68$ & $0.968$ & Linear\tabularnewline
\midrule 
(Ref \cite{Faizan_ur_rab}) & $121.5$ & $2.7$ & $3.15$ & - & $0.579$ & Linear\tabularnewline
\midrule 
(Ref \cite{MacDonald2016}) & $125.73$ & $3.73$ & $1.70$ & - & - & Piecewise constant\tabularnewline
\midrule 
(Ref \cite{SinghalMurtazaParvej}) & $224$ & $2.7$ & $2.22$ & $6.74$ & $0.759$ & Linear\tabularnewline
\midrule 
(Ref \cite{yin2014}) & $172.4$ & $2.7$ & $2.37$ & $6.74$ & $0.696$ & Linear\tabularnewline
\midrule 
(Ref \cite{Liebersbach2020}) & $186.1$ & $1.73$ & $2.94$ & $4.41$ & $0.8190$ & Linear\tabularnewline
\midrule 
(Ref \cite{CFDEMmodelling}) & $140$ & $2.7$ & $2.37$ & $\approx1.5$ & $0.929$ & Linear\tabularnewline
\midrule 
(Ref \cite{raoelison2020}) & $137$ & $2.55$ & $1.88$ & $3.14$ & $0.956$ & Piecewise constant\tabularnewline
\midrule 
(Ref \cite{Li2005}) & $50$ & $2$ & 1;2;3;4 & $4$ & $0.8$ & Linear\tabularnewline
\midrule 
\multirow{2}{*}{(Ref \cite{samareh2009})} & $69.9$ & $2.7$ & $3.10$ & $5.19$ & - & Linear\tabularnewline
\cmidrule{2-7} \cmidrule{3-7} \cmidrule{4-7} \cmidrule{5-7} \cmidrule{6-7} \cmidrule{7-7} 
 & $130.3$ & $2.7$ & $2.43$ & $5.19$ & - & Linear\tabularnewline
\midrule
(Ref \cite{Ozdemir_Widener2017}) & $120$ & $2$ & $2.0$ & $5.0$ & $0.933$ & Linear\tabularnewline
\midrule 
(Ref \cite{Sova2018}) & $20.0$ & $0.5$ & $2.0$ & - & - & Linear\tabularnewline
\midrule 
(Ref \cite{Yin2010}) & $70$ & $2.0$ & $2.0$ & $5.0$ & $0.571$ & Linear\tabularnewline
\midrule 
\multirow{4}{*}{(Ref \cite{Lupoi2011})} & $210$ & $2$ & $3.0$ & $11$ & $0.857$ & Linear\tabularnewline
\cmidrule{2-7} \cmidrule{3-7} \cmidrule{4-7} \cmidrule{5-7} \cmidrule{6-7} \cmidrule{7-7} 
 & $115$ & $2.7$ & $3.0$ & $6.67$ & $0.565$ & Linear\tabularnewline
\cmidrule{2-7} \cmidrule{3-7} \cmidrule{4-7} \cmidrule{5-7} \cmidrule{6-7} \cmidrule{7-7} 
 & $87$ & $4$ & $1.15$ & $5.0$ & $0.805$ & Linear\tabularnewline
\cmidrule{2-7} \cmidrule{3-7} \cmidrule{4-7} \cmidrule{5-7} \cmidrule{6-7} \cmidrule{7-7} 
 & $35$ & $2.2$ & $1.73$ & $9.09$ & $0.571$ & Linear\tabularnewline
\midrule 
(Ref \cite{Cui2019}) & $121.5$ & $2.7$ & $3.07$ & $3.07$ & $0.579$ & Linear\tabularnewline
\bottomrule
\end{tabular}
\par\end{centering}
\caption{List of the different nozzle models proposed by the authors. The \textquotedbl -\textquotedbl{}
sign means the information was not available in the article. When
$\zeta$ is not known, the length $L$ given is the length of the
divergent. The \textquotedbl linear\textquotedbl{} shape means that
the diameter changes linearly between the inlet and the neck and between
the neck and the outlet. The form \textquotedbl piecewise constant\textquotedbl{}
means that the diameter evolves by sections over which it is constant.
\label{tab:Liste-des-diff=0000E9rents}}
\end{table}

After all the phenomena being presented, we recall that many researchers
and engineers aimed at optimizing the nozzles and application conditions
to obtain the best performing coating. For example, in (Ref \cite{Yin2010}),
the authors focused on the impact of the spray angle, in (Ref \cite{yin2014}),
the authors studied the effect of the pressure with which the particles
were injected into the flow and, in (Ref \cite{Sova2018}), the
authors wanted to build a nozzle of reduced size in order to obtain
refined coating grooves. Empiricism and parametric studies dominate
the papers. In (Ref \cite{Sova2018}), the authors mentioned a
totally empirical rule of thumb for constructing a cold spray nozzle
with respect to Mach number and the ratio of divergent length to nozzle
outlet diameter. In order to give an overview of the multiplicity
of models, the \prettyref{tab:Liste-des-diff=0000E9rents} gathers
parameters characterizing the cold spray nozzles. A remark can be
made about the geometries mentioned in the \prettyref{tab:Liste-des-diff=0000E9rents}:
even if $\xi_{i}$, $\zeta$ and $L$ seem quite variable, $D$ and
$\xi_{o}$ are relatively stable and suggest either a tacit agreement
of the authors on the experimental conditions of application of cold
spray, or a lack of investigation of different geometries. Also, the
dominant shape of the nozzles is a linear one. The real added value
for the application of cold spray still seems to be unknown.

\subsection{Contents of this paper}

The main objective of this paper is to reach a new step in numerical
simulations comparing its fidelity with previous models. This kind
of comparison has never been done before in the field which adds originality
to this work. Even if the whole problem remains complex, it is now
possible to advance further into the comprehension of the problem.
The paper presents first a commonly used model to validate the approach
before increasing the fidelity with a new high fidelity model. This
gain in fidelity will also allow to present a new type of configuration
with an increased performance. Hence, the paper shows, in \prettyref{sec:Models,-geometries,-boundary},
the models with the hypothesis, the geometries, the boundary conditions
and the CFD conditions. Then, in \prettyref{sec:Numerical-results-and},
the results of the two previous models will be discussed. Finally,
in \prettyref{sec:Conclusion}, a conclusion will be drawn according
to which has been presented.

\section{Models, geometries, boundary conditions and CFD conditions}

\label{sec:Models,-geometries,-boundary}

As mentioned in the previous section, we propose two approaches. The
first is related to the common modeling of cold spray, referred as
the RANS model. Detailed analysis and comparisons with (Ref \cite{Lupoi2011})
will be presented for validations. Then, we will provide a more complex
model for the high fidelity simulations known as the Improved Delayed
Detached Eddy Simulation, and referred here as the IDDES model.

\subsection{Models}

\subsubsection{Theoretical model}

\label{subsec:Theoretical-model}

Assuming the gas as an ideal gas, neglecting any dissipation, i.e.
$\lambda=0$, $\mu=0$, $\kappa=0$, and assuming an isentropic flow
with a one dimension hypothesis, according to (Ref \cite{Leitz2017,CFDEMmodelling,Handbook,cours}),
we can write 
\begin{equation}
1+\frac{\gamma-1}{2}M^{2}=\frac{T_{0}}{T}=\left(\frac{p_{0}}{p}\right)^{\frac{\gamma-1}{\gamma}}=\left(\frac{\rho_{0}}{\rho}\right)^{\gamma-1},\quad M=\frac{\left\Vert \boldsymbol{u}\right\Vert }{\sqrt{\gamma rT}}\label{eq:Theory}
\end{equation}
which are called the Barr� de Saint-Venant's relations. The $0$ refers
to the values imposed at the beginning of the nozzle in the chamber
of stagnation. Hence, all the quantities are depending on the Mach
number $M$ which is determined by the following relation, 
\begin{align}
\left(\frac{D}{D_{t}}\right)^{4} & =\frac{1}{M^{2}}\left(\frac{2}{\gamma+1}\left(1+\frac{\gamma-1}{2}M^{2}\right)\right)^{\frac{\gamma+1}{\gamma-1}}\label{eq:saintvenat}
\end{align}
where $D_{t}$ is the diameter of the throat. Therefore, $M<1$ before
the throat and $M>1$ after the throat. The \prettyref{eq:Theory}
and the \prettyref{eq:saintvenat} will be used afterwards to compare
and to discuss the predictions of both numerical models.

\subsubsection{General hypothesis}

We consider an axisymmetric modeling of the nozzle neglecting gravity.
The gas considered is nitrogen and the Stokes hypothesis is applied,
i.e. $\kappa=0$. The dynamic viscosity $\mu$ evolves according to
(Ref \cite{Sutherland}) to agree with other papers (see \prettyref{eq:sutherland}),
\begin{equation}
\mu=\mu_{\mathrm{ref}}\left(\frac{T}{T_{\mathrm{ref}}}\right)^{3/2}\frac{T_{\mathrm{ref}}+T_{S}}{T+T_{S}}\label{eq:sutherland}
\end{equation}
where $\mu_{\mathrm{ref}}=\unit[1.663\times10^{-5}]{kg.m^{-1}.s^{-1}}$,
$T_{\mathrm{ref}}=\unit[273.11]{K}$ and $T_{S}=\unit[106.67]{K}$
the Sutherland\noun{ }temperature of the model. From a purely numerical
point of view, the resolution method used is a density-based method
which, according to (Ref \cite{Lupoi2011}), is supposed to approach
compressible phenomena such as shocks better than the pressure-based
resolution method. On each wall, the no-slip boundary condition is
applied. The particles are tracked according to a Lagrangian description
because of their low volume fraction in the flow. Their motion is
tracked with a stochastic walk to account for turbulence effects.
The temperature of the particles is also governed by \prettyref{eq:temperatureparticle},
\begin{equation}
\frac{\mathrm{d}T_{p}}{\mathrm{d}t}=\frac{6\lambda}{\rho_{p}c_{p}d_{p}^{2}}\left(T-T_{p}\right)\mathrm{Nu}\label{eq:temperatureparticle}
\end{equation}
with 
\begin{equation}
\mathrm{Nu}=2.0+0.6\mathrm{Re}_{p}^{1/2}\mathrm{Pr}^{1/3},\quad\mathrm{Re}_{p}=\frac{\rho d_{p}\left\Vert \boldsymbol{u}-\boldsymbol{u_{p}}\right\Vert }{\mu}
\end{equation}
according to (Ref \cite{RanzMarshall}). The size of the particles
is uniform with $d_{p}=\unit[20]{\text{\textmu}m}$. These particles
are made of copper.

\subsubsection{RANS model}

We consider the stationary RANS (see (Ref \cite[p.12-9]{Handbook,FluentManual}))
to model the fluid flow given by \prettyref{eq:RANScontinuity} to
\prettyref{eq:RANSenergy}, 
\begin{equation}
\frac{\mathrm{d}\overline{\rho}}{\mathrm{d}t}+\overline{\rho}\frac{\partial\overline{u_{i}}}{\partial x_{i}}=0\label{eq:RANScontinuity}
\end{equation}
\[
\overline{\rho}\frac{\mathrm{d}\overline{u_{i}}}{\mathrm{d}t}=-\frac{\partial\overline{p}}{\partial x_{i}}+\frac{\partial}{\partial x_{j}}\left(\mu\left(\frac{\partial\overline{u_{i}}}{\partial x_{j}}+\frac{\partial\overline{u_{j}}}{\partial x_{i}}-\frac{2}{3}\frac{\partial\overline{u_{k}}}{\partial x_{k}}\delta_{ij}\right)+R_{ij}\right),
\]
\begin{equation}
\overline{\rho}\frac{\mathrm{d}\overline{e}}{\mathrm{d}t}=-\overline{u_{i}}\frac{\partial\overline{p}}{\partial x_{i}}+\frac{\partial}{\partial x_{i}}\left(\lambda_{\mathrm{eff}}\frac{\partial\overline{T}}{\partial x_{i}}\right)+\frac{\partial}{\partial x_{i}}\left(\left(\mu+\mu_{t}\right)\left(\frac{\partial\overline{u_{i}}}{\partial x_{j}}+\frac{\partial\overline{u_{j}}}{\partial x_{i}}-\frac{2}{3}\frac{\partial\overline{u_{k}}}{\partial x_{k}}\delta_{ij}\right)\overline{u_{j}}\right),\label{eq:RANSenergy}
\end{equation}
where 
\[
R_{ij}=\mu_{t}\left(\frac{\partial\overline{u_{i}}}{\partial x_{j}}+\frac{\partial\overline{u_{j}}}{\partial x_{i}}-\frac{2}{3}\frac{\partial\overline{u_{l}}}{\partial x_{l}}\delta_{ij}\right)-\frac{2}{3}\overline{\rho}k\frac{\partial\overline{u_{l}}}{\partial x_{l}}\delta_{ij}
\]
according to Boussinesq's hypothesis (see (Ref \cite[p.12-9]{Handbook}\noun{))}
and 
\begin{equation}
\lambda_{\mathrm{eff}}=\lambda+\frac{r\gamma\mu_{t}}{\mathrm{Pr}_{t}\left(\gamma-1\right)}.
\end{equation}
The turbulence model used is the realizable $k-\varepsilon$ model
which can be achieved by taking into account the effects of compressibility
and modeling the gas as an ideal gas with \prettyref{eq:idealgas},
\begin{equation}
p=\rho rT.\label{eq:idealgas}
\end{equation}
 The turbulence model is presented in \prettyref{eq:Turbulentviscosity}
to \prettyref{eq:Lastk-e}, 
\begin{equation}
\mu_{t}=C_{\mu}\overline{\rho}\frac{k^{2}}{\varepsilon},\label{eq:Turbulentviscosity}
\end{equation}
\begin{equation}
\overline{\rho}\frac{\mathrm{d}k}{\mathrm{d}t}=\frac{\partial}{\partial x_{i}}\left(\left(\mu+\frac{\mu_{t}}{\sigma_{k}}\right)\frac{\partial k}{\partial x_{i}}\right)+\mu_{t}\tilde{S}^{2}-\overline{\rho}\varepsilon-\frac{2\overline{\rho}\varepsilon k}{\gamma rT},
\end{equation}
\begin{equation}
\overline{\rho}\frac{\mathrm{d}\varepsilon}{\mathrm{d}t}=\frac{\partial}{\partial x_{i}}\left(\left(\mu+\frac{\mu_{t}}{\sigma_{\varepsilon}}\right)\frac{\partial\varepsilon}{\partial x_{i}}\right)+\overline{\rho}C_{1}\tilde{S}\varepsilon-\frac{\overline{\rho}C_{2}\varepsilon^{2}}{k+\sqrt{\mu\varepsilon/\overline{\rho}}},
\end{equation}

\begin{gather}
\mathrm{Pr}_{t}=0.85,\quad C_{1}=\max\left(0.43,\frac{\tau}{\tau+5}\right),\quad\tau=\tilde{S}\frac{k}{\varepsilon},\\
C_{\mu}=\left(A_{0}+A_{s}\frac{kU^{*}}{\varepsilon}\right)^{-1},\quad U^{*}=\sqrt{S^{2}+\Omega^{2}},\\
A_{0}=4.04,\quad A_{s}=\sqrt{6}\cos\left(\frac{1}{3}\arccos\left(\sqrt{6}W\right)\right),\\
W=\frac{S_{ij}S_{jk}S_{ki}}{S^{3}},\quad\tilde{S}=\sqrt{2}S,\quad S_{ij}=\frac{1}{2}\left(\partial_{j}\overline{u_{i}}+\partial_{i}\overline{u_{j}}\right),\\
C_{2}=1.9,\quad\sigma_{k}=1.0,\quad\sigma_{\varepsilon}=1.2.\label{eq:Lastk-e}
\end{gather}
Heat transfers between the gas and the nozzle walls are neglected
as well as heat transfers with the substrate. Also, the quantities
sought are calculated with second-order elements. The particles do
not have any influence on the flow and their motion is governed by
\prettyref{eq:particlemotion},
\begin{equation}
\frac{\mathrm{d}\boldsymbol{u_{p}}}{\mathrm{d}t}=\frac{3\rho}{4\rho_{p}d_{p}}\left\Vert \boldsymbol{u}-\boldsymbol{u_{p}}\right\Vert \left(\boldsymbol{u}-\boldsymbol{u_{p}}\right)C_{D}\label{eq:particlemotion}
\end{equation}
where 
\begin{equation}
C_{D}=\left(a_{1}+\frac{a_{2}}{\mathrm{Re}_{p}}+\frac{a_{3}}{\mathrm{Re}_{p}^{2}}\right)C^{-1}\label{eq:dragcoefficient}
\end{equation}
according to (Ref \cite{CliftGraceWeber,MorsiAlexander}) taking
into account the compressible effects with 
\begin{equation}
C=1+\sqrt{\frac{\pi\gamma}{2}}\frac{M_{p}}{\mathrm{Re}_{p}}\left(2.514+0.8\exp\left(-0.55\sqrt{\frac{2}{\pi\gamma}}\frac{\mathrm{Re}_{p}}{M_{p}}\right)\right),
\end{equation}
\begin{equation}
M_{p}=\frac{\left\Vert \boldsymbol{u}-\boldsymbol{u_{p}}\right\Vert }{c},\quad c=\sqrt{\gamma\left(\frac{\partial p}{\partial\rho}\right)_{T}}=\sqrt{\gamma rT}\label{eq:Machspeedofsound}
\end{equation}
with \prettyref{eq:idealgas}, 
\begin{equation}
\gamma=1-\frac{T}{\rho^{2}}\left(\frac{\partial e}{\partial T}\right)_{\rho}^{-1}\left(\frac{\partial p}{\partial T}\right)_{\rho}\left(\frac{\partial\rho}{\partial T}\right)_{p}=\frac{7}{5}
\end{equation}
with \prettyref{eq:idealgas} for nitrogen and $\left(a_{i}\right)_{i\in\left\llbracket 1,3\right\rrbracket }$
given by \prettyref{tab:Valeurs-des-coefficients}. The particles
do not exchange any heat with the nozzle walls or the substrate.

\begin{table}
\begin{centering}
\begin{tabular}{cccc}
\toprule 
$\mathrm{Re}_{p}$ & $a_{1}$ & $a_{2}$ & $a_{3}$\tabularnewline
\midrule
\midrule 
$\left[0;0.1\right]$ & 0 & 24 & 0\tabularnewline
\midrule 
$\left[0.1;1\right]$ & 3 & 22.73 & 0.090,3\tabularnewline
\midrule 
$\left[1;10\right]$ & 1.222 & 29.166,7 & -3.888,9\tabularnewline
\midrule 
$\left[10;100\right]$ & 0.616,7 & 46.50 & -116.67\tabularnewline
\midrule 
$\left[100;1,000\right]$ & 0.364,4 & 98.33 & -2,778\tabularnewline
\midrule 
$\left[1,000;5,000\right]$ & 0.357 & 148.62 & -47,500\tabularnewline
\midrule 
$\left[5,000;10,000\right]$ & 0.46 & -490.546 & 578,700\tabularnewline
\midrule 
$\left[10,000;+\infty\right[$ & 0.519,1 & -1,662.5 & 5,416,700\tabularnewline
\bottomrule
\end{tabular}
\par\end{centering}
\caption{Values of the parameters $\left(a_{i}\right)_{i\in\left\llbracket 1,3\right\rrbracket }$
according to Ref \cite{MorsiAlexander} and Ref \cite{FluentManual}\label{tab:Valeurs-des-coefficients}}
\end{table}

\subsubsection{IDDES model}

In this section, we present the details about the new proposed model
which aims to increase the fidelity of the simulation with more representative
assumptions and governing equations.

Therefore, we propose the implementation of the Improved Delayed Detached
Eddy Simulation (or IDDES) to model the fluid flow. The IDDES equations
are given by \prettyref{eq:IDDESbegin} to \prettyref{eq:IDDESend}(see
(Ref \cite{FluentManual,ShurSpalartStreletsTravin,GritskevichGarbarukSchutzeMenter})),
\begin{equation}
\rho\frac{\mathrm{d}k}{\mathrm{d}t}=\frac{\partial}{\partial x_{i}}\left(\left(\mu+\sigma_{k}'\mu_{t}\right)\frac{\partial k}{\partial x_{i}}\right)+P_{k}-\frac{\rho k^{3/2}}{l_{\mathrm{IDDES}}}\label{eq:IDDESbegin}
\end{equation}
\[
\rho\frac{\mathrm{d}\omega}{\mathrm{d}t}=\frac{\partial}{\partial x_{i}}\left(\left(\mu+\sigma_{\omega}\mu_{t}\right)\frac{\partial\omega}{\partial x_{i}}\right)+\left(1-F_{1}\right)\frac{2\rho\sigma_{\omega2}\partial_{i}k\partial_{i}\omega}{\omega}+\frac{\alpha\rho}{\mu_{t}}P_{k}-\beta\rho\omega^{2}
\]
where 
\begin{gather}
\mu_{t}=\frac{\rho A_{1}k}{\max\left(A_{1}\omega,F_{2}S\right)},\quad F_{1}=\tanh\left(\arg_{1}^{4}\right)\\
\arg_{1}=\min\left(\max\left(\frac{\sqrt{k}}{C_{\mu}\omega d_{w}},\frac{500\mu}{d_{w}^{2}\omega\rho}\right),\frac{4\rho\sigma_{\omega2}k}{CD_{k\omega}d_{w}^{2}}\right)\\
CD_{k\omega}=\max\left(\frac{2\rho\sigma_{\omega2}\partial_{i}k\partial_{i}\omega}{\omega},10^{-10}\right),
\end{gather}
\begin{gather}
\alpha=\frac{5}{9}F_{1}+0.44\left(1-F_{1}\right),\quad\beta=0.075F_{1}+0.0828\left(1-F_{1}\right),\\
\sigma_{k}'=0.85F_{1}+1-F_{1},\quad\sigma_{\omega}=0.5F_{1}+\sigma_{\omega2}\left(1-F_{1}\right),\\
\sigma_{\omega2}=0.856,\quad F_{2}=\tanh\left(\max\left(\frac{2\sqrt{k}}{C_{\mu}\omega d_{w}},\frac{500\mu}{d_{w}^{2}\omega\rho}\right)^{2}\right),
\end{gather}
\begin{gather}
P_{k}=\min\left(\mu_{t}S^{2},10\rho k\omega\right),\\
l_{\mathrm{IDDES}}=f_{d}\left(1+f_{e}\right)l_{\mathrm{RANS}}+\left(1-f_{d}\right)l_{\mathrm{LES}},\label{eq:lIDDES}\\
l_{\mathrm{LES}}=C_{\mathrm{DES}}\min\left(C_{w}\max\left(d_{w},h_{\max}\right),h_{\max}\right),
\end{gather}
\begin{gather}
l_{\mathrm{RANS}}=\frac{\sqrt{k}}{C_{\mu}\omega},\quad C_{\mathrm{DES}}=C_{\mathrm{DES}1}F_{1}+C_{\mathrm{DES}2}\left(1-F_{1}\right),\\
f_{d}=\max\left(1-f_{dt},f_{b}\right),\quad f_{dt}=1-\tanh\left(\left(C_{dt1}r_{dt}\right)^{C_{dt2}}\right),\\
r_{dt}=\frac{\mu_{t}}{\varkappa^{2}\rho d_{w}\sqrt{\left(S^{2}+\Omega^{2}\right)/2}},
\end{gather}
\begin{gather}
f_{b}=\min\left(2\mathrm{e}^{-9\alpha'^{2}},1.0\right),\quad\alpha'=\frac{1}{4}-\frac{d_{w}}{h_{\max}},\\
f_{e}=f_{e2}\max\left(f_{e1}-1.0,0.0\right),\quad f_{e1}=\begin{cases}
2\mathrm{e}^{-11.09\alpha'^{2}} & \alpha'\geq0\\
2\mathrm{e}^{-9.0\alpha'^{2}} & \alpha'<0
\end{cases},\\
f_{e2}=1.0-\tanh\left(\max\left(\left(C_{t}^{2}r_{dt}\right)^{3},\left(C_{l}^{2}r_{dl}\right)^{10}\right)\right),
\end{gather}
\begin{gather}
r_{dl}=\frac{\mu}{\varkappa^{2}\rho d_{w}\sqrt{\left(S^{2}+\Omega^{2}\right)/2}},\quad C_{w}=0.15,\quad C_{dt1}=20,\\
C_{dt2}=3,\quad C_{l}=5.0,\quad C_{t}=1.87,\quad C_{\mu}=0.09,\quad\varkappa=0.41,\\
A_{1}=0.31,\quad C_{\mathrm{DES}1}=0.78,\quad C_{\mathrm{DES}2}=0.61.\label{eq:IDDESend}
\end{gather}
This model of turbulence is designed to take advantage of both Large
Eddy Simulation (or LES) and RANS simulation. The \prettyref{eq:lIDDES}
and the \prettyref{eq:IDDESbegin} show length scales which are coming
from either LES or RANS simulation. In the detached areas, the model
calculates completely the flow until the smallest local cell size
and simulates the lower scales. Otherwise, near the walls, to properly
get the boundary layer and its effects, the RANS model takes the dominance.
The whole IDDES model purpose is to manage the transition between
the detached areas and the boundary layer. We recall that the steady
state solution of the previously described RANS model can be used
as an initialization for the IDDES model. The gas is modeled with
the Redlich-Kwong-Aungier's law (see (Ref \cite{Aungier,FluentManual})
) given by \prettyref{eq:Aungierlaw} to \prettyref{eq:Aungierlawend}
\begin{equation}
p=\frac{\rho rT}{1+\rho\left(c-b\right)}-\frac{\alpha_{0}\rho^{2}}{1+\rho b}\left(\frac{T}{T_{c}}\right)^{-n}\label{eq:Aungierlaw}
\end{equation}
where 
\begin{gather}
c=rT_{c}\left(p_{c}+\frac{\alpha_{0}\rho_{c}^{2}}{1+\rho_{c}b}\right)^{-1}+b-\frac{1}{\rho_{c}},\\
\alpha_{0}=\frac{0.42747r^{2}T_{c}^{2}}{p_{c}},\quad b=\frac{0.08664rT_{c}}{p_{c}},\\
n=0.4986+1.1735\upsilon+0.4754\upsilon^{2}.\label{eq:Aungierlawend}
\end{gather}
The wall of the nozzle and the substrate are modeled as $\unit[3]{mm}$
thick steel walls with an external temperature of $\unit[300]{K}$.
Also, the quantities sought are calculated with third-order elements.
The particles have an influence on the flow \textemdash{} a two-way
coupled system \textemdash{} and their motion is governed by \prettyref{eq:particlemotion-1},
\begin{equation}
\left(1+\frac{C_{\mathrm{vm}}\rho}{\rho_{p}}\right)\frac{\mathrm{d}\boldsymbol{u_{p}}}{\mathrm{d}t}=\frac{3\rho}{4\rho_{p}d_{p}}\left\Vert \boldsymbol{u}-\boldsymbol{u_{p}}\right\Vert \left(\boldsymbol{u}-\boldsymbol{u_{p}}\right)C_{D}+\left(1+C_{\mathrm{vm}}\right)\frac{\rho}{\rho_{p}}\boldsymbol{u_{p}}\mathrm{div}\left(\boldsymbol{u}\right)\label{eq:particlemotion-1}
\end{equation}
where $C_{\mathrm{vm}}=0.5$ is the virtual mass factor assuming spherical
particles (see (Ref \cite{CEAreport})) and $C_{D}$ is defined
by \prettyref{eq:dragcoefficient} with $c$ given by \prettyref{eq:Machspeedofsound}
with the new state law \prettyref{eq:Aungierlaw}.

\subsection{Geometries, boundary conditions and mesh}

\subsubsection{Geometries}

\begin{figure}
\begin{centering}
\subfloat[Configuration A]{\begin{centering}
\includegraphics[viewport=14.99499bp 0bp 742.252bp 142bp,clip,width=15cm]{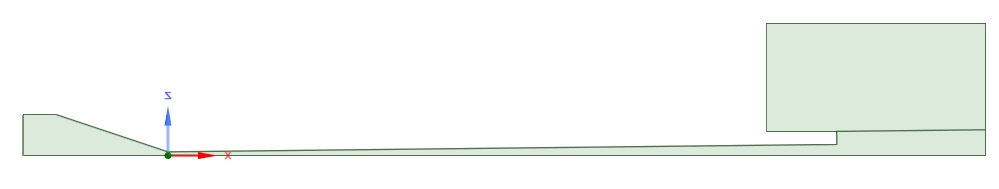}
\par\end{centering}
}
\par\end{centering}
\begin{centering}
\subfloat[Configuration B]{\begin{centering}
\includegraphics[viewport=10bp 0bp 876bp 213bp,clip,width=13.32cm]{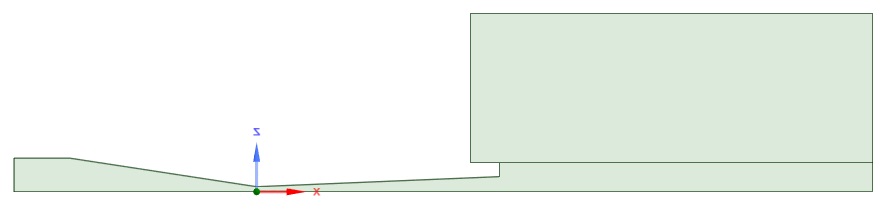}
\par\end{centering}
}
\par\end{centering}
\begin{centering}
\subfloat[Configuration C]{\begin{centering}
\includegraphics[clip,width=6.3417cm]{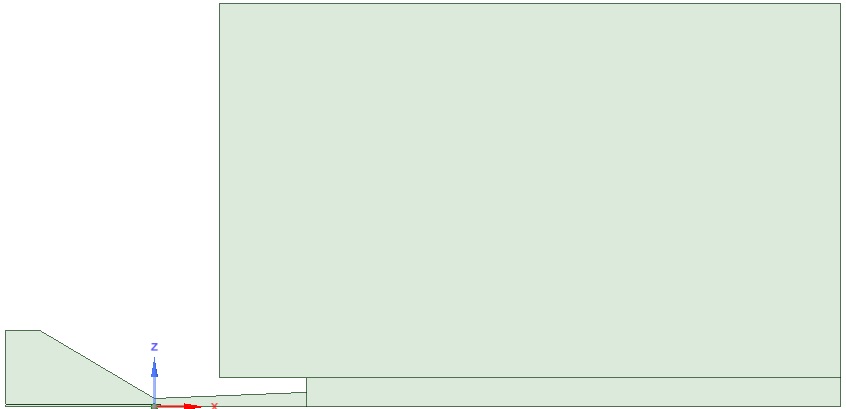}
\par\end{centering}
}
\par\end{centering}
\begin{centering}
\subfloat[Configuration D. The convex part in the convergent is visible on the
left.]{\begin{centering}
\includegraphics[width=15cm]{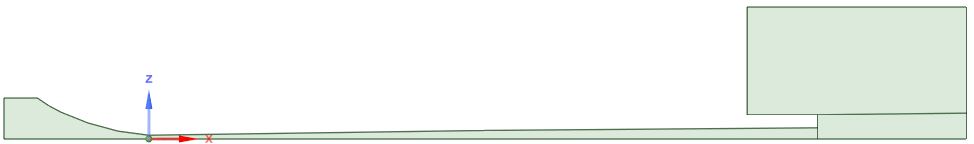}
\par\end{centering}
}
\par\end{centering}
\caption{Geometries of the four configurations. The scale is $150/259$.\label{fig:Geometries-use-by}}
\end{figure}

\begin{table}
\begin{centering}
\begin{tabular}{cccc>{\centering}p{4cm}}
\toprule 
Configuration & A & B & C & D\tabularnewline
\midrule
\midrule 
$L\unit{(mm)}$ & $210$ & $115$ & $35$ & $210$\tabularnewline
\midrule 
$\unit[D]{(mm)}$ & $2$ & $2.7$ & $2.2$ & $2$\tabularnewline
\midrule 
$\zeta$ & $0.857$ & $0.565$ & $0.571$ & $0.857$\tabularnewline
\midrule 
$\xi_{i}$ & $3.0$ & $3.0$ & $1.73$ & $3.0$\tabularnewline
\midrule 
$\xi_{o}$ & $11$ & $6.67$ & $9.09$ & $11$\tabularnewline
\midrule 
$\eta$ & $0.3$ & $0.3$ & $0.3$ & $0.3$\tabularnewline
\midrule 
Injector diameter (mm) & $2$ & $2$ & $2$ & $2$\tabularnewline
\midrule 
Injector position (mm) & S & S & $+0.8$ & S\tabularnewline
\midrule 
Injector pressure (bar) & $30.4$ & $30.4$ & $10$ & $30.4$\tabularnewline
\midrule 
Standoff distance (mm) & $40$ & $100$ & $70$ & $40$\tabularnewline
\midrule 
Inlet pressure (bar) & $30$ & $30$ & $20$ & $30$\tabularnewline
\midrule 
Outlet pressure (bar) & $1$ & $1$ & $1$ & $1$\tabularnewline
\midrule 
Inlet temperature (K) & $300$ & $300$ & $300$ & $300$\tabularnewline
\midrule 
Outlet temperature (K) & $300$ & $300$ & $300$ & $300$\tabularnewline
\midrule 
Shape & Linear & Linear & Linear & Convex in the convergent and linear in the divergent\tabularnewline
\bottomrule
\end{tabular}
\par\end{centering}
\caption{Dimensions and boundary conditions of the configurations. ``S''
means in the stagnation chamber and $+x$ means after the throat of
$x\unit{mm}$. The configuration A to C corresponds to the configurations
1, 2 and 4 presented by Ref \cite{Lupoi2011}\label{tab:Dimensions-and-boundary}}
\end{table}

The geometries used in this paper are given on \prettyref{fig:Geometries-use-by}.
The dimensions of the various configurations are given by \prettyref{tab:Dimensions-and-boundary}.
$\eta$ is kept between the four configurations. The outlet boundaries
are set as far as possible from the nozzle exit to set the pressure
to the atmospheric one and the temperature to the ambient one. The
configurations A to C correspond to three configurations proposed
in Ref \cite{Lupoi2011}. However, the last case D corresponds
to a new configuration keeping the same characteristics as A but changing
the shape of the channel in the convergent.

\subsubsection{Boundary conditions}

\begin{figure}
\begin{centering}
\includegraphics[viewport=0bp 90bp 910bp 470bp,clip,width=8cm]{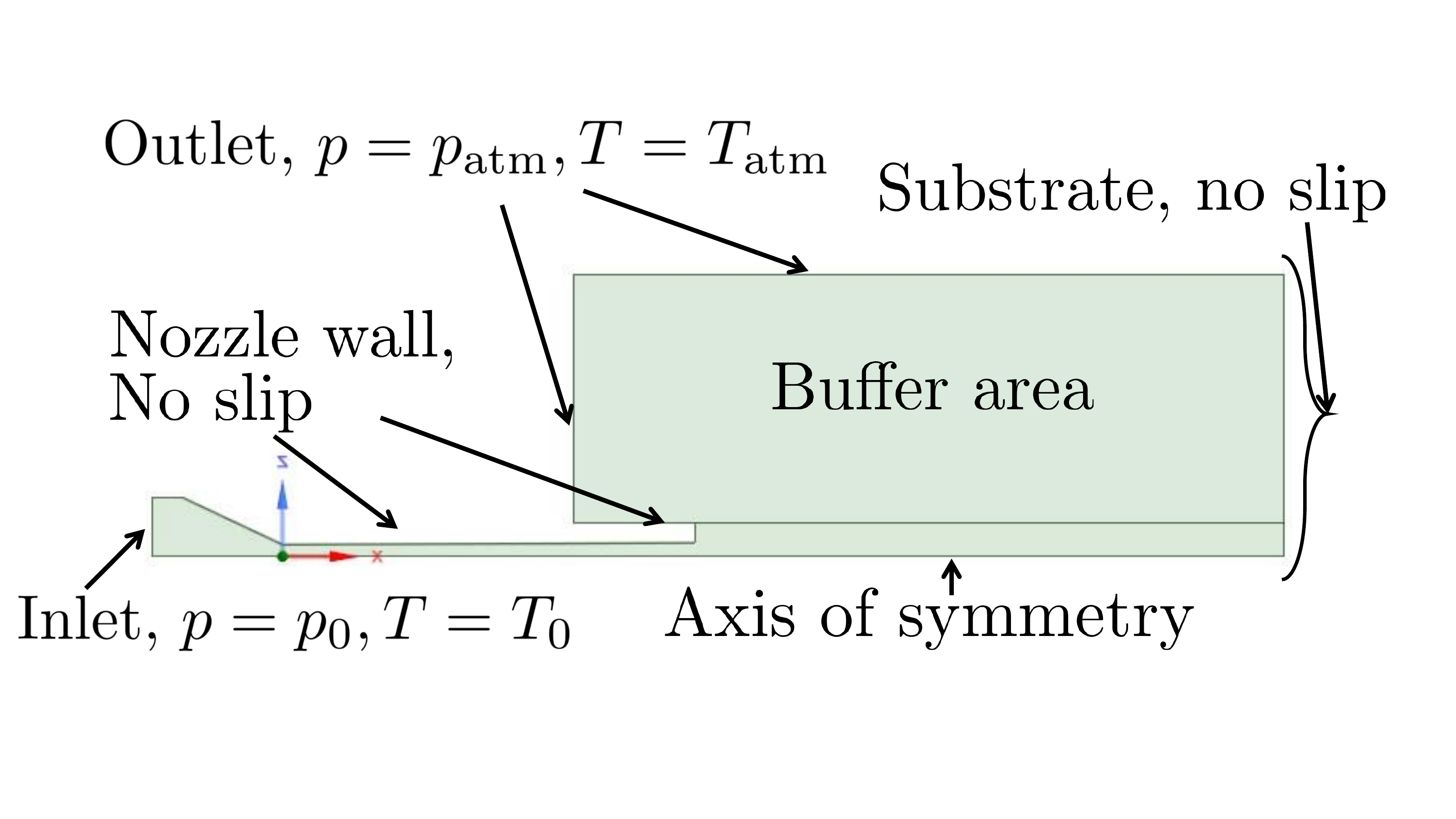}
\par\end{centering}
\caption{Sketch of the boundary conditions\label{fig:Sketch-of-the}}
\end{figure}

The boundary conditions, sketched on \prettyref{fig:Sketch-of-the},
are summarized in \prettyref{tab:Dimensions-and-boundary}. It is
interesting to point out that the exact position of the injector when
placed in the stagnation does not matter because the major acceleration
occurs in the divergent part. For configurations A, B and D, the particles
enter into the nozzle at the inlet where the pressure and the temperature
are fixed. For configuration C, the particles have their own injector
and start at the inlet of the injector where the pressure and the
temperature are fixed. The mass flow rate of the particles is fixed
for each configuration at $\unit[0.3]{g.s^{-1}}$ to agree with common
values found in the literature.

\subsubsection{Mesh}

\begin{figure}
\begin{centering}
\subfloat[Global mesh]{\begin{centering}
\includegraphics[viewport=30bp 80bp 1297bp 300bp,clip,width=15cm]{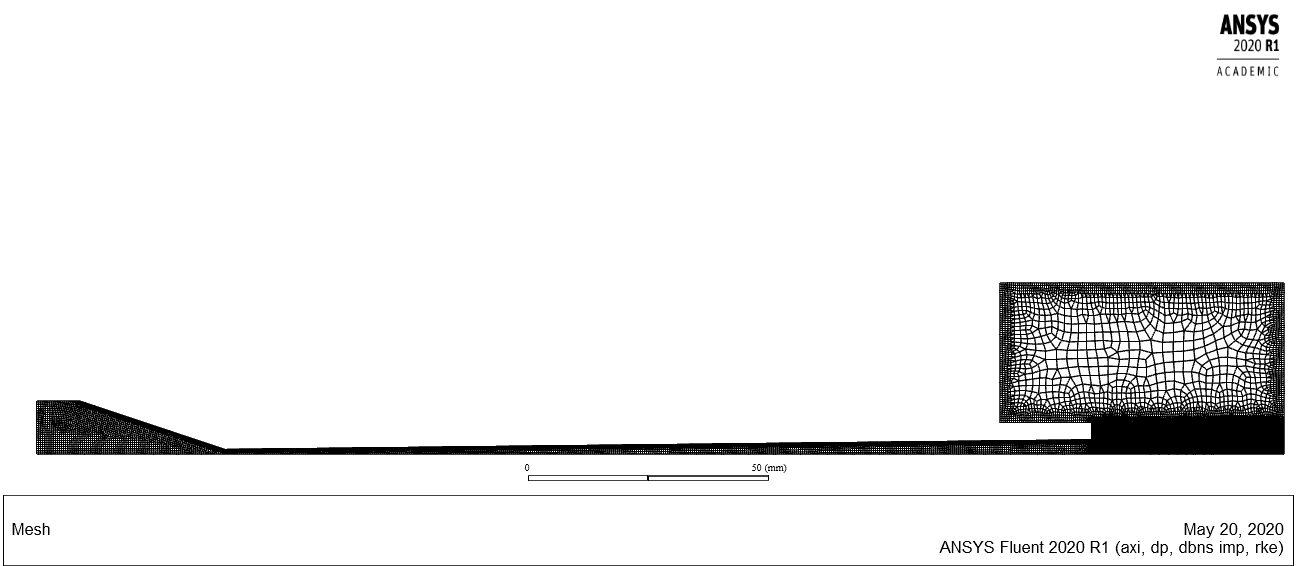}
\par\end{centering}
}
\par\end{centering}
\begin{centering}
\subfloat[Zoom in the convergent]{\begin{centering}
\includegraphics[viewport=80bp 80bp 1297bp 450bp,clip,width=5cm]{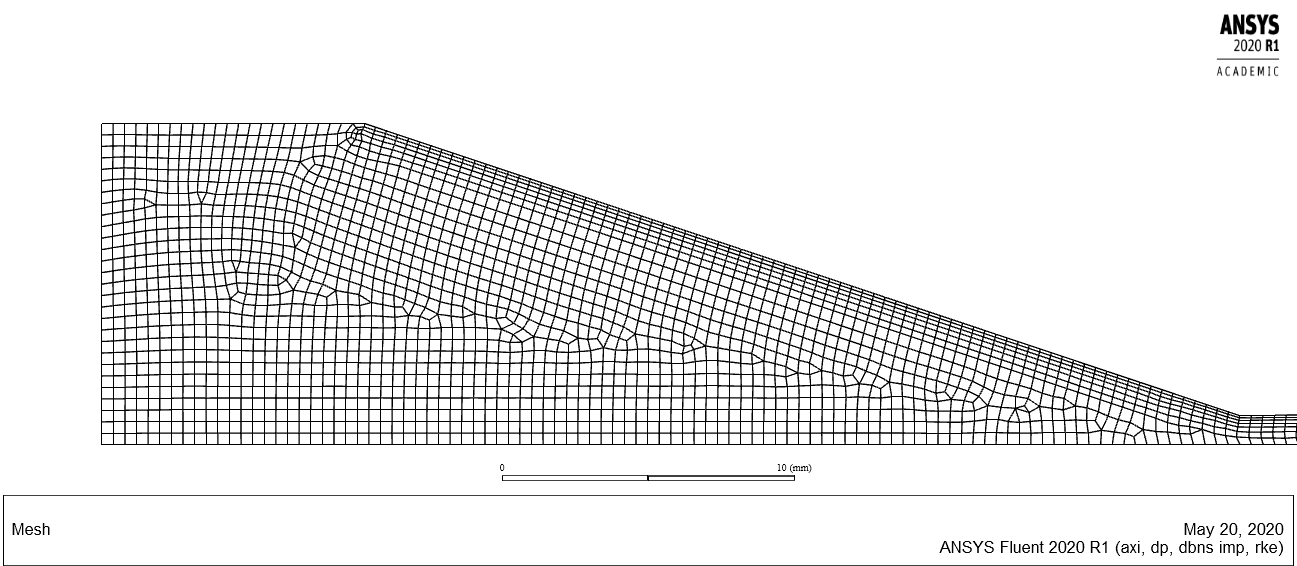}
\par\end{centering}
}\subfloat[Zoom in the divergent]{\begin{centering}
\includegraphics[viewport=400bp 80bp 900bp 200bp,clip,width=5cm]{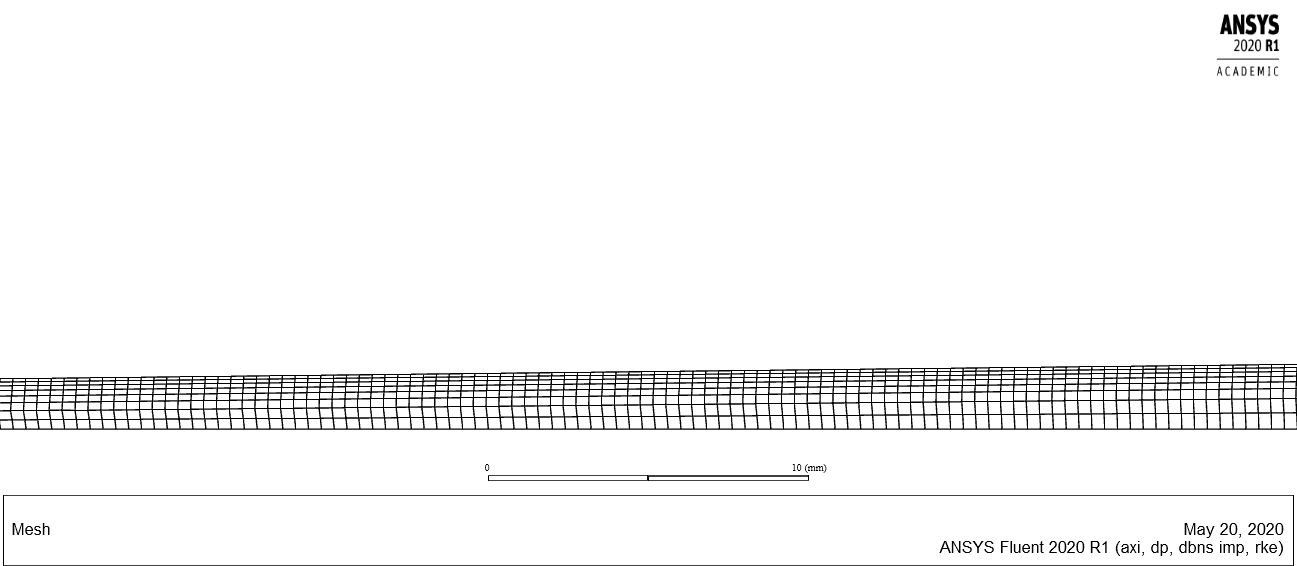}
\par\end{centering}
}\subfloat[Zoom in the exit]{\begin{centering}
\includegraphics[viewport=0bp 80bp 800bp 500bp,clip,width=5cm]{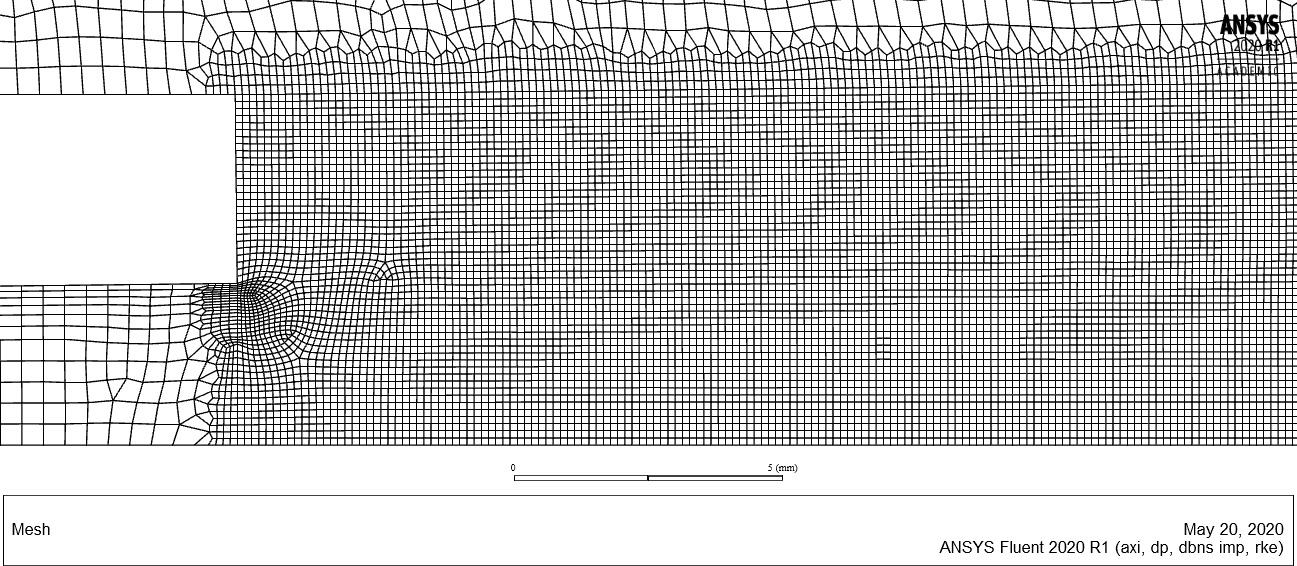}
\par\end{centering}
}
\par\end{centering}
\caption{Screenshots of the mesh used\label{fig:Screenshots-of-the}}
\end{figure}

The mesh, depicted on \prettyref{fig:Screenshots-of-the}, is 1,000
times wider in the buffer area to gain in computation time and refined
along the outside the nozzle to be more accurate around the shocks
and the substrate (the factor compared to the inside of the nozzle
is around 4 times smaller). Attempts were made with a refined mesh
in the inside of the nozzle but there is not a significant difference
in terms of the performance of the nozzle. Moreover, the mesh is refined
near the nozzle, which is not very relevant for the RANS model but
useful for the IDDES model. Therefore, there are between 20,000 and
35,000 elements according to the configuration.

\section{Numerical results and discussions}

\label{sec:Numerical-results-and}

\subsection{RANS model}

\subsubsection{Velocity magnitude of the flow}

\begin{figure}
\begin{centering}
\subfloat[Configuration A]{\begin{centering}
\includegraphics[viewport=130bp 150bp 1300bp 350bp,clip,width=6cm]{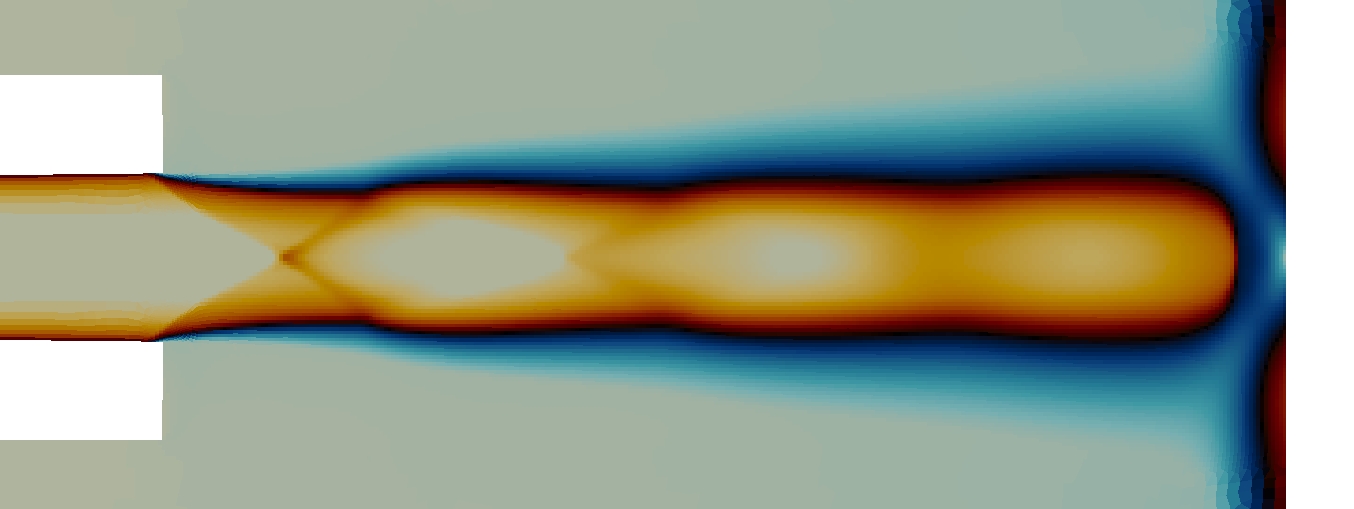}
\par\end{centering}
}
\par\end{centering}
\begin{centering}
\subfloat[Configuration B]{\begin{centering}
\includegraphics[viewport=100bp 180bp 1330bp 320bp,clip,width=15cm]{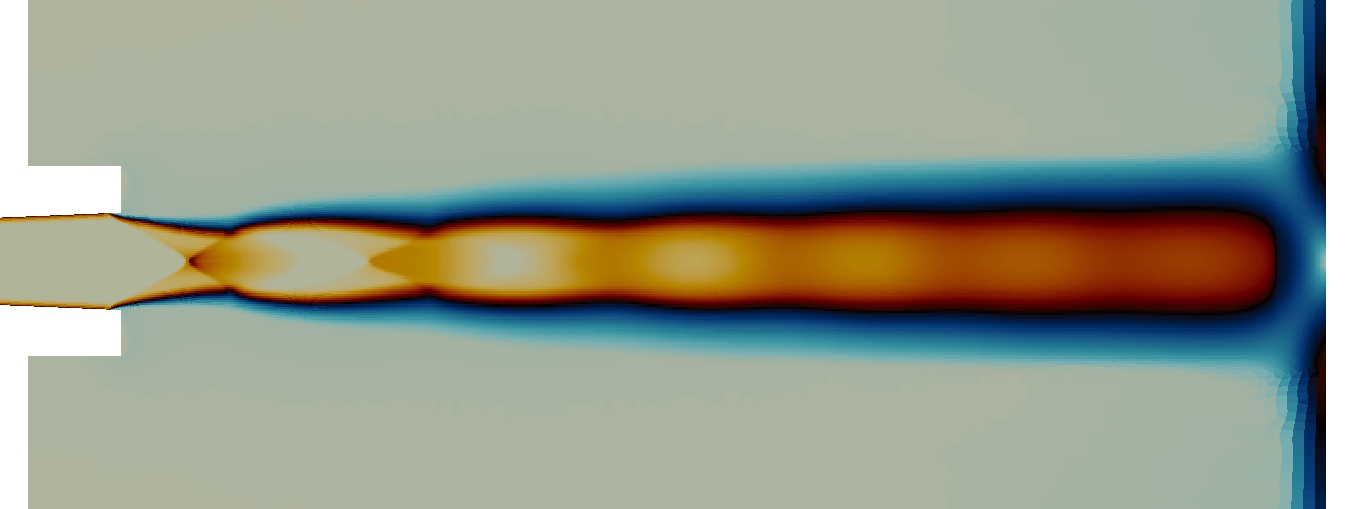}
\par\end{centering}
}
\par\end{centering}
\begin{centering}
\subfloat[Configuration C]{\begin{centering}
\includegraphics[viewport=20bp 210bp 1330bp 360bp,clip,width=10.5cm]{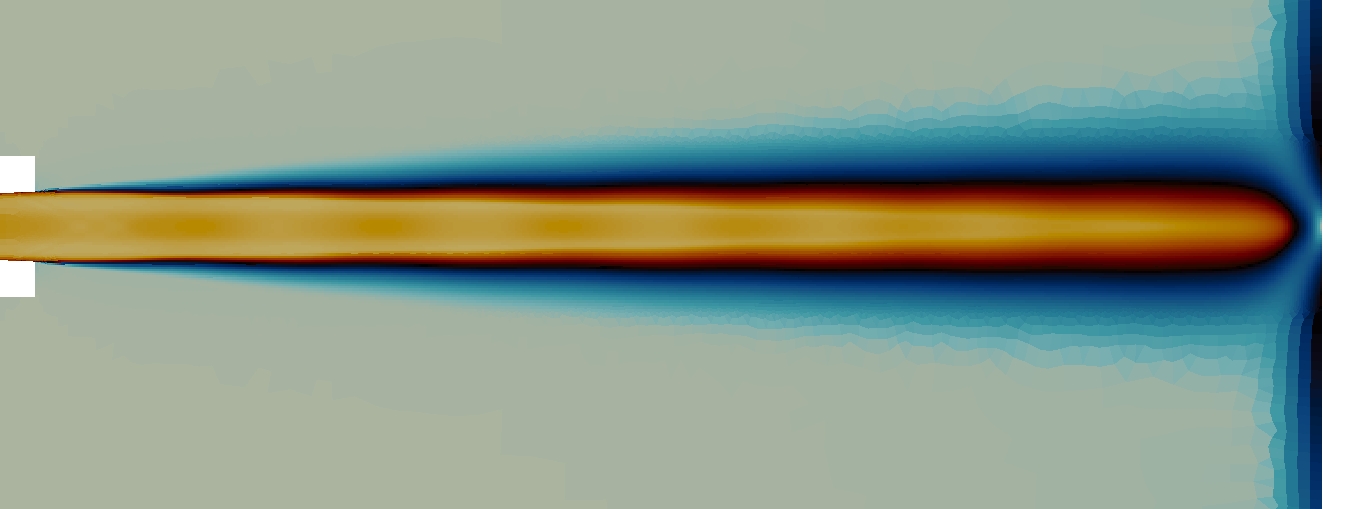}
\par\end{centering}
}
\par\end{centering}
\begin{centering}
\includegraphics[viewport=0bp 200bp 1354bp 400bp,clip,width=15cm]{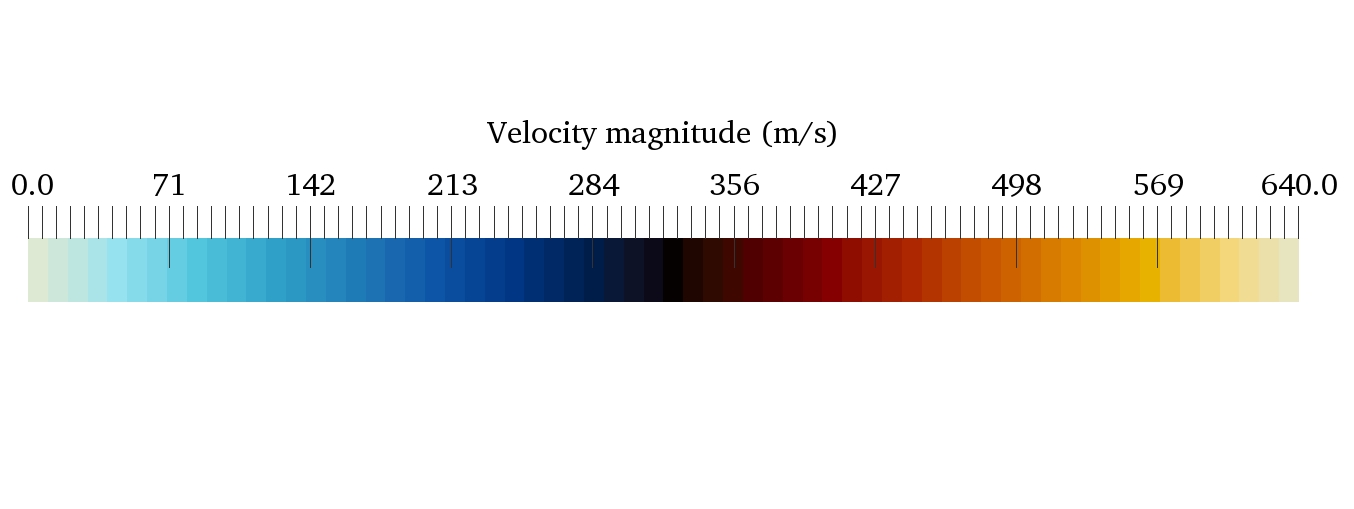}
\par\end{centering}
\caption{Velocity magnitude of the flow between the exit of the nozzle and
the substrate for the configuration A to C. The scale is $3/2$. \label{fig:Velocity-magnitude-between-1}}
\end{figure}

\begin{figure}
\begin{centering}
\subfloat[RANS model]{\begin{centering}
\includegraphics[viewport=0bp 310bp 1160bp 500bp,clip,width=15cm]{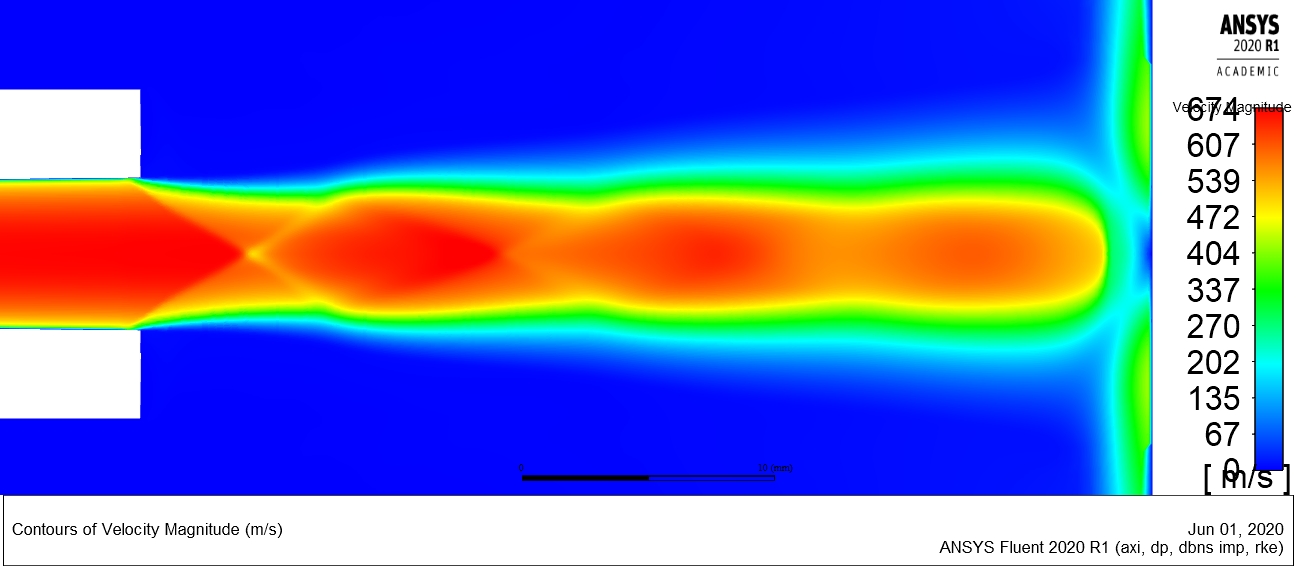}
\par\end{centering}
}
\par\end{centering}
\begin{centering}
\subfloat[(Ref \cite{Lupoi2011}) results]{\begin{centering}
\includegraphics[viewport=74.9686bp 45bp 543.522bp 103.5bp,clip,width=15cm]{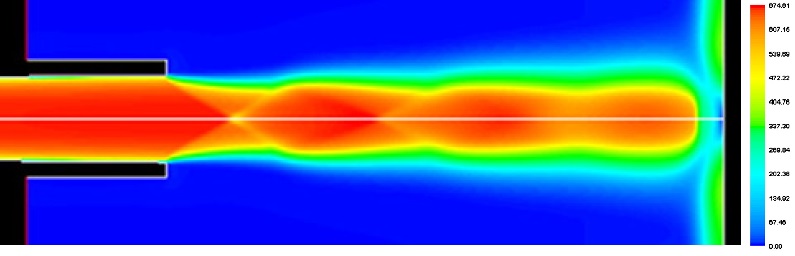}
\par\end{centering}
}
\par\end{centering}
\begin{centering}
\includegraphics[width=15cm]{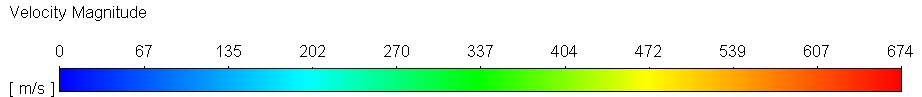}
\par\end{centering}
\caption{Velocity magnitude of the flow from the exit to the substrate for
configuration A with RANS model comparing with the results in (Ref
\cite{Lupoi2011})\label{fig:Comparaison-des-r=0000E9sultats}}
\end{figure}

The results to compare with those of (Ref \cite{Lupoi2011}) are
presented in this subsection. \prettyref{fig:Velocity-magnitude-between-1}
presents the obtained velocity magnitude of configuration A to C.
As expected, we note a strong agreement between the results and we
show that the RANS simulations have succeeded to well capture the
shocks in the flow, the global acceleration of the flow and its shape.
The velocity magnitude increases progressively from the throat to
the exit until the first shock.

A close comparison between both configurations A shows approximately
1\% of error. Nevertheless, even if the RANS model has tried to mimic
the model of (Ref \cite{Lupoi2011}), there are several approximations
which are important to take into account because they create differences
with real experiments. Looking at the model purely, the gas law is
a first step to analyze because if the Van der Waals' gas law is used
(see (Ref \cite{Berthelot})), comparing the density given by the
ideal gas law $\rho_{\mathrm{ig}}$ and the density given by the Van
der Waals' law $\rho_{\mathrm{VdW}}$, we get approximately 
\begin{equation}
\frac{\left|\rho_{\mathrm{VdW}}-\rho_{\mathrm{ig}}\right|}{\rho_{\mathrm{VdW}}}\approx\frac{pT_{c}}{8p_{c}T}\left|\frac{27T_{c}}{8T}-\left(1-\frac{pT_{c}}{8p_{c}T}\right)^{-1}\right|=1.7\%
\end{equation}
with $p=\unit[30]{bar}$, $T=\unit[300]{K}$ and for nitrogen, $p_{c}=\unit[34]{bar}$
and $T_{c}=\unit[126]{K}$. Therefore, looking for the lowest error
between the simulations is not relevant because even a calculation
with orders of magnitude gives slight differences. It is also particularly
relevant to compare the flow without adding particles because those
ones do not have influence on the flow.

\subsubsection{Turbulent kinetic energy}

\begin{comment}
\begin{figure}
\begin{centering}
\includegraphics[width=15cm]{ArXivImages/Energiecinetique,config1,axesymetrie,english,RANS.pdf}
\par\end{centering}
\caption{$k$ as a function of the position along the nozzle on the axis of
symmetry from RANS model results for configuration A. Position 0 corresponds
to the nozzle throat\label{fig:-as-a}}
\end{figure}
\end{comment}

\begin{figure}
\begin{centering}
\includegraphics[width=15cm]{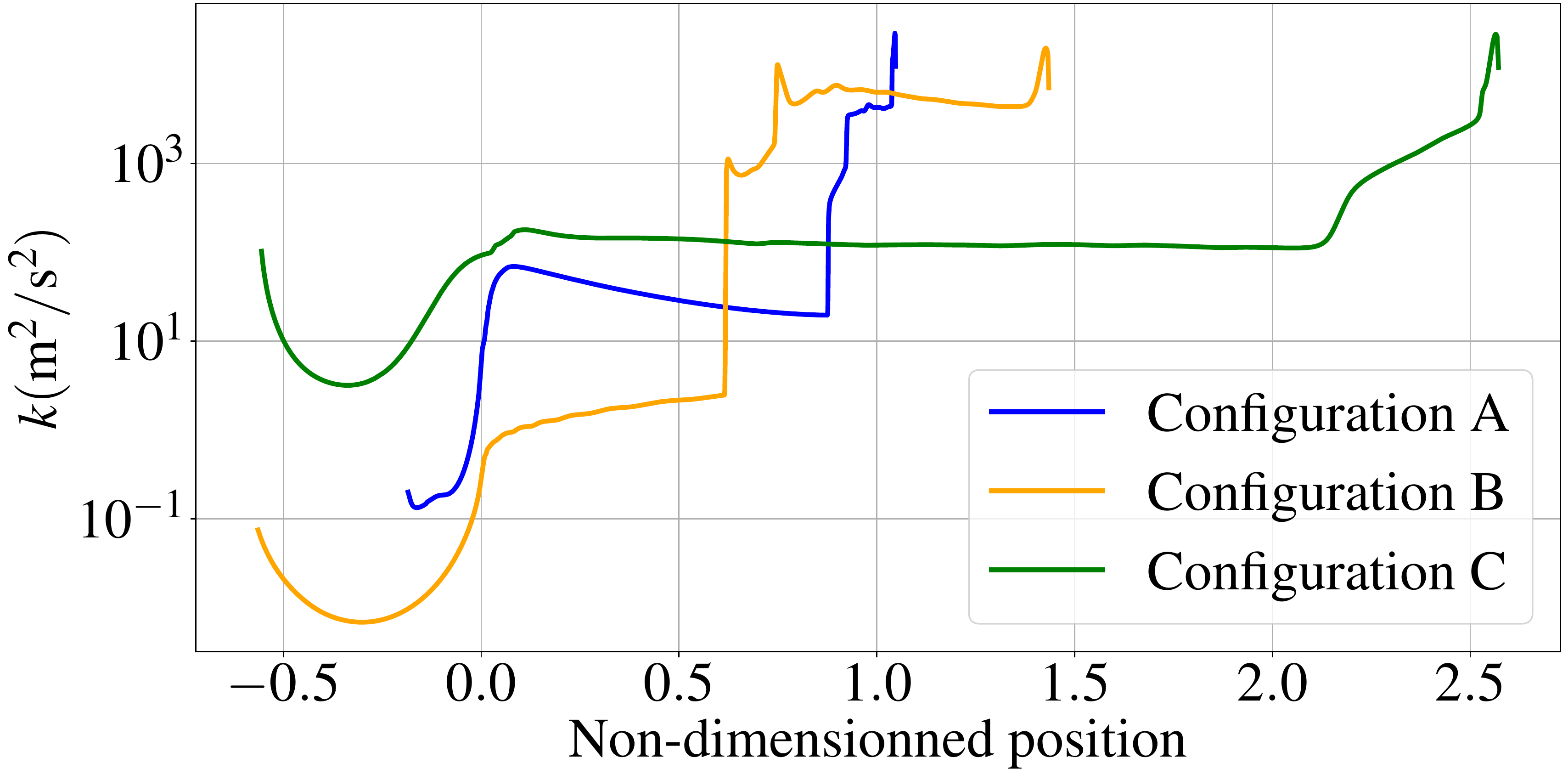}
\par\end{centering}
\caption{$k$ as a function of the dimensionless position along the nozzle
on the axis of symmetry from RANS model results on configuratons A
to C. The dimensionless position corresponds to the position divided
by $L$. Dimensionless position 0 corresponds to the nozzle throat\label{fig:-as-a-1}}
\end{figure}

In this sub section, we focus on plotting the turbulent kinetic energy
for configuration A. It is in good agreement with (Ref \cite{Lupoi2011}).
The shape are rather similar between both results. Precisely, there
is a peak of turbulent kinetic energy just after the throat where
the flow becomes supersonic and where shocks may appear. This increase
of turbulent kinetic energy is therefore an increase of turbulent
viscosity which enhance dissipation from turbulent effects. It is
also a sign there is a lot of mixing around the throat and confirms
the fact that the particle beam would completely fill the nozzle divergent
during the particles fly. To analyze furthermore the behavior of these
configurations, the \prettyref{fig:-as-a-1} compares configurations
A to C in terms of turbulent kinetic energy $k$.

The U shape in the convergent is clearly visible on \prettyref{fig:-as-a-1}.
Near the throat, it starts to increase because of the supersonic flow
beginning. Then, in the diverging part, it is approximately constant,
demonstrating that there is no additional phenomena to dissipate like
shocks or something else. For each configuration, after the diverging
part, there is a brutal increase of turbulent kinetic energy. Using
the \prettyref{fig:Velocity-magnitude-between-1}, these brutal increases
occur on the location of the first oblique shock. Continuing along
the axis of the nozzle, the fluctuations of the turbulent kinetic
energy occur exactly where there is a shock. Finally, near the substrate,
there is a strong increase of the turbulent kinetic energy caused
by the bow shock at the impact. Considering the values of $k$, this
bow shock has a important effect on the dissipation and on turbulent
effect. It has therefore a non negligible influence on the particles
impact velocity because they are slowed down by this discontinuity.
It confirms then the conclusion drawn in the literature that lighter
particles will be easily slowed down in comparison to heavier particles
with bigger inertia. However, these heavier ones gain less speed during
their acceleration thanks to the same reason. An optimum must be found
to reach the highest impact velocity with a given configuration. Comparing
configurations A to C, it happens configuration B has more fluctuations
along the flow which can be problematic for the nozzle in use and
all the machinery nearby. We would then prefer to use configuration
C which has less steep variations of turbulent kinetic energy to avoid
breaking the material in use.

\subsubsection{Particle tracks}

\begin{figure}
\begin{centering}
\subfloat[RANS model]{\begin{centering}
\includegraphics[viewport=60bp 190bp 1120bp 385bp,clip,width=15cm]{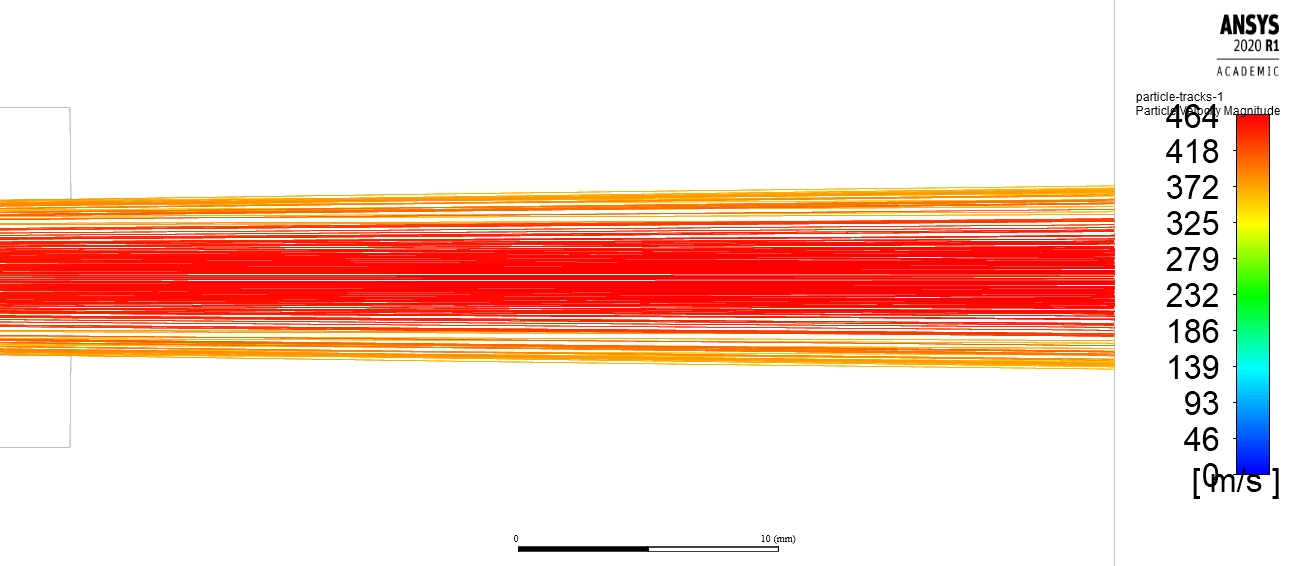}
\par\end{centering}
}
\par\end{centering}
\begin{centering}
\subfloat[(Ref \cite{Lupoi2011}) results]{\begin{centering}
\includegraphics[viewport=5bp 50bp 310bp 105bp,clip,width=15cm]{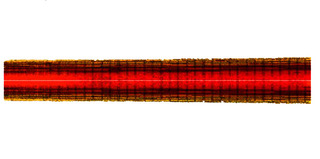}
\par\end{centering}
}
\par\end{centering}
\begin{centering}
\includegraphics[width=15cm]{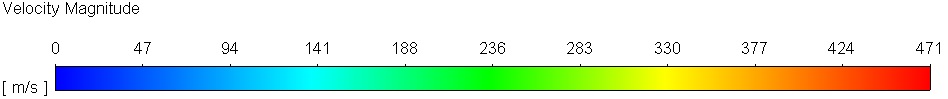}
\par\end{centering}
\caption{Particles track colored according to their velocity magnitude between
the exit of the nozzle and the substrate for configuration A. The
scale is $15/4$.\label{fig:Velocity-magnitude-between-3}}
\end{figure}

\begin{figure}
\begin{centering}
\subfloat[RANS model]{\begin{centering}
\includegraphics[viewport=120bp 305bp 1140bp 410bp,clip,width=15cm]{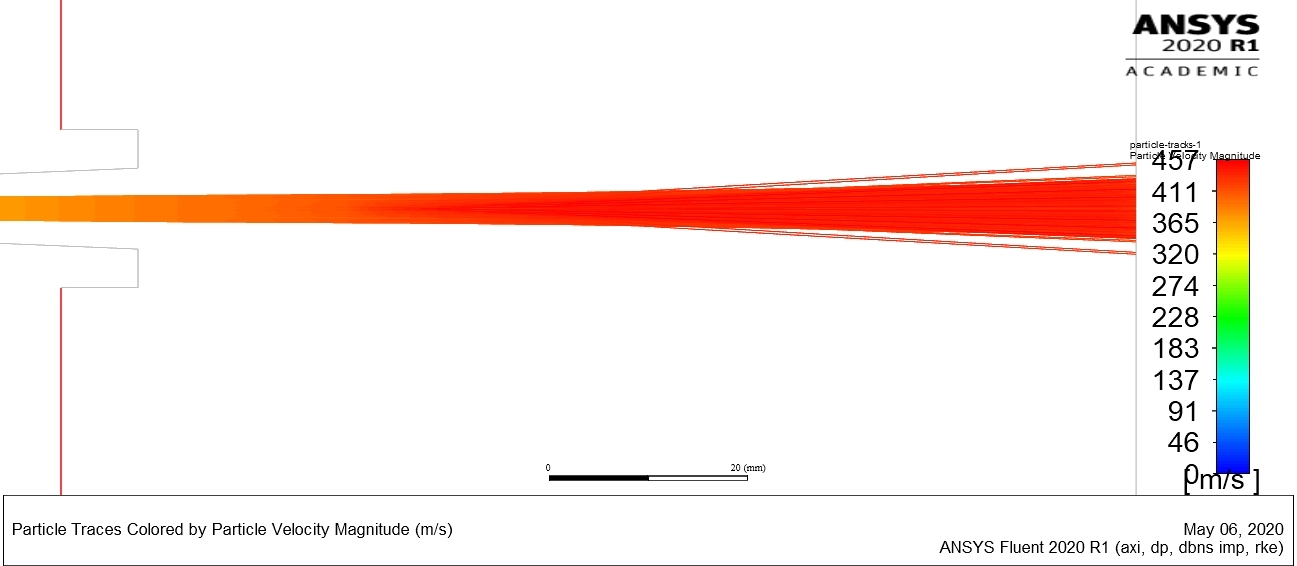}
\par\end{centering}
}
\par\end{centering}
\begin{centering}
\subfloat[(Ref \cite{Lupoi2011}) results]{\begin{centering}
\includegraphics[viewport=20bp 28bp 310bp 57bp,clip,width=15cm]{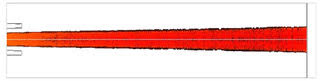}
\par\end{centering}
}
\par\end{centering}
\begin{centering}
\includegraphics[width=15cm]{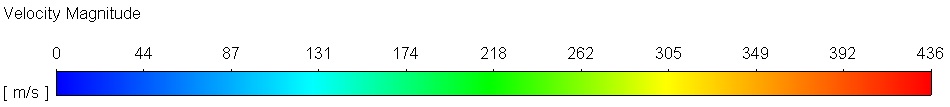}
\par\end{centering}
\caption{Particles track colored according to their velocity magnitude between
the exit of the nozzle and the substrate for configuration B. The
scale is $3/2$\label{fig:Velocity-magnitude-between-2}}
\end{figure}

\begin{figure}
\begin{centering}
\subfloat[RANS model]{\begin{centering}
\includegraphics[viewport=40bp 330bp 1190bp 400bp,clip,width=15cm]{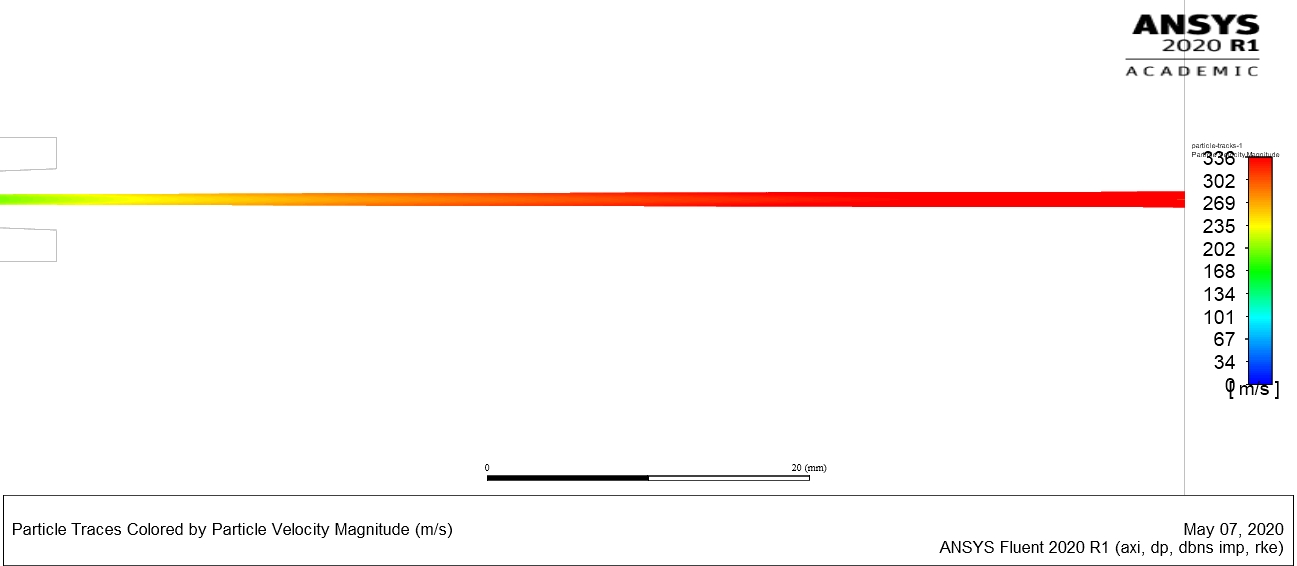}
\par\end{centering}
}
\par\end{centering}
\begin{centering}
\subfloat[(Ref \cite{Lupoi2011}) results]{\begin{centering}
\includegraphics[viewport=30bp 38bp 310bp 68bp,clip,width=15cm]{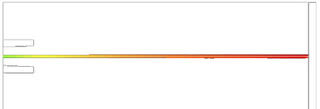}
\par\end{centering}
}
\par\end{centering}
\begin{centering}
\includegraphics[width=15cm]{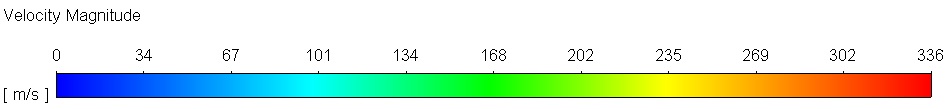}
\par\end{centering}
\caption{Particles track colored according to their velocity magnitude between
the exit of the nozzle and the substrate for configuration C. The
scale is $15/7$\label{fig:Velocity-magnitude-between}}
\end{figure}

Pursuing the comparison of the results, \prettyref{fig:Velocity-magnitude-between-3}
presents the velocity magnitude of the particles during the fly outside
the nozzle for configuration A\footnote{The behaviour is similar between the configurations}.
The shapes of all the particle tracks is conserved between the RANS
model results and those in (Ref \cite{Lupoi2011}). In terms of
values, the error reaches a maximum among all the configurations of
4.5\% which is pretty good knowing all the reasons mentioned above
about the model, the mesh, the approximations, etc. The distribution
of the velocities across the particle beam looks rather similar between
the two types of results.

\subsubsection{Particle impact velocity}

\begin{figure}
\begin{centering}
\includegraphics[width=12cm]{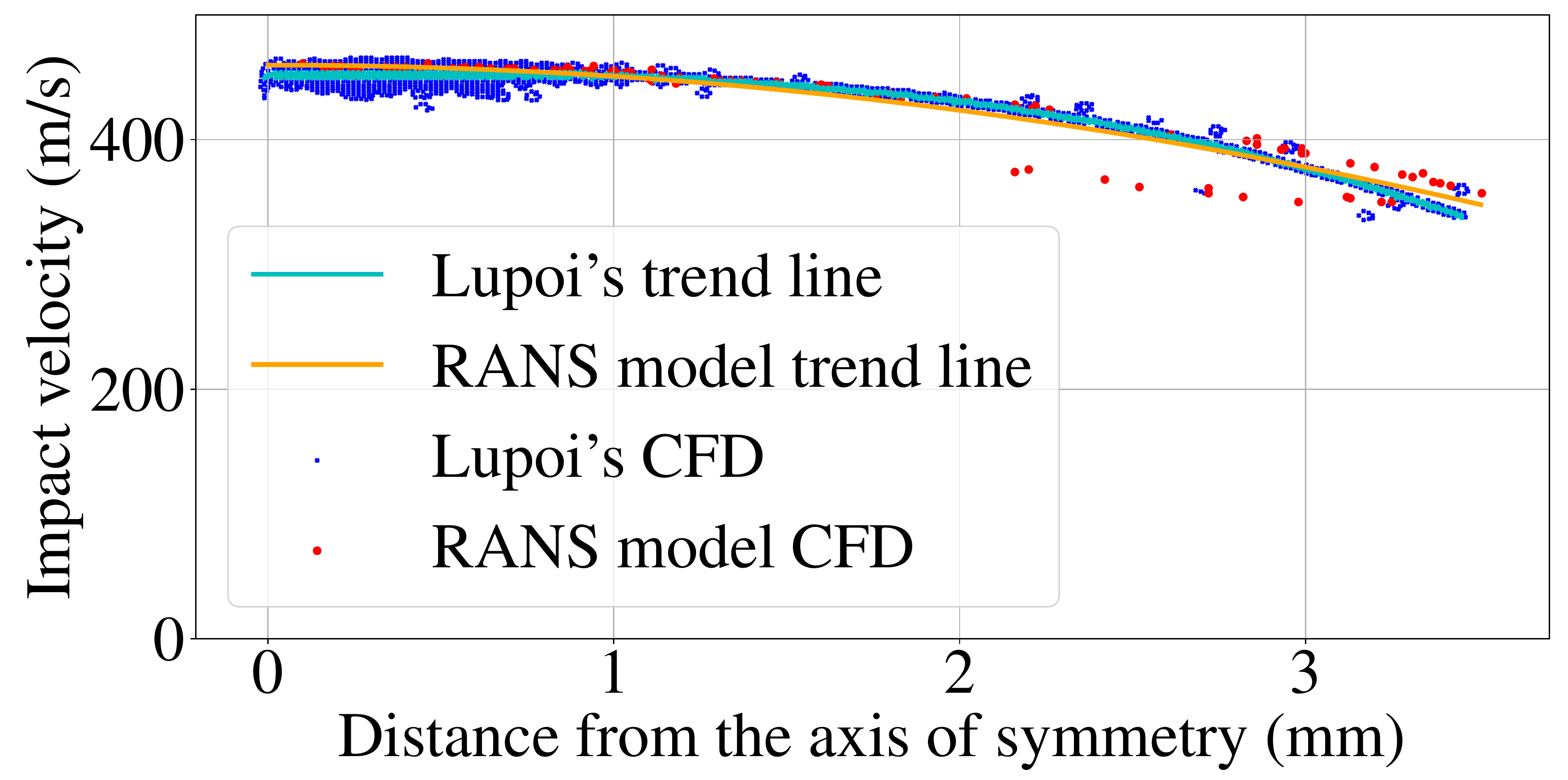}
\par\end{centering}
\caption{Particle impact velocity as a function of the distance from the axis
of symmetry for configuration A with RANS model\label{fig:Particle-impact-velocity-3}}
\end{figure}

\begin{comment}
\begin{figure}
\begin{centering}
\subfloat[Configuration A]{\begin{centering}
\includegraphics[width=12cm]{ArXivImages/Particleimpact,config1,100particle,english.pdf}
\par\end{centering}
}
\par\end{centering}
\begin{centering}
\subfloat[Configuration B]{\begin{centering}
\includegraphics[width=12cm]{ArXivImages/Particleimpact,config2,7000particle,english.pdf}
\par\end{centering}
}
\par\end{centering}
\begin{centering}
\subfloat[Configuration C]{\begin{centering}
\includegraphics[width=12cm]{ArXivImages/Particleimpact,config4,1200particle,english.pdf}
\par\end{centering}
}
\par\end{centering}
\caption{Particle impact velocity as a function of the distance from the axis
of symmetry for configuration A to C with RANS model\label{fig:Particle-impact-velocity}}
\end{figure}
\end{comment}

Finally, it is possible to compare the particle impact velocities
against their distance from the axis of symmetry. All theses results
are presented in \prettyref{fig:Particle-impact-velocity-3}. Again,
the results look very similar between the configurations. The maximum
velocity in each case of the RANS model is close to the maximum velocity
of the results in (Ref \cite{Lupoi2011}) with an error close to
1\%. Moreover, the distribution of particles are identical in terms
of proportion; that is to say the configurations A and B have a denser
proportion of particles near the center for both results and the configuration
C has a more uniform distribution of particles in both results. The
width of the band in configuration A is precisely respected in both
results. Besides, for configuration B, the width seems a little narrower
whereas configuration C has a wider width. Configuration C does not
seem to be problematic because the authors of (Ref \cite{Lupoi2011})
carried out experiment with each configuration and showed that the
width of the band for configuration C is experimentally wider than
its predictions. The RANS model results are closer to the experimental
results than those in (Ref \cite{Lupoi2011}). About configuration
B, the RANS model results show $\unit[5]{mm}$-wide band instead of
the $\unit[8]{mm}$-wide one in (Ref \cite{Lupoi2011}). Hence,
the RANS model disagree with (Ref \cite{Lupoi2011}) on configuration
B for particle impact velocity against their distribution.

\subsection{IDDES model}

In this section, we present all the new results obtained using the
proposed IDDES model. For comparisons, the previous RANS model will
be used in most of the cases.

\subsubsection{Velocity magnitude of the flow}

\begin{sidewaysfigure}
\begin{centering}
\subfloat[{Configuration A at $t=\unit[10]{ms}$}]{\begin{centering}
\includegraphics[width=7cm]{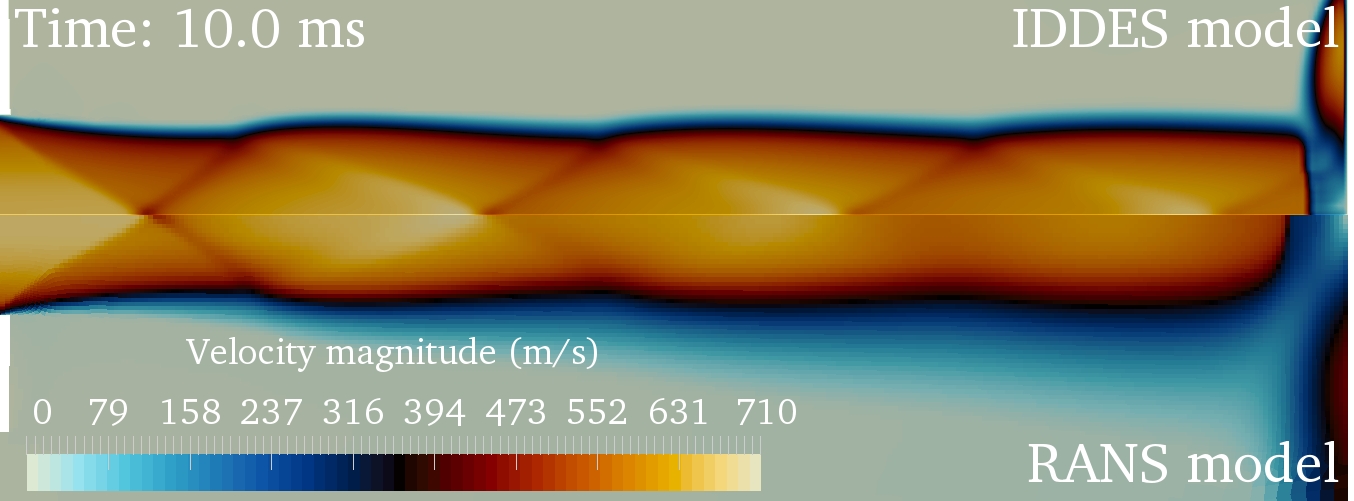}
\par\end{centering}
}\subfloat[{Configuration A at $t=\unit[20]{ms}$}]{\begin{centering}
\includegraphics[width=7cm]{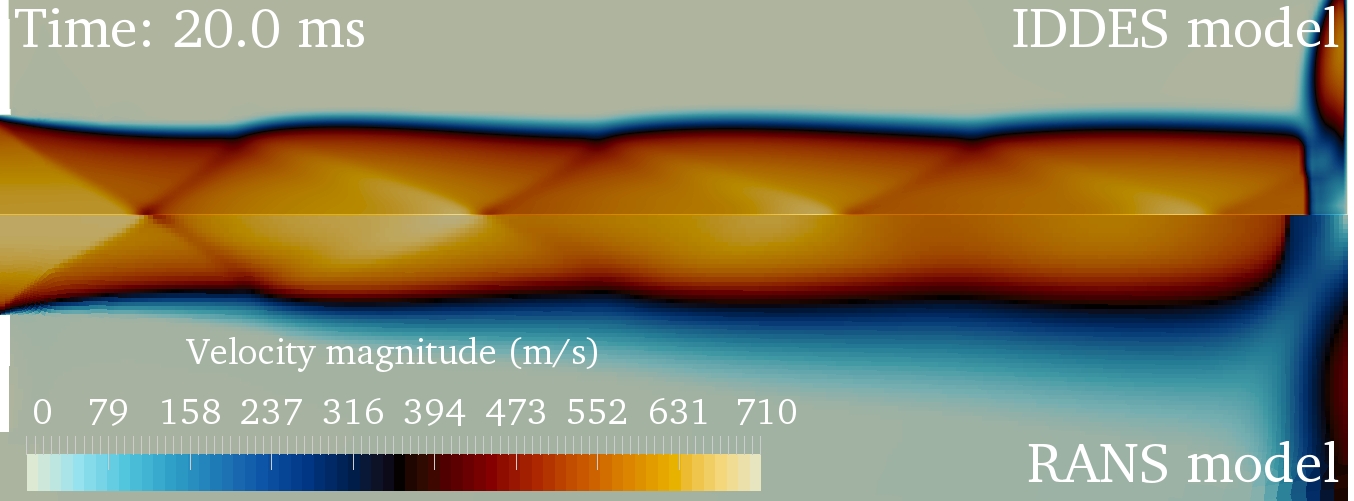}
\par\end{centering}
}\subfloat[{Configuration A at $t=\unit[30]{ms}$}]{\begin{centering}
\includegraphics[width=7cm]{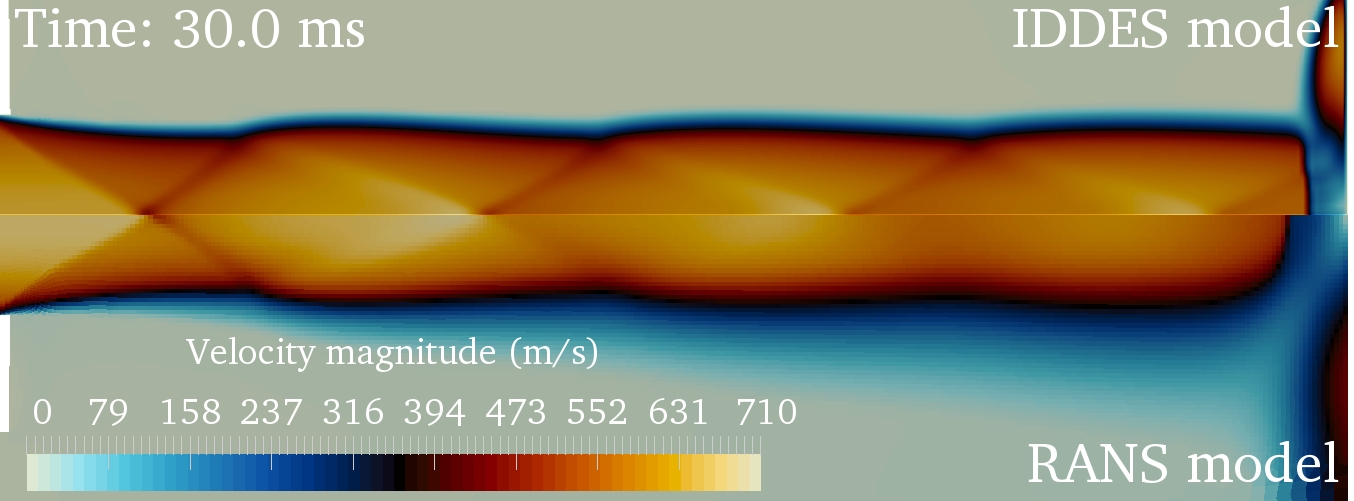}
\par\end{centering}
}
\par\end{centering}
\begin{centering}
\subfloat[{Configuration B at $t=\unit[10]{ms}$}]{\begin{centering}
\includegraphics[width=7cm]{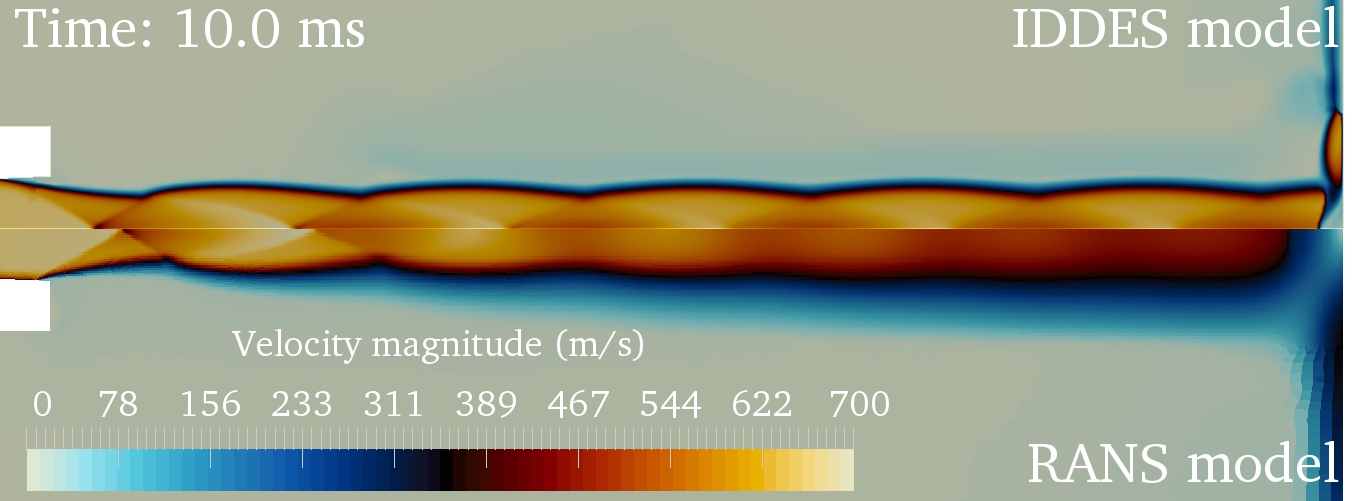}
\par\end{centering}
}\subfloat[{Configuration B at $t=\unit[20]{ms}$}]{\begin{centering}
\includegraphics[width=7cm]{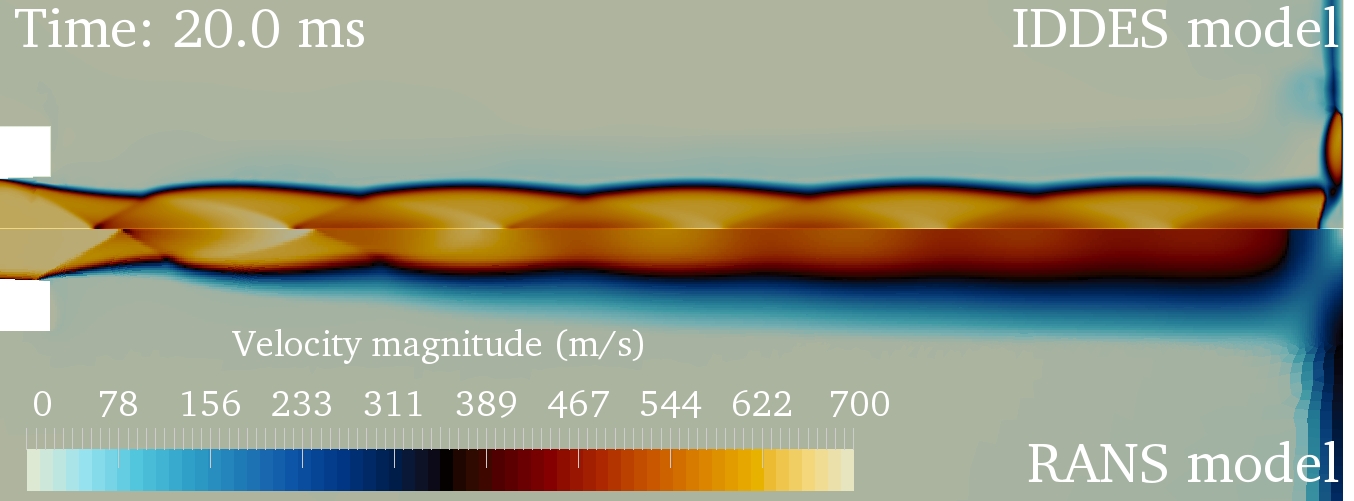}
\par\end{centering}
}\subfloat[{Configuration B at $t=\unit[30]{ms}$}]{\begin{centering}
\includegraphics[width=7cm]{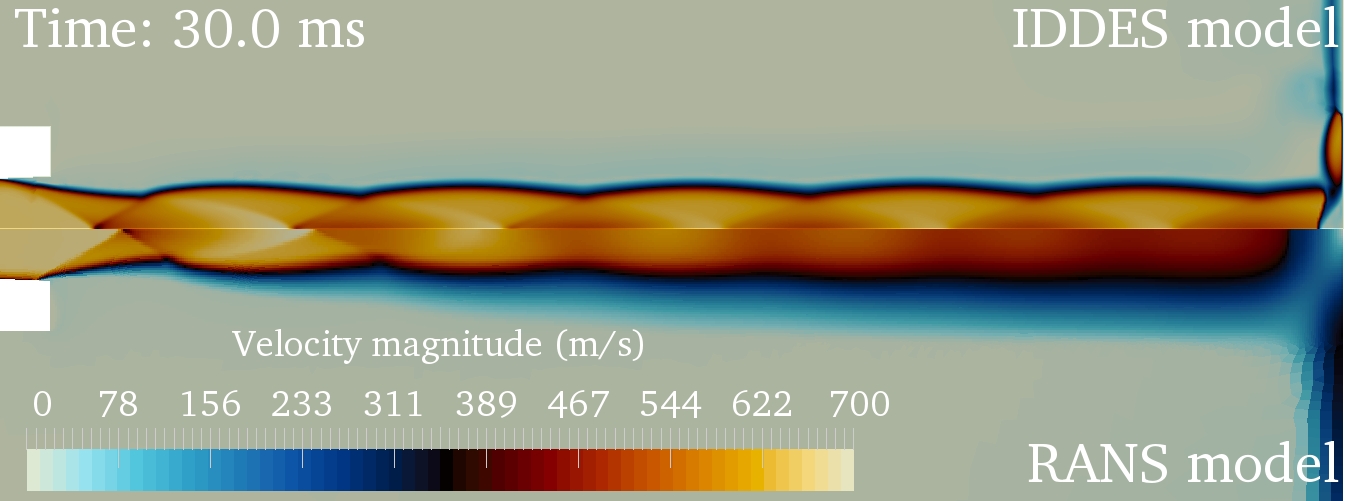}
\par\end{centering}
}
\par\end{centering}
\begin{centering}
\subfloat[{Configuration C at $t=\unit[10]{ms}$}]{\begin{centering}
\includegraphics[width=7cm]{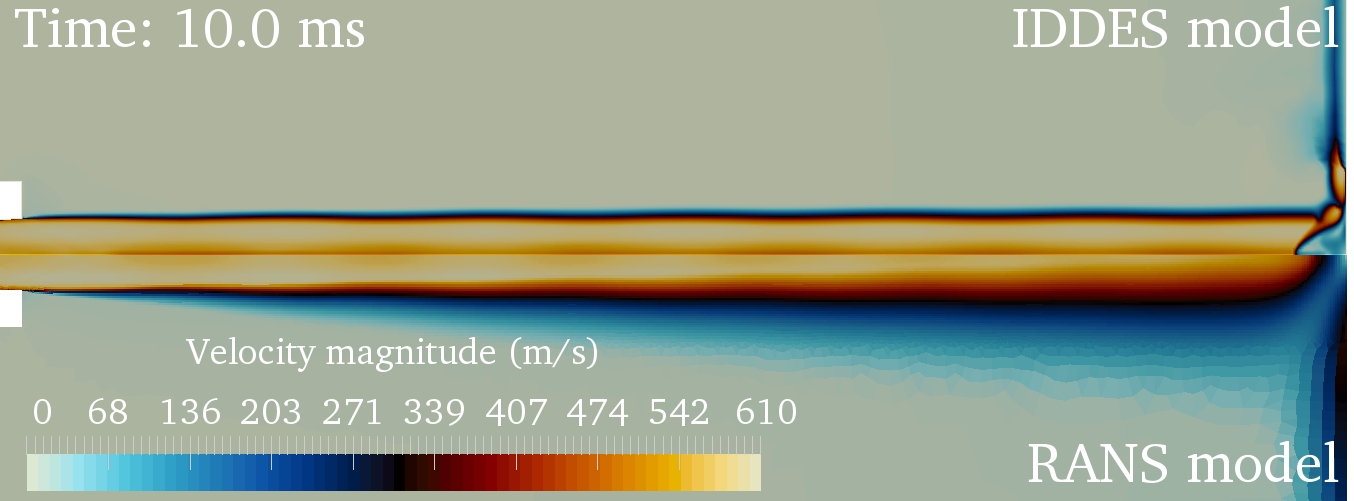}
\par\end{centering}
}\subfloat[{Configuration C at $t=\unit[20]{ms}$}]{\begin{centering}
\includegraphics[width=7cm]{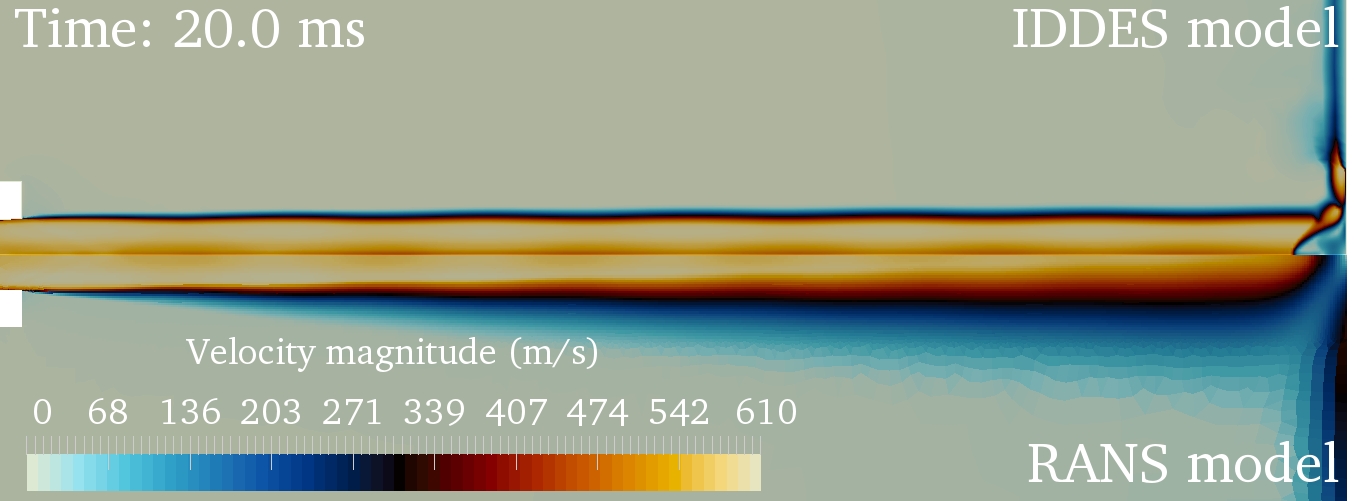}
\par\end{centering}
}\subfloat[{Configuration C at $t=\unit[30]{ms}$}]{\begin{centering}
\includegraphics[width=7cm]{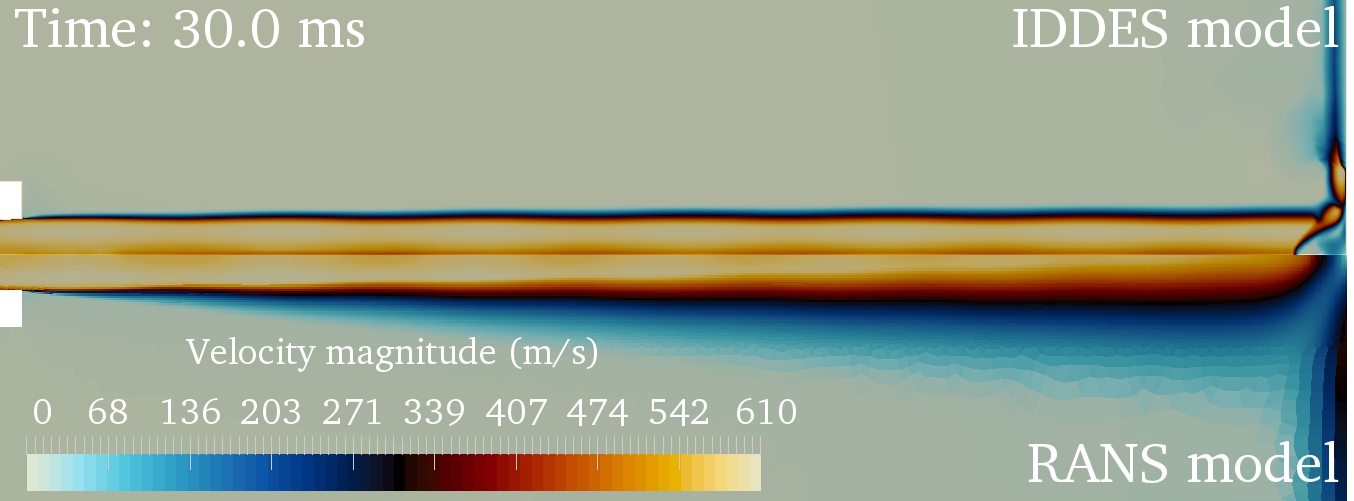}
\par\end{centering}
}
\par\end{centering}
\caption{Velocity magnitude for configuration A to C at different time steps
from the exit of the nozzle to the substrate with the RANS model and
the IDDES model. The behavior inside the nozzle is rather similar
between the models. The IDDES model is in the upper half part of each
picture and the RANS model is in the lower half part of each picture.
\label{fig:Velocity-magnitude-for-1}}
\end{sidewaysfigure}

We recall that the IDDES simulation is transient. Therefore, \prettyref{fig:Velocity-magnitude-for-1}
shows the comparison of the velocity magnitude between the RANS model
and the IDDES model at different time steps for configuration A\footnote{The behaviour is similar between the configurations}.
Globally, thanks to the increased accuracy for the space discretization
and the improved turbulent model, the results in this case seem less
diffusive and capture precisely the flow.

The slow radial decrease of velocity around the substrate is now a
clear difference, like a jump. It is also visible that a steady state
of the flow could be approximately reached.

\subsubsection{Comparison between the theoretical model, RANS model and IDDES model}

We studied a detailed comparison of all the previous models with the
plots of $M$, $p/p_{0}$ and $T/T_{0}$.

The general behavior of the plots are rather similar. Even if the
theoretical model needs stronger assumptions, it gives rather relevant
orders magnitude for the different variables under study. Furthermore,
on each configuration and each variable, the range of evolution is
wider for the theoretical model presented in \prettyref{subsec:Theoretical-model}
because we assumed no dissipation by turbulence, shocks or thermal
conduction. Thus, the flow is allowed to gain more velocity with a
higher $M$ lowering then drastically $p$ and $T$. Between the RANS
model and the IDDES model, the latter tends to lead to more dissipation
with a lower range of values for $M$, $p$ and $T$. It can be surprising
with what has been said previously but the turbulent effects are more
controlled by the IDDES model and allow a more accurate simulation.

\subsubsection{Bow shock near the substrate}

\begin{sidewaysfigure}
\begin{centering}
\subfloat[{Configuration A at $t=\unit[10]{ms}$}]{\begin{centering}
\includegraphics[viewport=0bp 0bp 1240bp 440bp,clip,width=7cm]{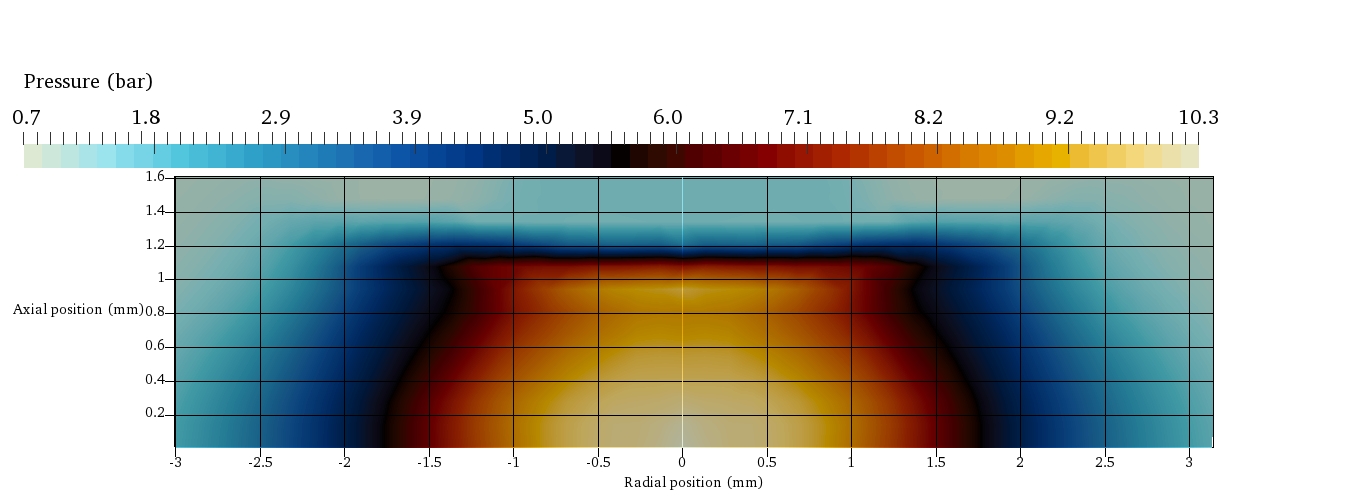}
\par\end{centering}
}\subfloat[{Configuration A at $t=\unit[20]{ms}$}]{\begin{centering}
\includegraphics[viewport=0bp 0bp 1240bp 440bp,clip,width=7cm]{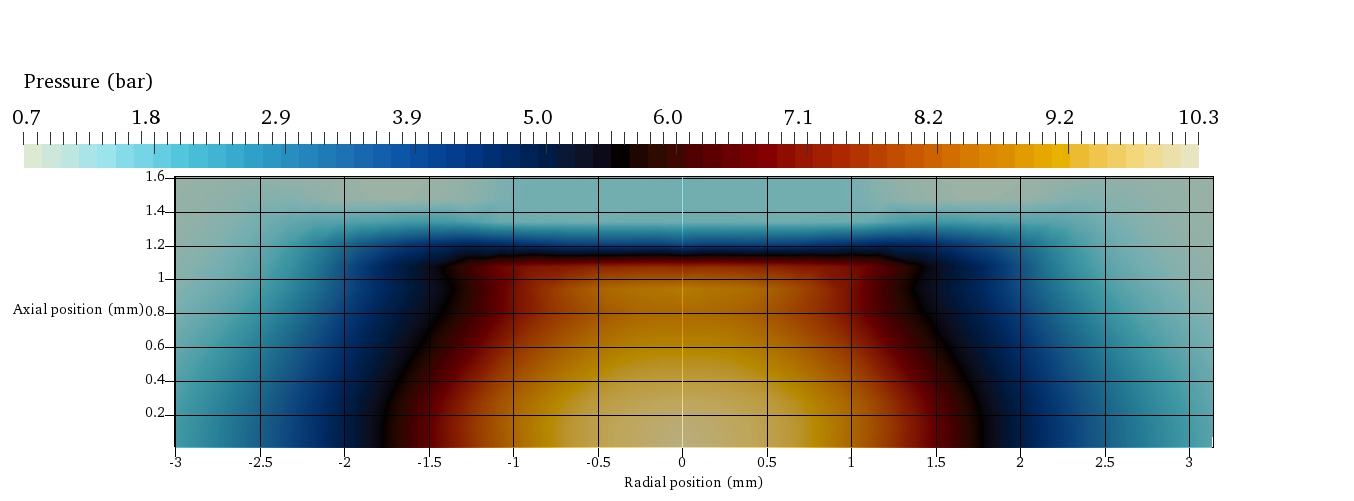}
\par\end{centering}
}\subfloat[{Configuration A at $t=\unit[30]{ms}$}]{\begin{centering}
\includegraphics[viewport=0bp 0bp 1240bp 440bp,clip,width=7cm]{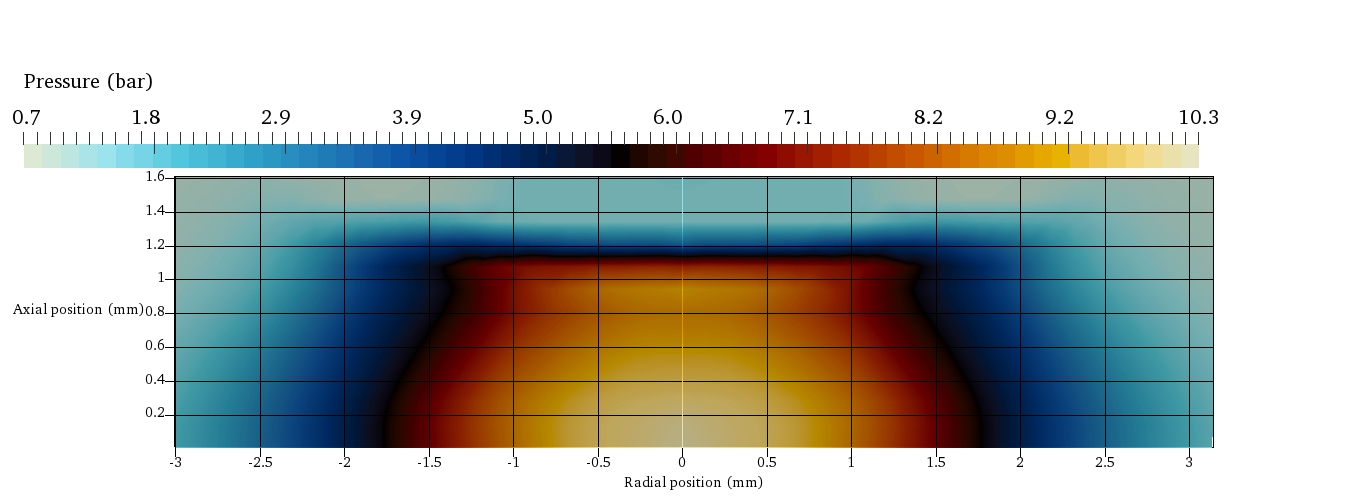}
\par\end{centering}
}
\par\end{centering}
\begin{centering}
\subfloat[{Configuration B at $t=\unit[10]{ms}$}]{\begin{centering}
\includegraphics[width=7cm]{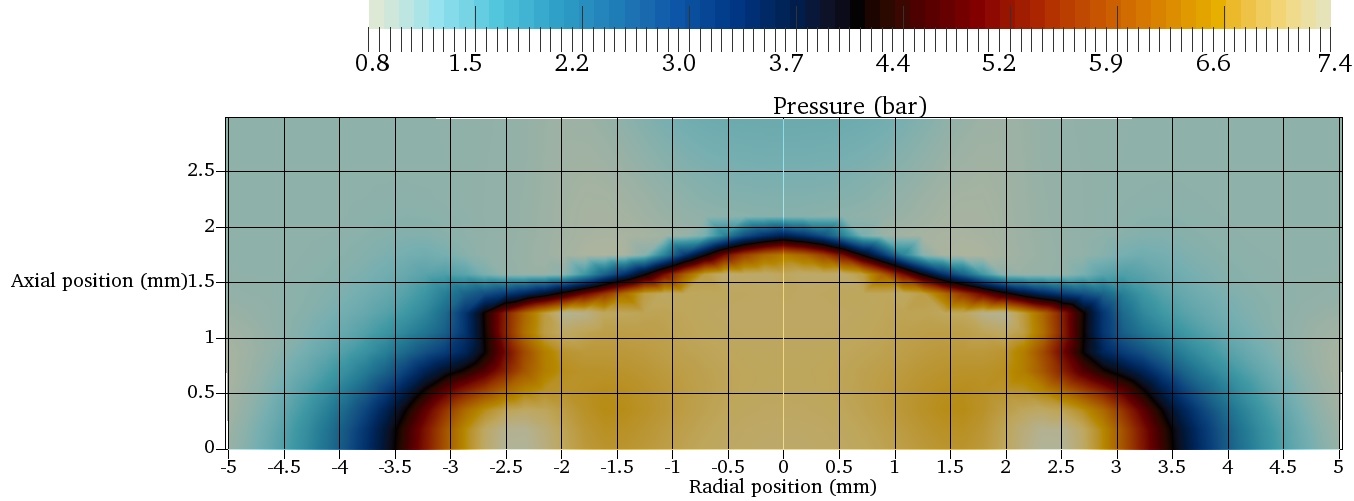}
\par\end{centering}
}\subfloat[{Configuration B at $t=\unit[20]{ms}$}]{\begin{centering}
\includegraphics[width=7cm]{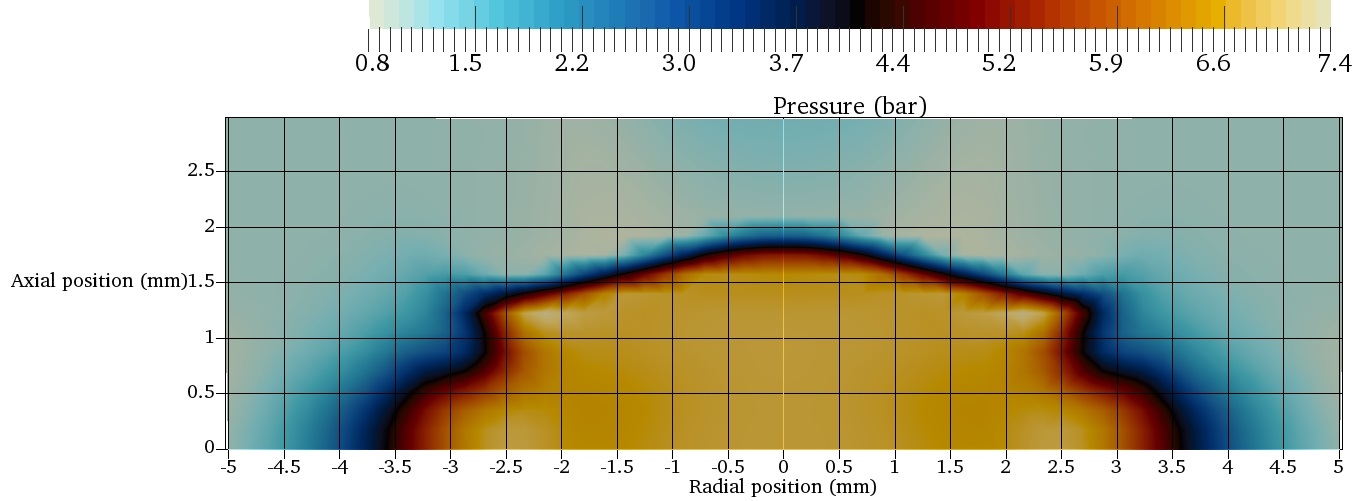}
\par\end{centering}
}\subfloat[{Configuration B at $t=\unit[30]{ms}$}]{\begin{centering}
\includegraphics[width=7cm]{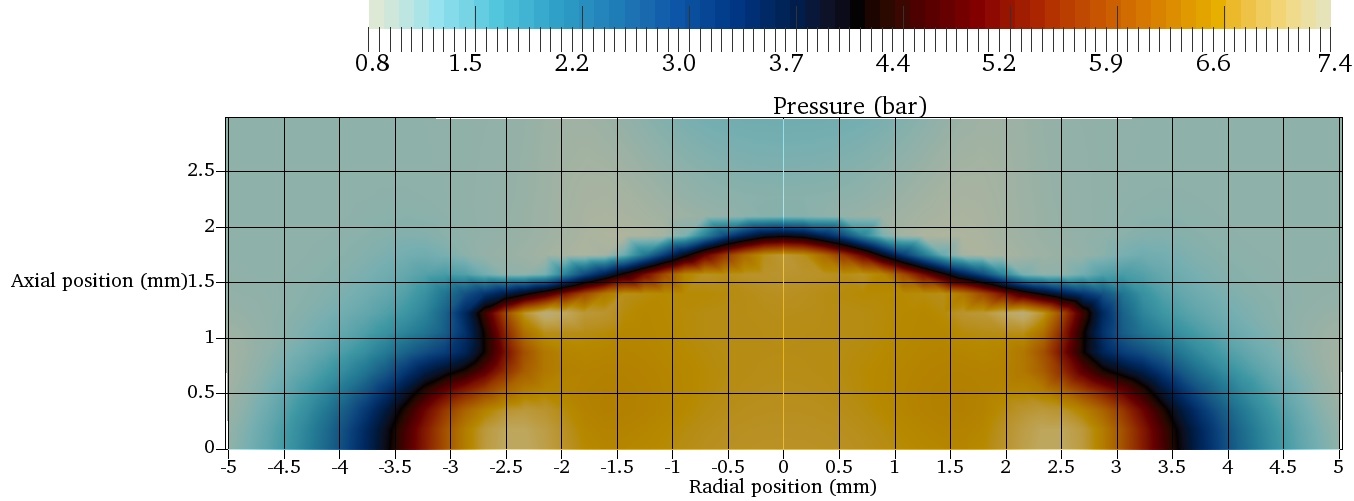}
\par\end{centering}
}
\par\end{centering}
\begin{centering}
\subfloat[{Configuration C at $t=\unit[10]{ms}$}]{\begin{centering}
\includegraphics[viewport=165.0812bp 0bp 862.925bp 376bp,clip,width=7cm]{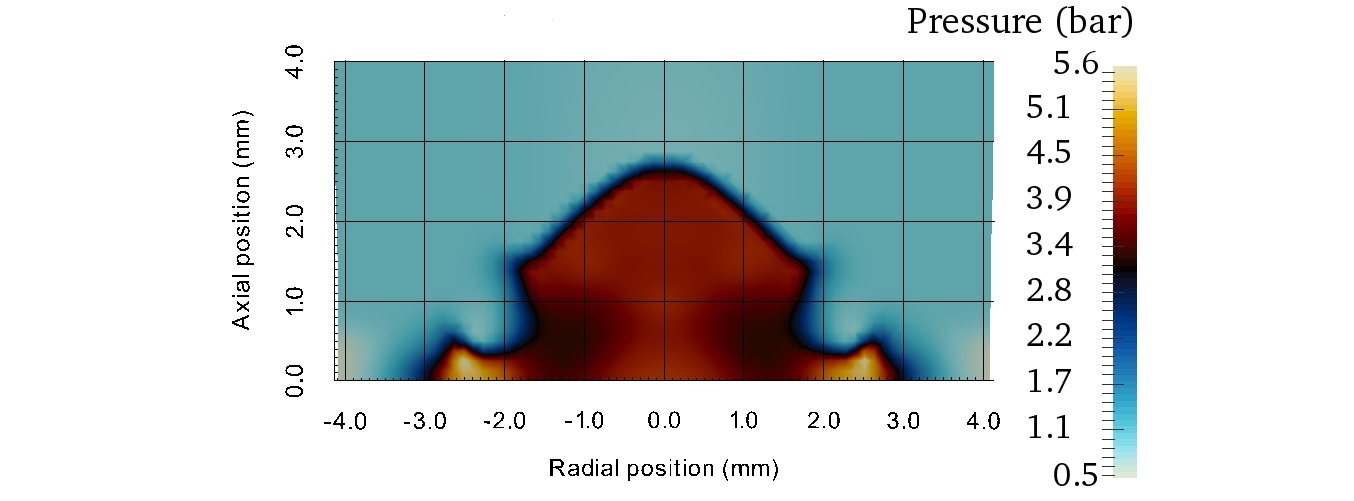}
\par\end{centering}
}\subfloat[{Configuration C at $t=\unit[20]{ms}$}]{\begin{centering}
\includegraphics[viewport=165.0812bp 0bp 862.925bp 376bp,clip,width=7cm]{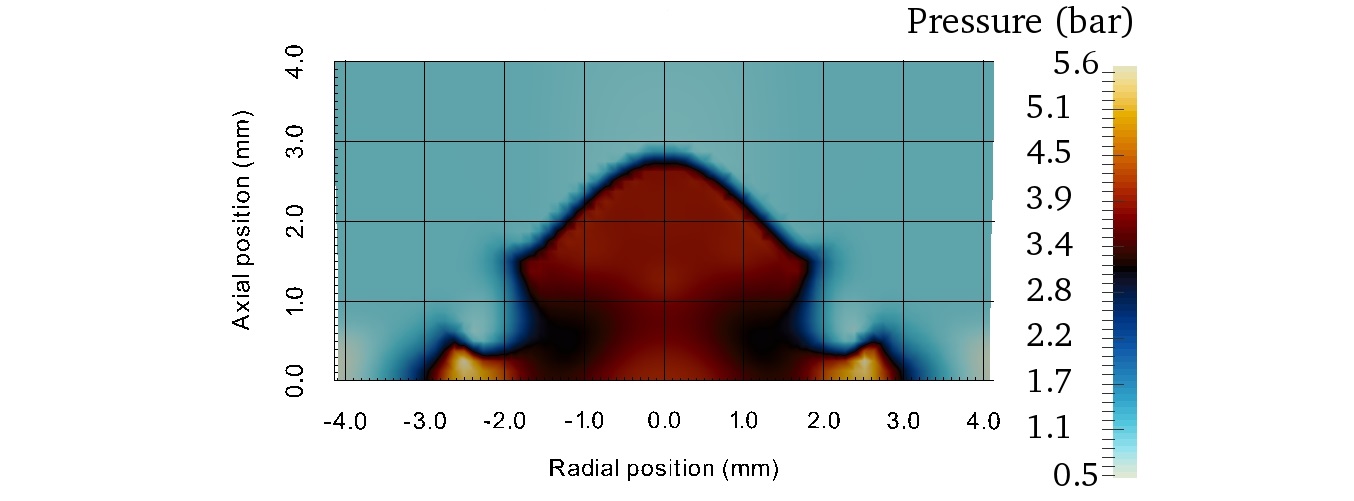}
\par\end{centering}
}\subfloat[{Configuration C at $t=\unit[30]{ms}$}]{\begin{centering}
\includegraphics[viewport=165.0812bp 0bp 862.925bp 376bp,clip,width=7cm]{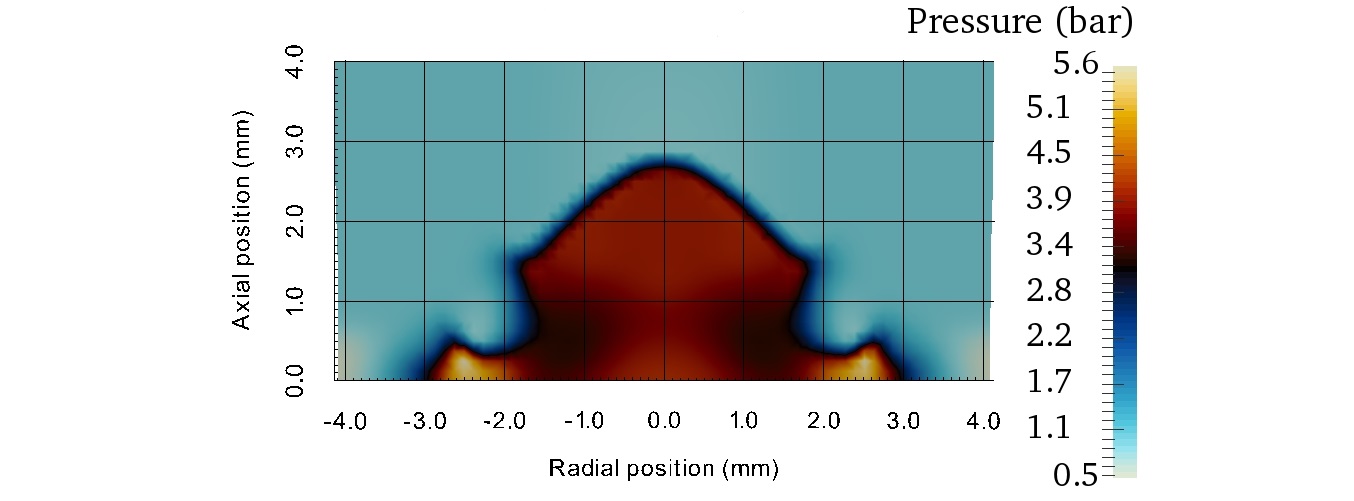}
\par\end{centering}
}
\par\end{centering}
\caption{Pressure for configuration A to C with the IDDES model at different
time steps near the substrate. The substrate is at bottom of each
image and the axial position corresponds to the distance from the
substrate.\label{fig:Pressure-for-configuration-1}}
\end{sidewaysfigure}

Compared to the RANS model, the IDDES model allows to catch the phenomena
occurring near the substrate like the bow shock. This shock, visible
in \prettyref{fig:Pressure-for-configuration-1} in terms of pressure\footnote{the same behavior exists for density with the same relative orders
of magnitude}, creates a huge deceleration of the flow which impacts directly the
particles coming onto the substrate. Because the IDDES model is less
dissipative, the bow shock near the substrate is strong and has a
more important effect on the small particles because of their lower
inertia. The shape of the bow shock depends on the topology of the
whole flow as presented by the \prettyref{fig:Pressure-for-configuration-1},
but the global phenomenon is pretty similar between configuration
A to C.

\subsubsection{Turbulent kinetic energy}

\begin{figure}
\begin{centering}
\includegraphics[width=15cm]{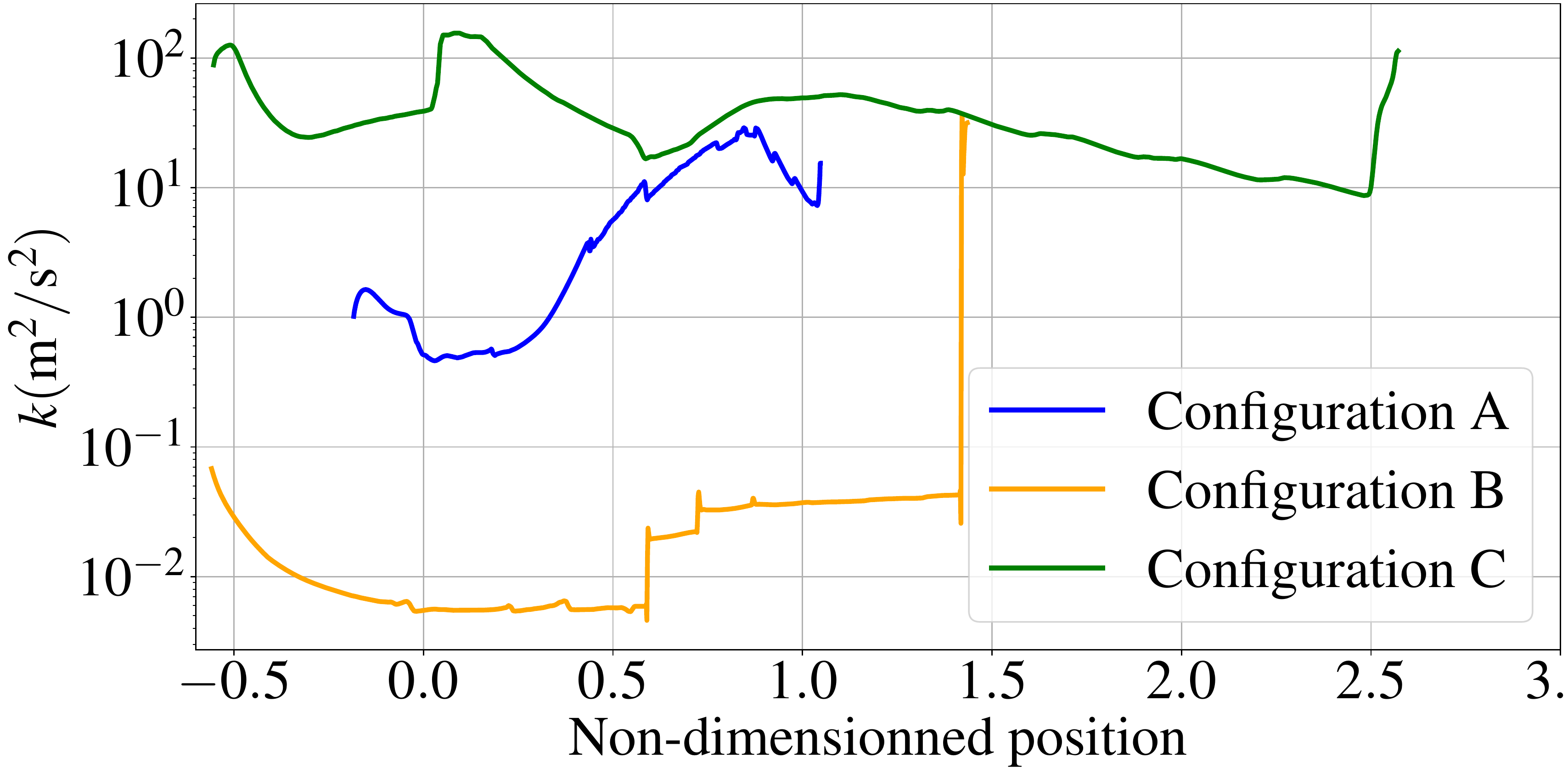}
\par\end{centering}
\caption{$k$ as a function of the dimensionless position along the nozzle
on the axis of symmetry from IDDES model on configuratons A to C.
The dimensionless position corresponds to the position divided by
$L$. Dimensionless position 0 corresponds to the nozzle throat\label{fig:-as-a-2}}
\end{figure}

One of the steps to compare both models is to compare $k$ on the
axis of symmetry as shown in \prettyref{fig:-as-a-2}. Talking only
about the differences, $k$ obtained with the IDDES is clearly lower
than $k$ got by the RANS model. The lines follow approximately the
same path that is to say there are constant in the diverging part
and increase suddenly where there is a shock.

\subsubsection{Pressure and turbulent fluctuations}

\begin{figure}
\begin{centering}
\includegraphics[width=15cm]{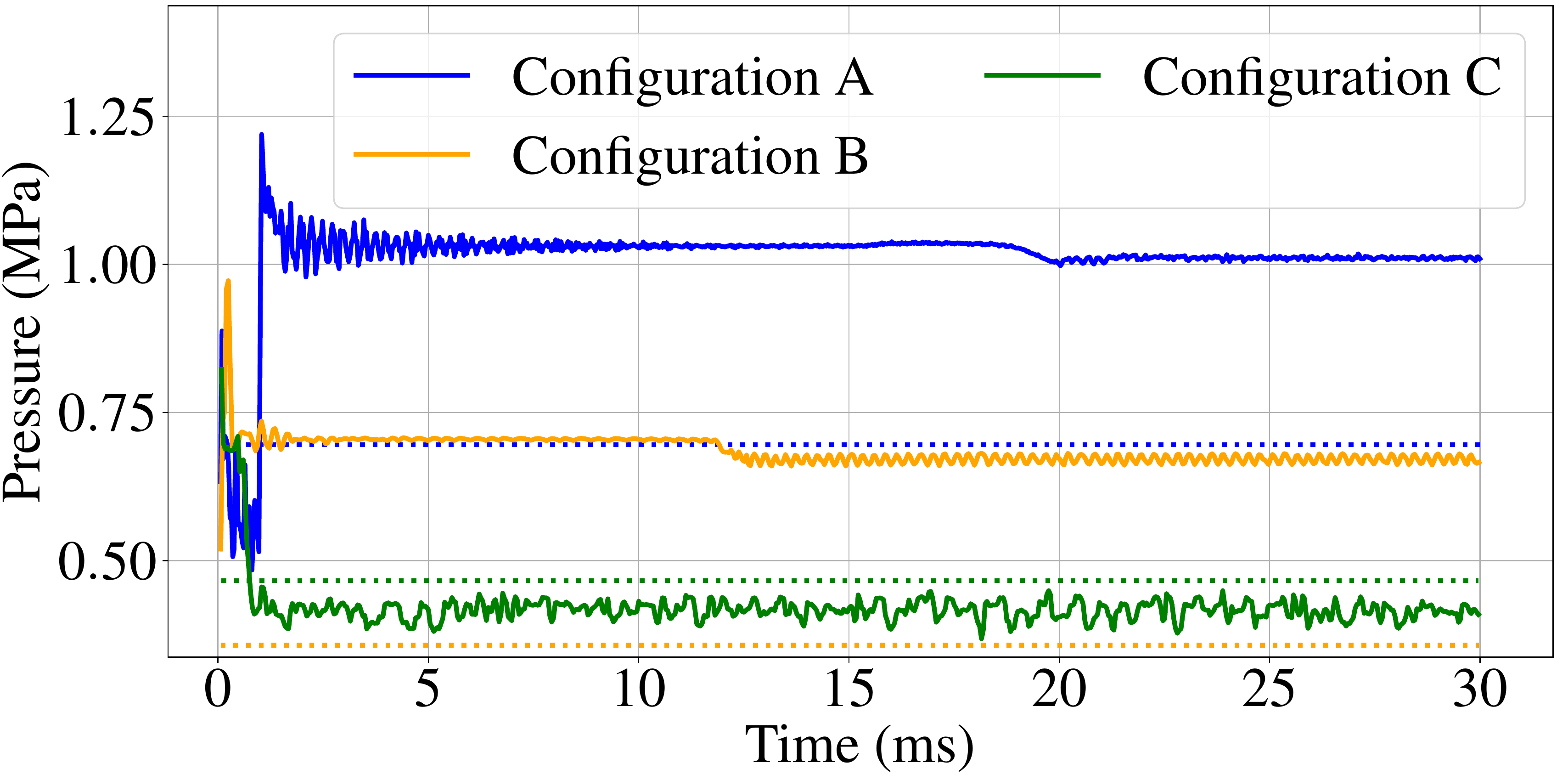}
\par\end{centering}
\caption{Pressure as a function of time at the intersection of the axis of
symmetry and the substrate. The solid curves correspond to the IDDES
model and the dotted lines correspond to the RANS value constant along
time. \label{fig:Pressure-as-a}}
\end{figure}

Furthermore, we propose in \prettyref{fig:Pressure-as-a} to compare
the pressure at the intersection of the axis of symmetry and the substrate
between the two models. Generally, the pressure obtained with the
IDDES model is higher than the one resulting from the RANS model except
for configuration C. It must be related to the less dissipative turbulent
effects modeled by the IDDES model compared to the RANS model. Also,
the fluctuations due to the turbulence is clearly visible of the solid
lines. These fluctuations oscillate around an average value and there
is a change in the evolution of the pressure from $t=\unit[20]{ms}$
for configuration A and $t=\unit[12]{ms}$ for configuration B. This
change is related to the arrival of huge amount of particles. In fact,
the first particles start to fly at $t=\unit[0]{ms}$ and must travel
along the nozzle before reaching the substrate. Therefore, because
there is a two-way coupling between the flow and the particles, the
effect of the particles on the turbulence appears when there are enough
particles colliding with the substrate.

\subsubsection{Particle tracks}

\begin{sidewaysfigure}
\begin{centering}
\subfloat[{Configuration A at $t=\unit[10]{ms}$}]{\begin{centering}
\includegraphics[viewport=0bp 30bp 1354bp 350bp,clip,width=7cm]{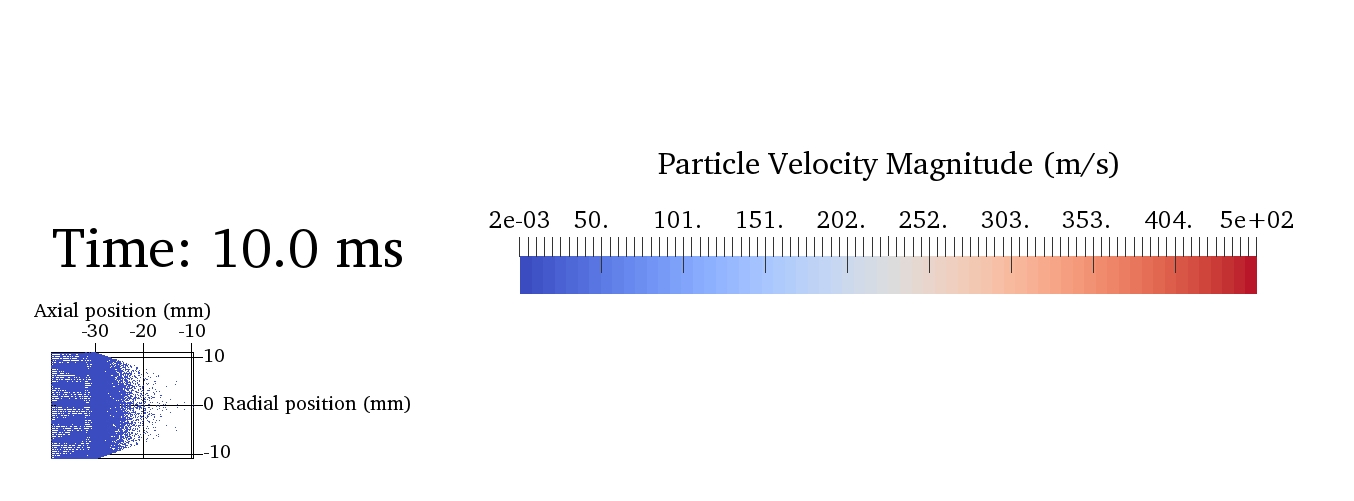}
\par\end{centering}
}\subfloat[{Configuration A at $t=\unit[20]{ms}$}]{\begin{centering}
\includegraphics[viewport=0bp 30bp 1354bp 350bp,clip,width=7cm]{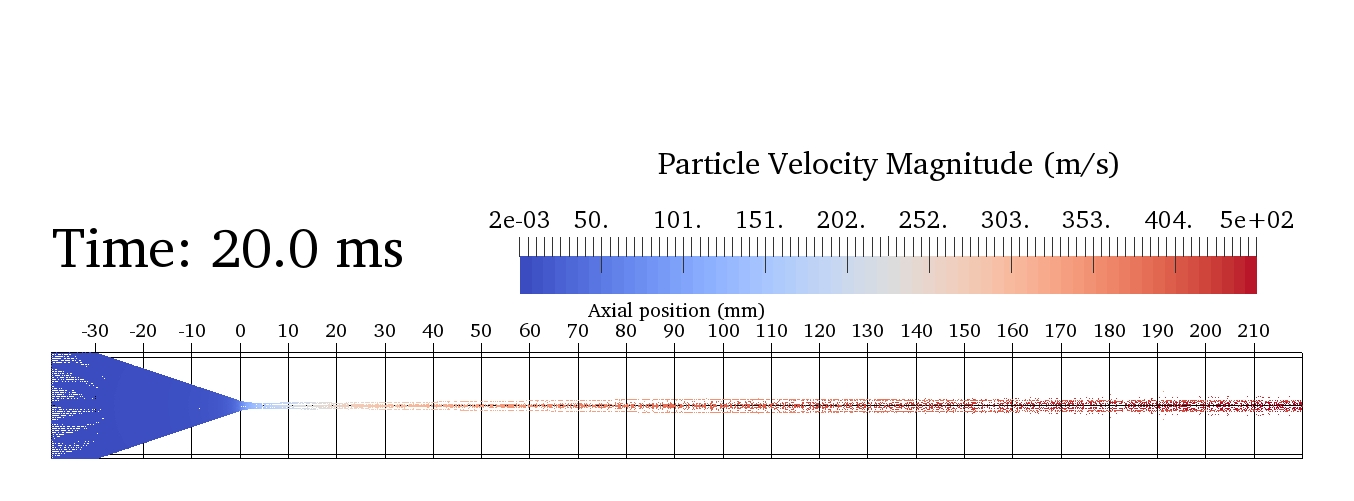}
\par\end{centering}
}\subfloat[{Configuration A at $t=\unit[30]{ms}$}]{\begin{centering}
\includegraphics[viewport=0bp 30bp 1354bp 350bp,clip,width=7cm]{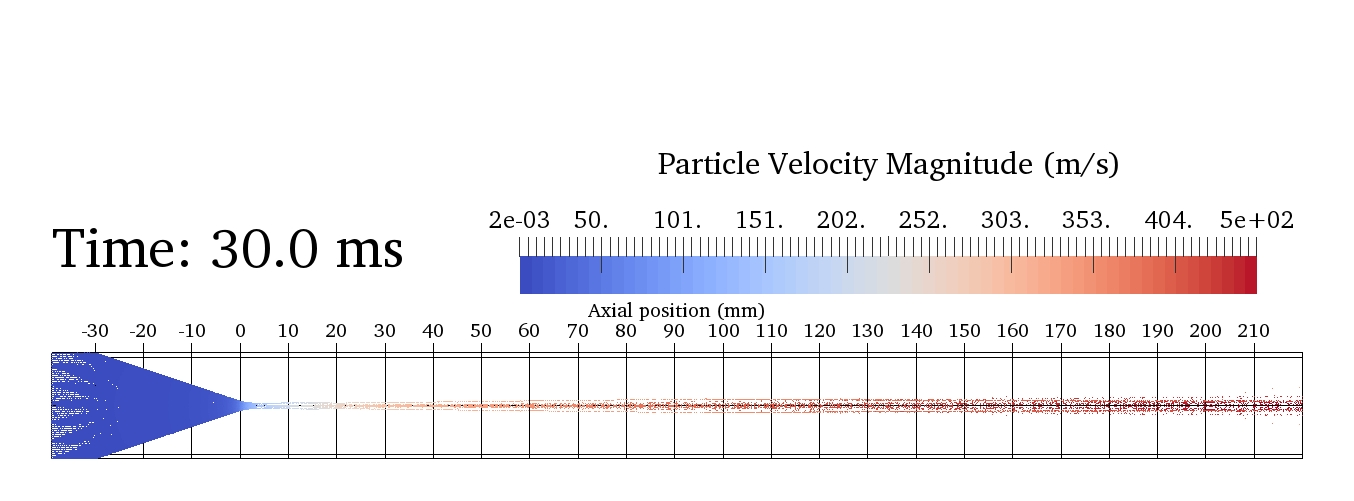}
\par\end{centering}
}
\par\end{centering}
\begin{centering}
\subfloat[{Configuration B at $t=\unit[10]{ms}$}]{\begin{centering}
\includegraphics[viewport=0bp 30bp 1354bp 350bp,clip,width=7cm]{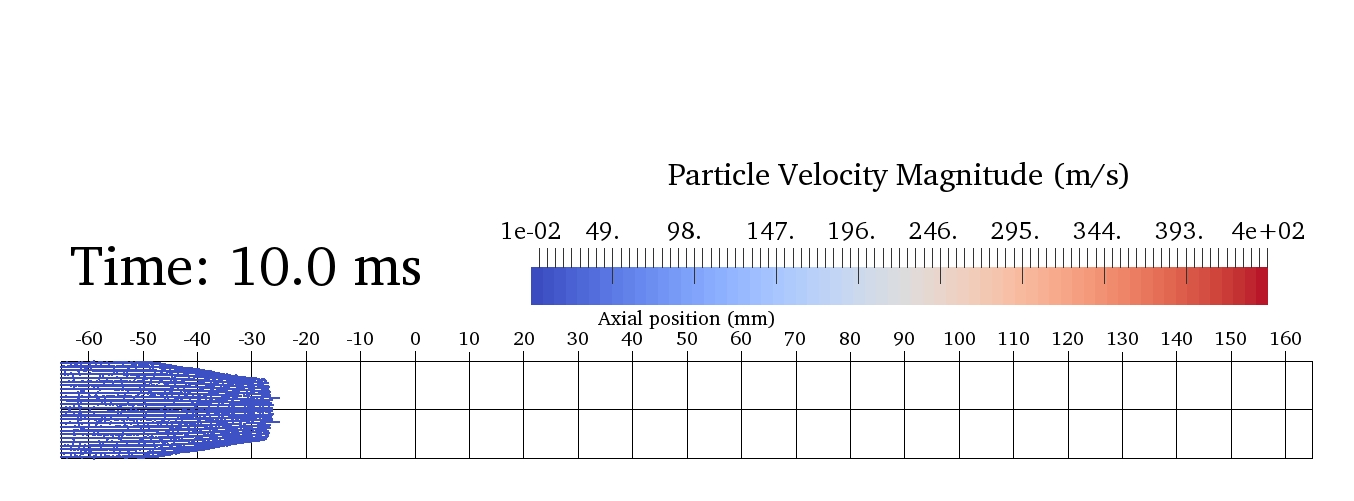}
\par\end{centering}
}\subfloat[{Configuration B at $t=\unit[20]{ms}$}]{\begin{centering}
\includegraphics[viewport=0bp 30bp 1354bp 350bp,clip,width=7cm]{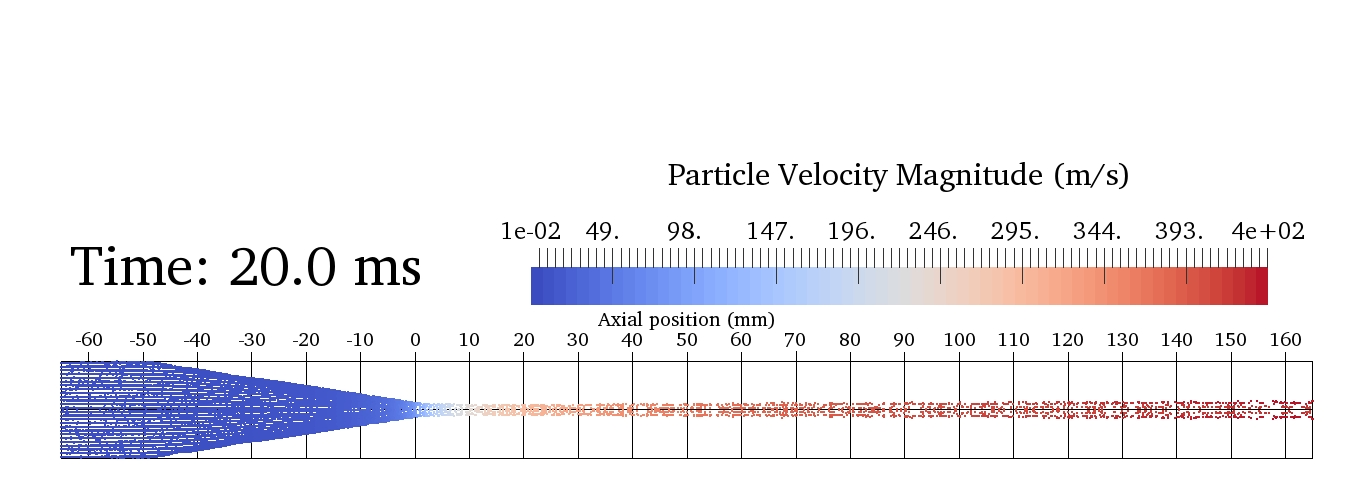}
\par\end{centering}
}\subfloat[{Configuration B at $t=\unit[30]{ms}$}]{\begin{centering}
\includegraphics[viewport=0bp 30bp 1354bp 350bp,clip,width=7cm]{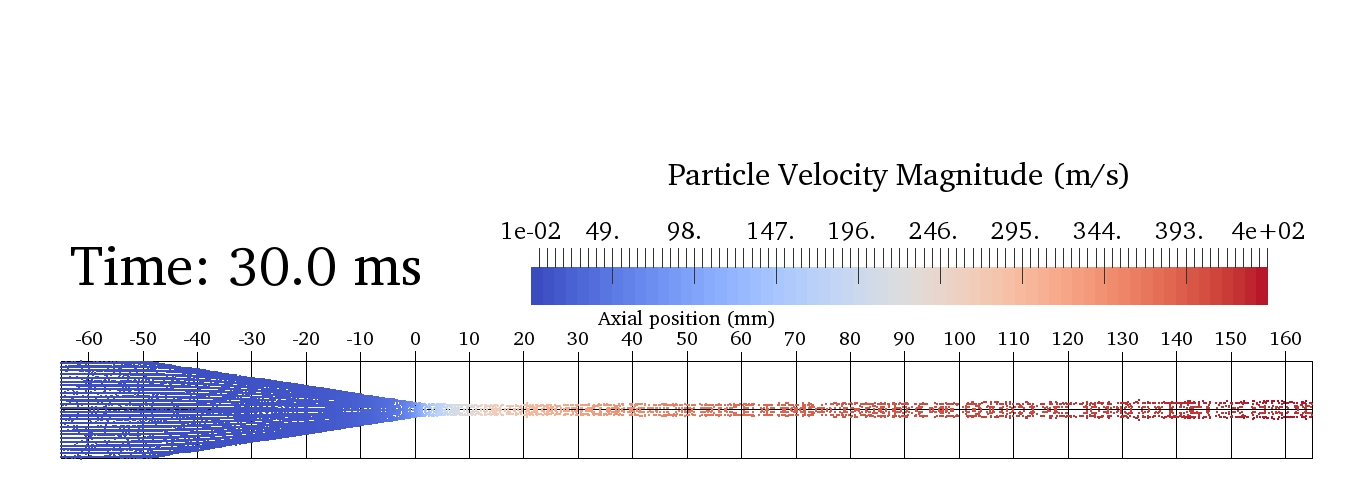}
\par\end{centering}
}
\par\end{centering}
\begin{centering}
\subfloat[{Configuration C at $t=\unit[10]{ms}$}]{\begin{centering}
\includegraphics[viewport=0bp 60bp 1354bp 300bp,clip,width=7cm]{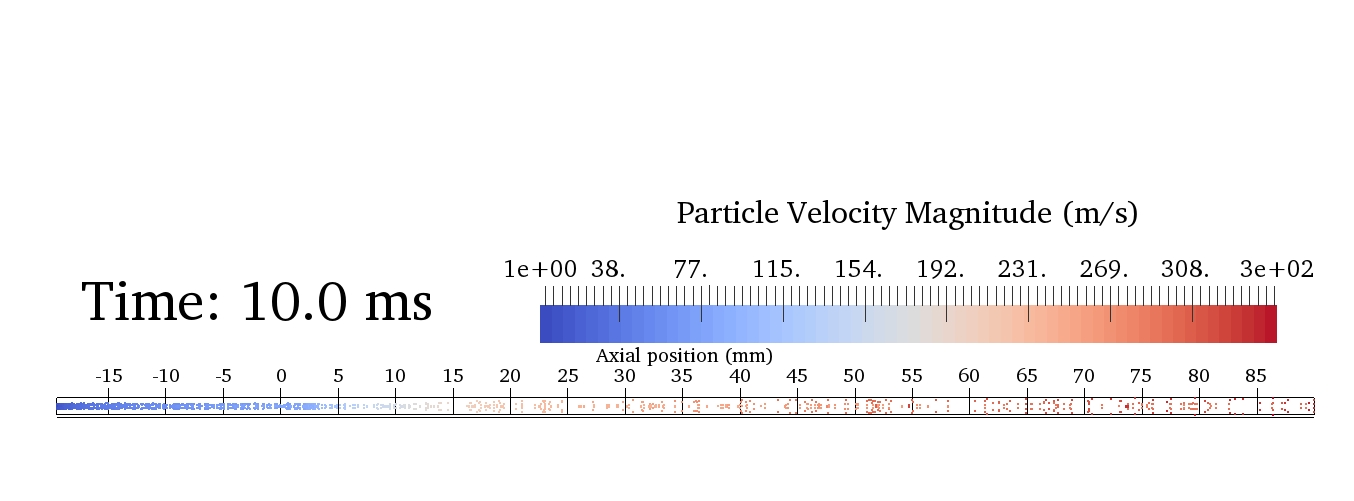}
\par\end{centering}
}\subfloat[{Configuration C at $t=\unit[20]{ms}$}]{\begin{centering}
\includegraphics[viewport=0bp 60bp 1354bp 300bp,clip,width=7cm]{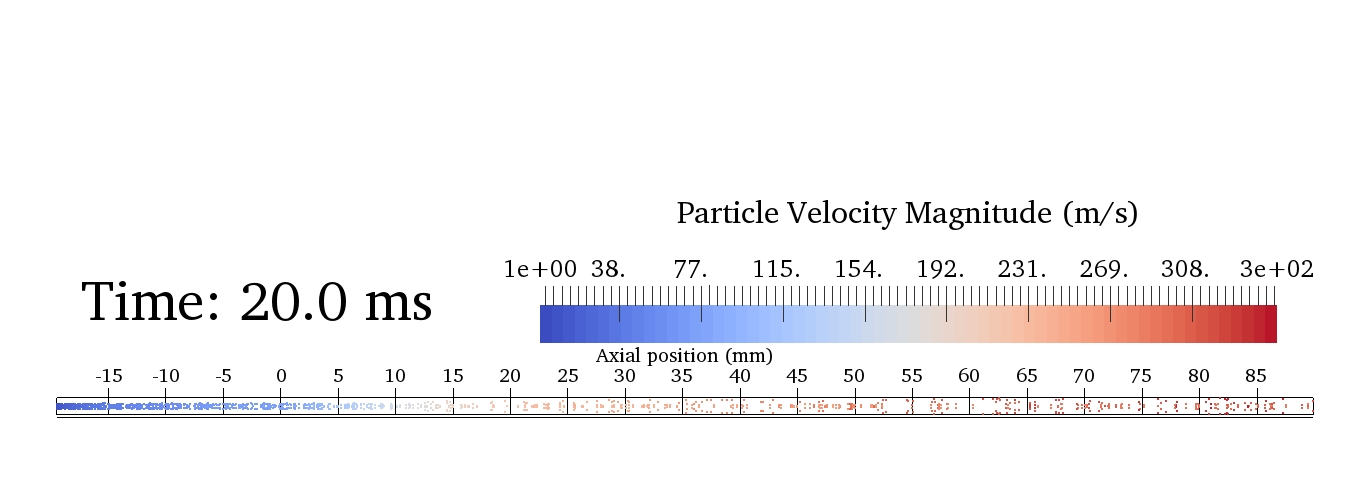}
\par\end{centering}
}\subfloat[{Configuration C at $t=\unit[30]{ms}$}]{\begin{centering}
\includegraphics[viewport=0bp 60bp 1354bp 300bp,clip,width=7cm]{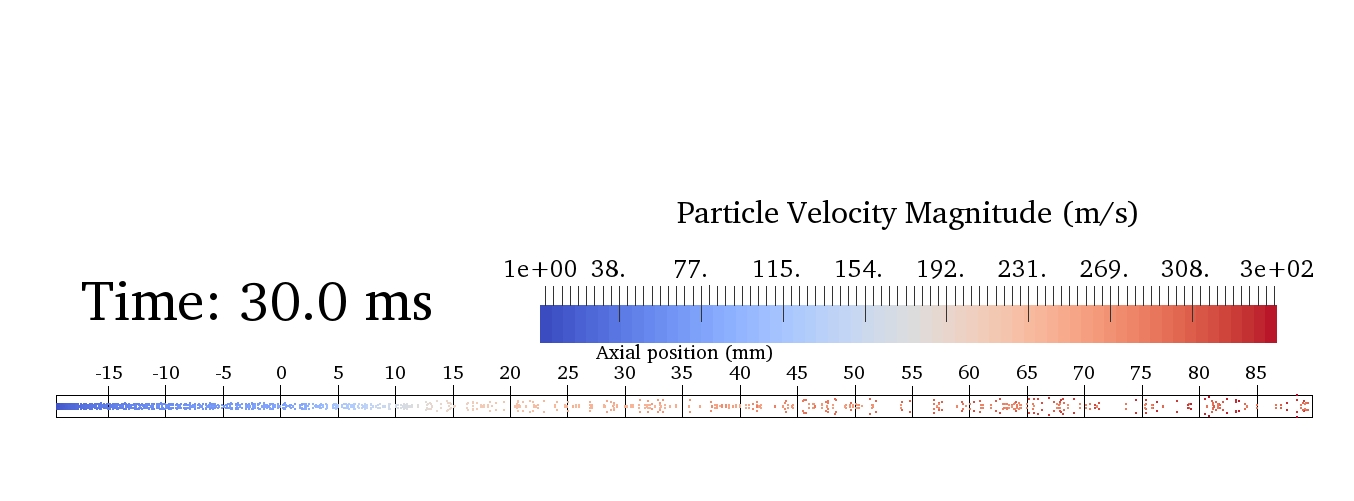}
\par\end{centering}
}
\par\end{centering}
\caption{Particle velocity magnitude for configuration A to C with IDDES model
at different time steps. \label{fig:Particle-velocity-magnitude-2}}
\end{sidewaysfigure}

\prettyref{fig:Particle-velocity-magnitude-2} shows the particles
coloured according to their velocity magnitude. It is visible that
the particles have not reached yet the substrate at $t=\unit[10]{ms}$
for configuration A and B which was confirmed by \prettyref{fig:Pressure-as-a}
with the fluctuations of pressure on the substrate. The configuration
C has already particles onto the substrate at $t=\unit[10]{ms}$ as
the fluctuations in \prettyref{fig:Pressure-as-a} have shown. The
particles are quite organized in the convergent but tend to collide
and to deviate from the throat. From this point, the turbulent effects
are strong and produce a lot of mixing in the particles, as expected.
The distribution of the particles along the radial position is then
strongly dependent on the path of each particle which will be discussed
afterwards.

\subsubsection{Particle impact velocity}

\begin{figure}
\begin{centering}
\includegraphics[width=12cm]{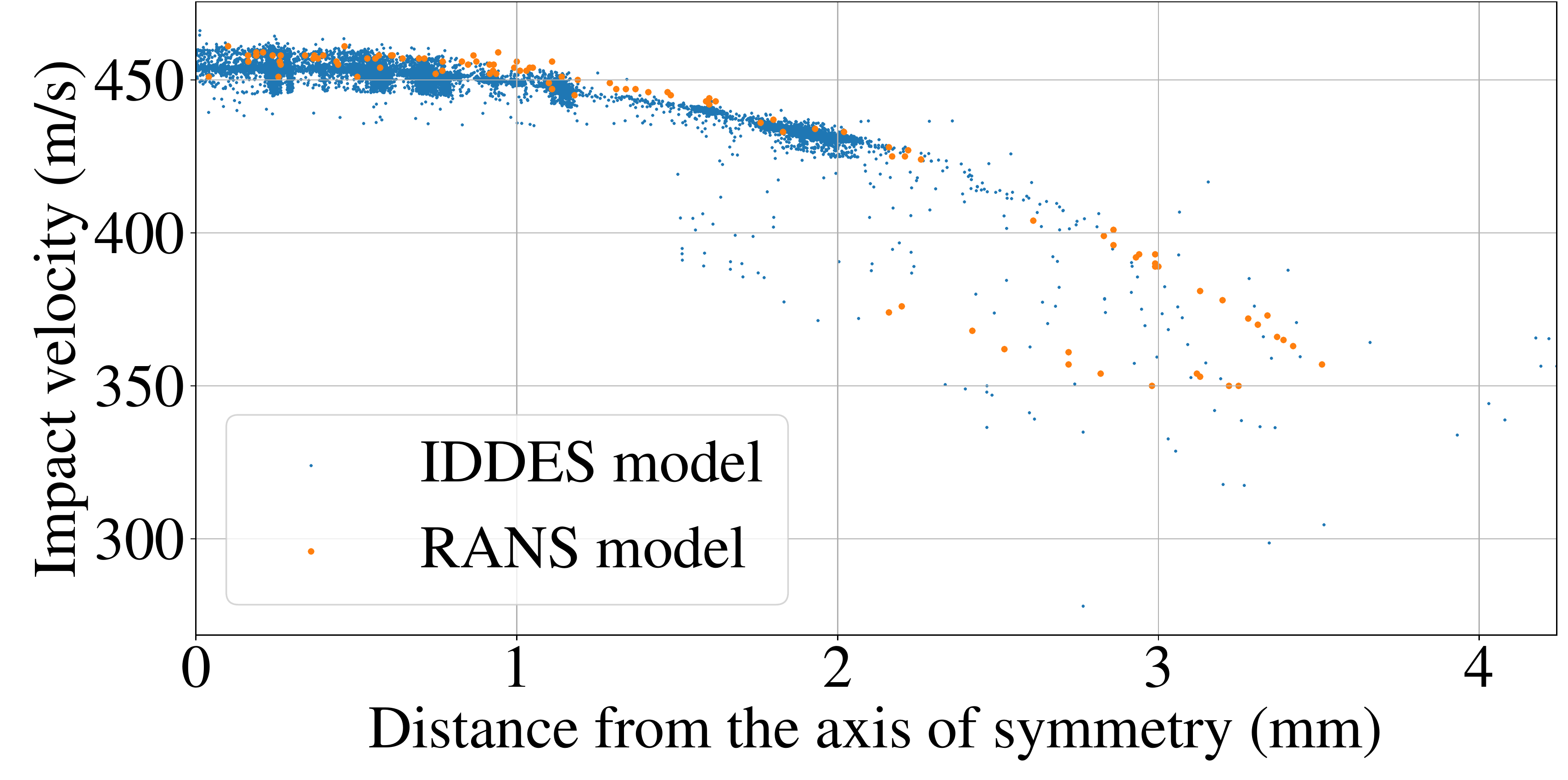}
\par\end{centering}
\caption{Particle impact velocity as a function of the distance from the axis
of symmetry for configuration A with IDDES model and RANS model\label{fig:Particle-impact-velocity-1}}
\end{figure}

\begin{comment}
\begin{figure}
\begin{centering}
\subfloat[Configuration A]{\begin{centering}
\includegraphics[width=12cm]{ArXivImages/Particleimpact,Lupoi,config1,RANSIDDES,english.pdf}
\par\end{centering}
}
\par\end{centering}
\begin{centering}
\subfloat[Configuration B]{\begin{centering}
\includegraphics[width=12cm]{ArXivImages/Particleimpact,Lupoi,config2,RANSIDDES,english.pdf}
\par\end{centering}
}
\par\end{centering}
\begin{centering}
\subfloat[Configuration C]{\begin{centering}
\includegraphics[width=12cm]{ArXivImages/Particleimpact,Lupoi,config4,RANSIDDES,english.pdf}
\par\end{centering}
}
\par\end{centering}
\caption{Particle impact velocity as a function of the distance from the axis
of symmetry for configuration A to C with IDDES model and RANS model\label{fig:Particle-impact-velocity-1-2}}
\end{figure}
\end{comment}

Finally, to compare the two models, we can present the \prettyref{fig:Particle-impact-velocity-1}
showing the particle impact velocity as a function of the distance
from the axis of symmetry according to the IDDES model and the RANS
model. On configuration A, the impacts of the particles are really
similar in terms of distribution of velocity and radial position.
Thanks to the transient simulation, the IDDES model has succeeded
to capture accurately the distribution of particles. On configuration
B, which slightly differs from the results of (Ref \cite{Lupoi2011}),
the impacts of particles are rather different. One of the reasons
can be the strong presence of the bow shock near the substrate as
illustrated by \prettyref{fig:Pressure-for-configuration-1}. Inside
the bow shock, the pressure and the density of the fluid are much
higher and the bow shock is deeper near the center. Hence, the particles
close to the center spend more time in a fluid which decelerates them
and have a lower impact velocity than those far from the center with
a thinner bow shock. Compared to the RANS model which does not show
any bow shock for configuration B, the distribution of velocity is
less uniform but the width of the band is kept thanks to the similar
flow topology between both models. On configuration C, this discussion
is close to the one for configuration B. The presence of the bow shock
in the IDDES model creates a gradient of impact velocity from the
center because the bow shock gets deeper when approaching the center.
The width of the track seems higher because of the strong turbulent
effects in the flow. Nevertheless, the IDDES simulation has succeeded
to catch the width of the track given by the experimental results
of (Ref \cite{Lupoi2011}) which gives more credit again to the
IDDES model against the RANS model.

As a partial conclusion, the IDDES model is confirmed to be a more
accurate simulation than the RANS model which gives more insight of
the phenomena appearing in cold spray such as turbulence, oblique
shocks, bow shocks, fluctuations, particles motion and particles impacts.
All these results can be used now to assess the performances of the
last new proposed configuration.

\subsection{Configuration D}

While relying on the obtained results with the configurations A to
C modeled first with the RANS model and then with the IDDES model,
it is now possible to propose and to study the new configuration D.
The topology of the flow is rather similar in terms of velocity magnitude,
series of shocks, bow shock near the substrate, etc. Nevertheless,
there is a strong difference for the particles and their impact on
the substrate which are going to be discussed in the following.

\subsubsection{Particle tracks}

\begin{figure}
\begin{centering}
\includegraphics[viewport=80bp 200bp 1245bp 385bp,clip,width=15cm]{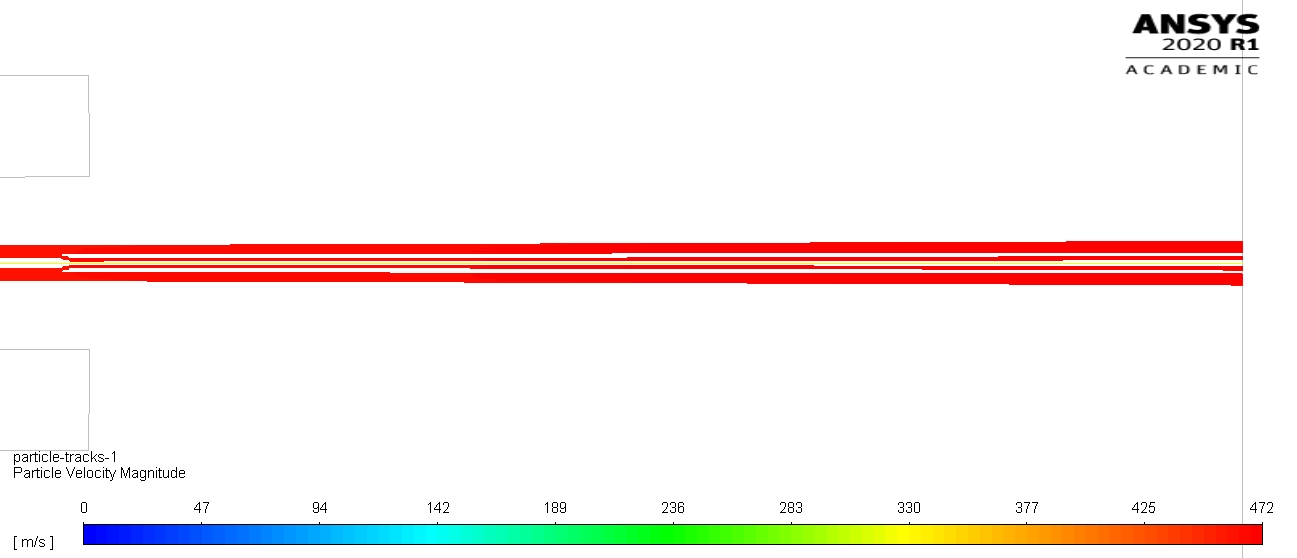}
\par\end{centering}
\begin{centering}
\includegraphics[width=15cm]{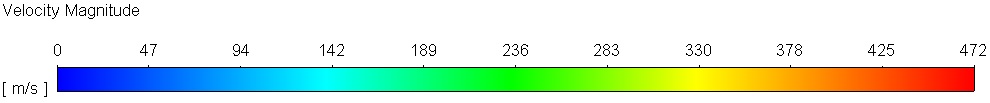}
\par\end{centering}
\caption{Particle tracks colored according to their velocity magnitude between
the exit of the nozzle and the substrate for configuration D with
RANS model. The scale is $15/4$.\label{fig:Particles-track-colored}}
\end{figure}

\begin{figure}
\begin{centering}
\subfloat[{$t=\unit[10]{ms}$}]{\begin{centering}
\includegraphics[viewport=0bp 160bp 1354bp 450bp,clip,width=16cm]{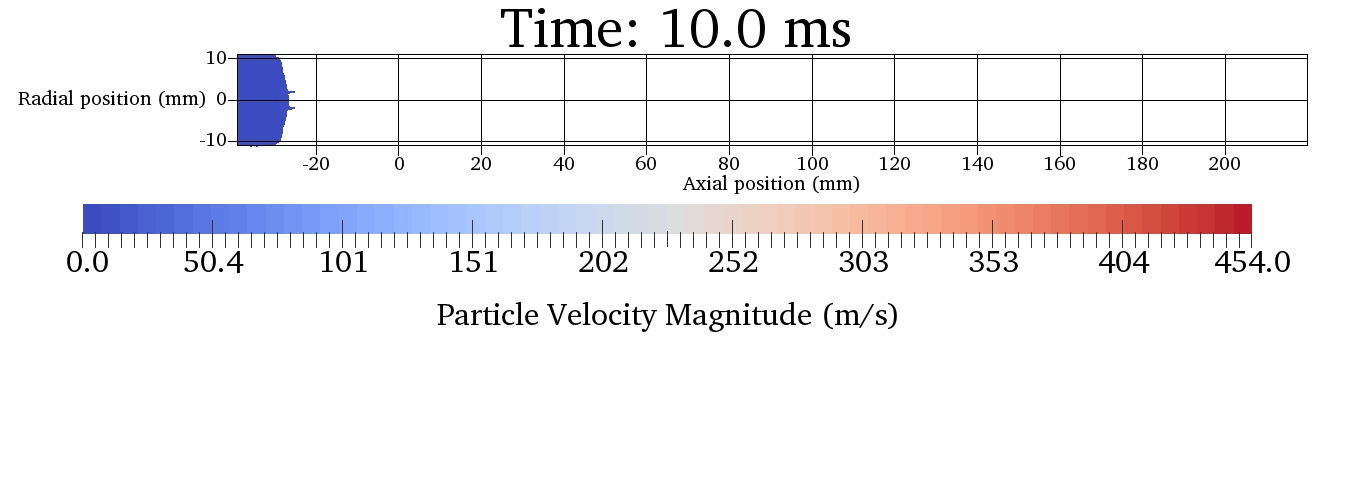}
\par\end{centering}
}
\par\end{centering}
\begin{centering}
\subfloat[{$t=\unit[20]{ms}$}]{\begin{centering}
\includegraphics[viewport=0bp 160bp 1354bp 450bp,clip,width=16cm]{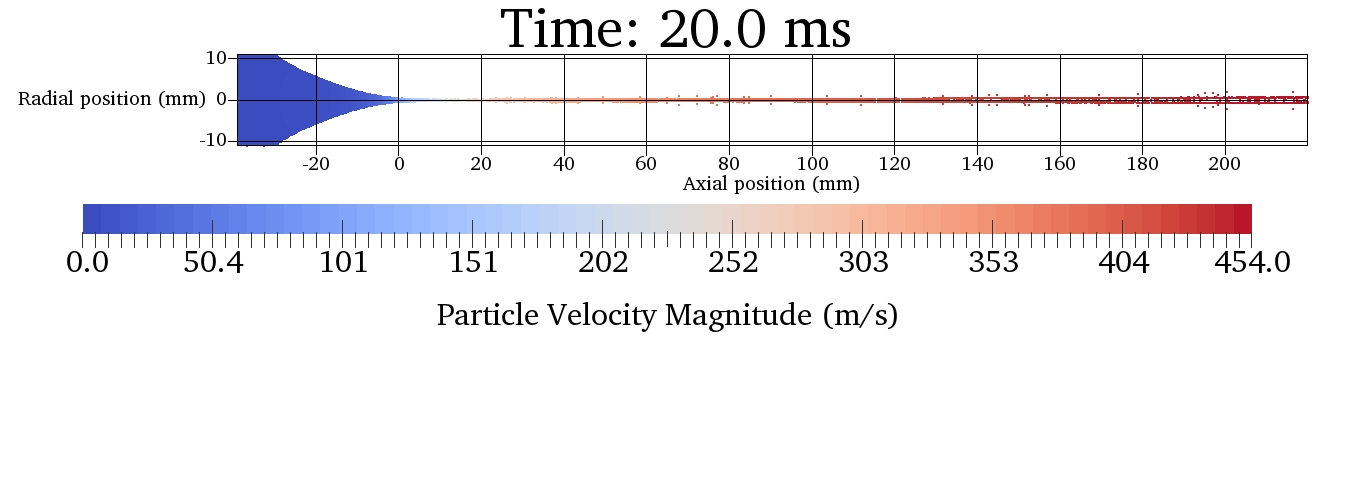}
\par\end{centering}
}
\par\end{centering}
\begin{centering}
\subfloat[{$t=\unit[30]{ms}$}]{\begin{centering}
\includegraphics[viewport=0bp 160bp 1354bp 450bp,clip,width=16cm]{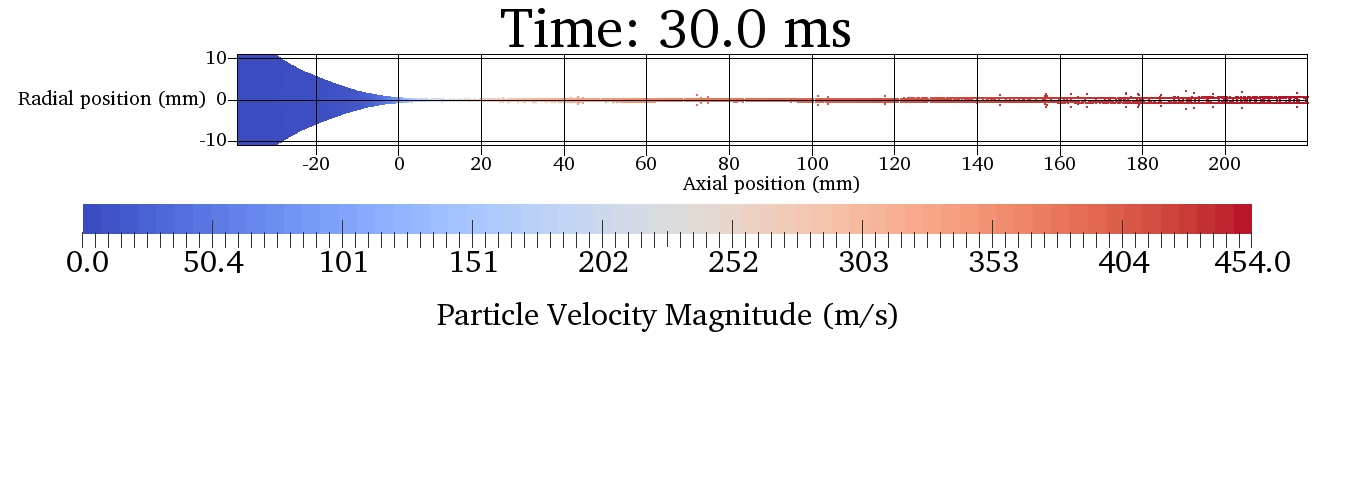}
\par\end{centering}
}
\par\end{centering}
\caption{Particle velocity magnitude for configuration D with IDDES model at
different time steps.\label{fig:Particle-velocity-magnitude-1}}
\end{figure}

\prettyref{fig:Particles-track-colored} shows the particle tracks
coloured according to the velocity magnitude of the particles with
RANS model. Comparing this results with \prettyref{fig:Velocity-magnitude-between-3},
the velocity magnitude of the particles seems rather similar between
both pictures. However, the width of the particle jet is much narrower.
Thanks to the convex convergent, the particles did not suffer of the
strong turbulence occurring at the throat. The convergent has succeeded
to maintain the jet as narrow as possible. Also, because the particles
remain in the center of the flow, they take advantage of the maximum
flow velocity to accelerate. Then, considering only the RANS model,
configuration D seems to be a good improvement to carry on accurate
cold spraying on the substrate.

\prettyref{fig:Particle-velocity-magnitude-1} shows the particles
coloured according to their velocity magnitude at different time steps
obtained using the IDDES model. Again, comparing \prettyref{fig:Velocity-magnitude-between-3}
and \prettyref{fig:Particle-velocity-magnitude-1}, the behavior remains
the same but the width of the tracks looks narrower for configuration
D. The orders of magnitude for the velocity are pretty similar and
do not change between the configurations or the models.

\subsubsection{Particle impact velocity}

\begin{figure}
\begin{centering}
\includegraphics[width=15cm]{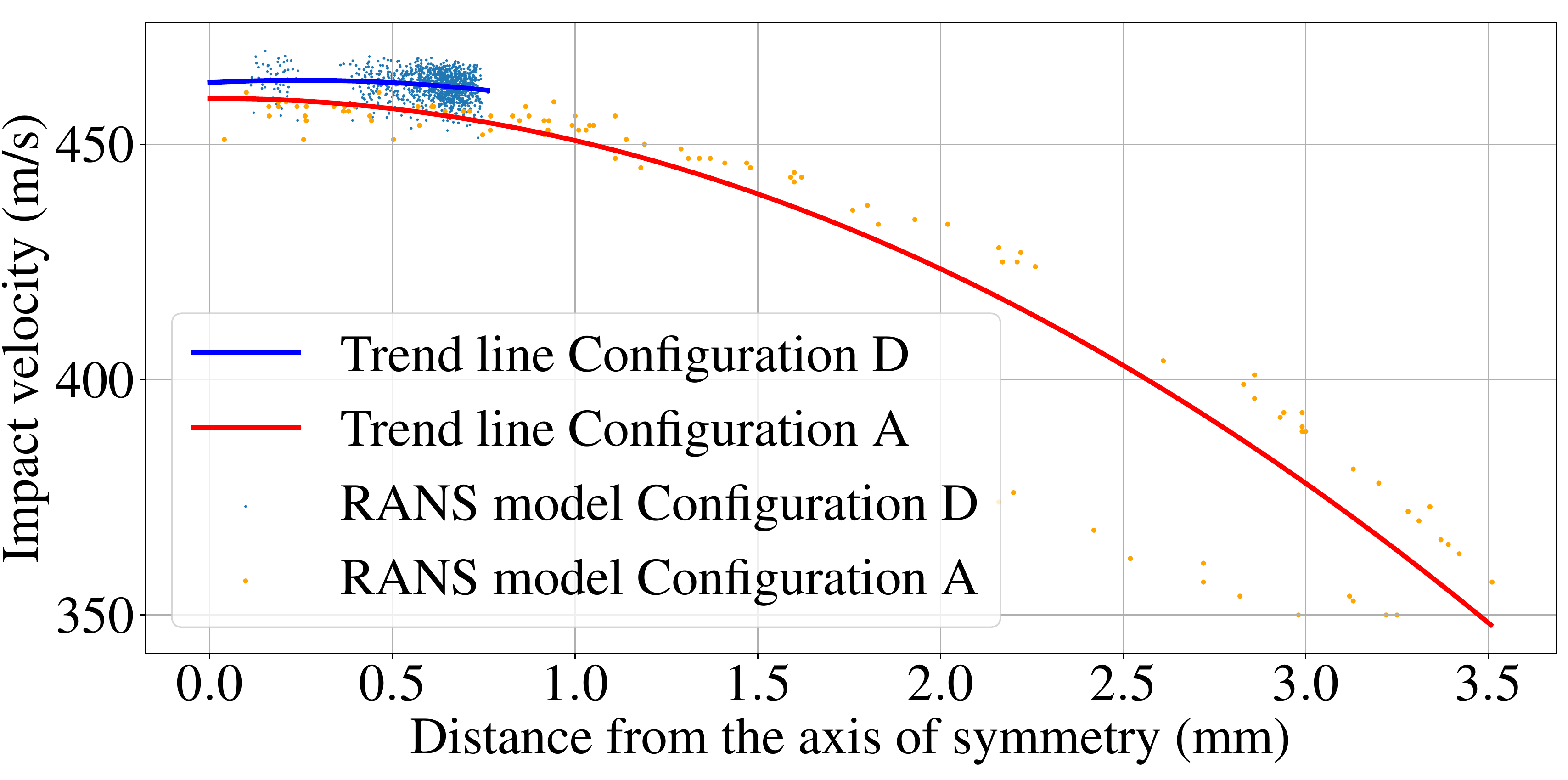}
\par\end{centering}
\caption{Particle impact velocity as a function of the distance from the axis
of symmetry for configuration A and D with RANS model\label{fig:Particle-impact-velocity-1-1}}
\end{figure}

\begin{figure}
\begin{centering}
\includegraphics[width=15cm]{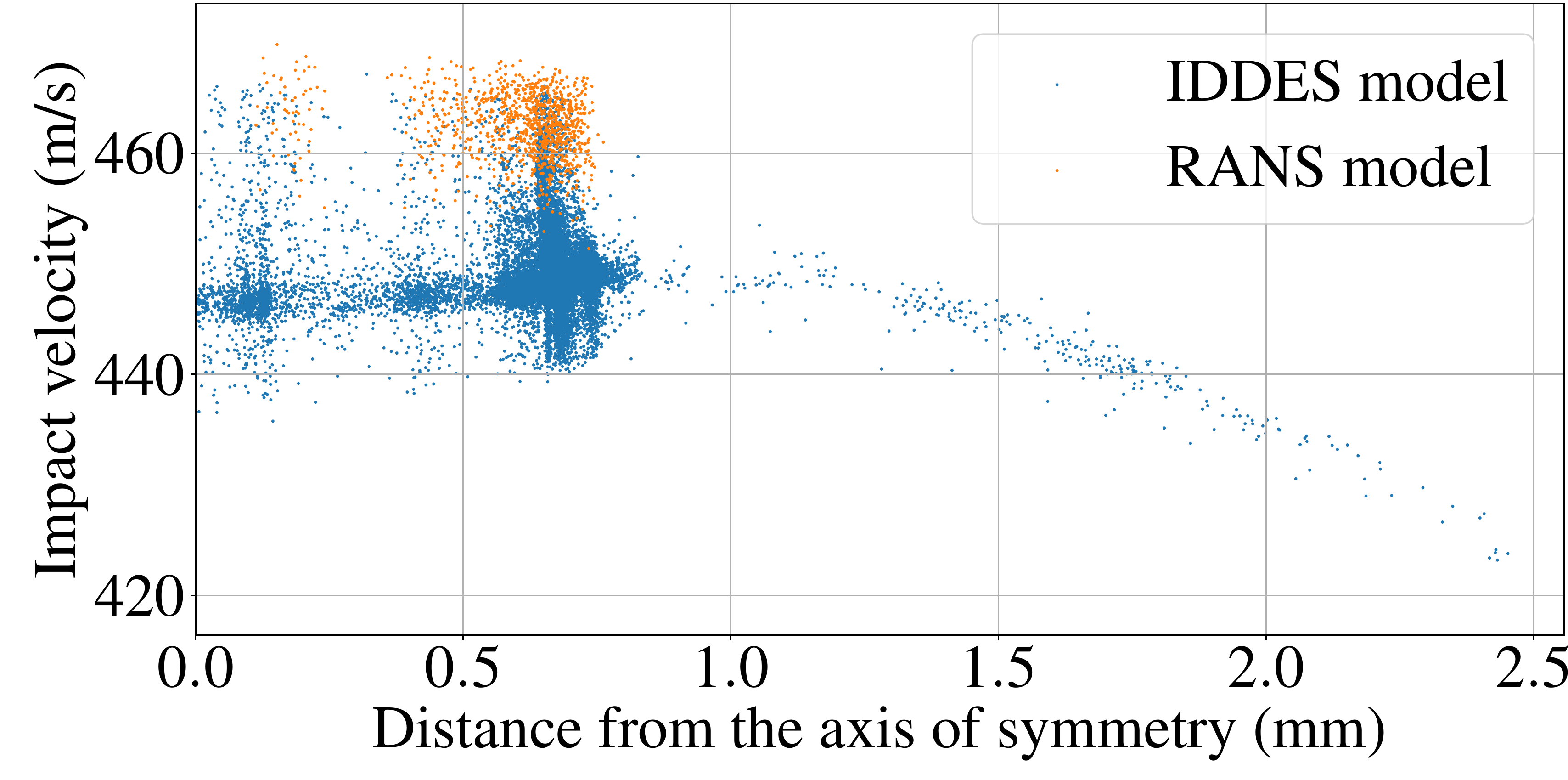}
\par\end{centering}
\caption{Particle impact velocity as a function of the distance from the axis
of symmetry for configuration D for IDDES model and RANS model\label{fig:Particle-impact-velocity-2}}
\end{figure}

To carry on an accurate comparison of the configurations A and D,
it is now interesting to plot the evolution of the particle impact
velocity versus the distance from the axis of symmetry. These data
are given in \prettyref{fig:Particle-impact-velocity-1-1} for RANS
model. It is clear that the particles coming from configuration D
have a higher impact velocity than those of configuration A with RANS
model with a much narrower width of the track on the substrate. The
comments given previously are verified here thus the configuration
D gives improved results for the impact velocity magnitude and the
width of the track according to RANS model. However, the particles
are not distributed as usual: the particles are more present near
$\unit[0.5]{mm}$ from the center than in the center. The flow creates
a hole of particles in the center which can be problematic under certain
conditions. It is important to point out that these conclusions are
valid in the context of these simulations. Hence, using a real powder
with different sizes of particles or adding some fluctuations in the
model will change drastically the results. That is why a comparison
with the IDDES model helps to understand.

Therefore, in \prettyref{fig:Particle-impact-velocity-2}, the evolution
of the particle impact velocity versus the distance from the axis
of symmetry is given with the RANS model and the IDDES model. The
particle impact velocity is a little bit lower with the IDDES model
and reaches the same values as in (Ref \cite{Lupoi2011}). Nevertheless,
the width of the track and the distribution of particles is kept with
the RANS model. There are more particles in the center than in the
RANS model which avoid the hole presented before. Some particles appear
outside the track due to turbulent effects but the main track is much
narrower than the one given by configuration A. With the IDDES model,
because a cold spray system is used with a certain motion to create
a coating, the small number of particles outside the thin track will
not appear significantly and the user will get a narrower track with
similar performance in terms of solidity, porosity or resistance.

Both models conclude that configuration D has an improved performance
against configuration A in terms of width of track. Because the particle
impact velocities are kept, the adhesion of the particles remains
the same but with a narrower track that is to say a more accurate
coating system.

\section{Conclusion}

\label{sec:Conclusion}

In this paper, a new high fidelity modeling for the cold spray process
is presented. A review on several existing methods is also proposed
showing the capability of the new model by giving more insight of
the phenomena appearing in cold spray such as turbulence, oblique
shocks, bow shocks, fluctuations, particles motion and particles impacts.
Several test cases and comparisons were presented and discussed. The
improvements of this proposed model are clearly visible on several
fields of the problem from the topology of the flow, the fluctuations
in time to the two-way coupled influence of the particles.

These results underline the potential of this approach, and shall
be pursued for new configurations. The extension of the current work
to three-dimensional can also be considered, using parallel computing
and mesh adaptation. Also, exploiting a Large Eddy Simulation (LES)
may lead to even better results.

\bibliographystyle{plain}

\end{document}